\documentclass[aps,showpacs,preprint,epsf,epsfig, nofootinbib]{revtex4-2}
\usepackage{mathrsfs}
\usepackage{float}
\usepackage{dcolumn}
\usepackage{bm}
\usepackage{hyperref}
\usepackage{epsfig,graphicx,color,appendix}
\usepackage{placeins}
\usepackage{multirow}
\usepackage{tabularx}
\usepackage{capt-of}
\usepackage{makecell} 
\usepackage{booktabs}
\usepackage{subcaption}
\usepackage[countmax]{subfloat}

\usepackage{threeparttable}
\usepackage{array}
\usepackage{amsmath}
\usepackage[format=hang]{caption}

\graphicspath{ ./Figures/ }

\usepackage[english]{babel}[2020/02/03]
\makeatletter\AtBeginDocument{\let\@elt\relax}\makeatother
\begin{document}
	\count\footins=1000
	\def\ben{\begin{eqnarray}}
		\def\ben{\end{eqnarray}}
	\def\non{\nonumber}
	\def\obtp{\frac{1^{\prime}}{2}}
	\def\la{\langle}
	\def\ra{\rangle}
	\def\t{\times}
	\def\ve{\varepsilon}
	\def\p{{\prime}}
	\def\pp{{\prime\prime}}
	\def\nc{N_c^{\rm eff}}
	\def\hep{\hat{\varepsilon}}
	\def\J{{J/\psi}}
	\def\ov{\overline}
	\def\vp{\varepsilon}
\def\vma{{_{V-A}}}
\def\vpa{{_{V+A}}}
\def\m{\hat{m}}
\def\ov{\overline}
	\def\q2{q^2}
	\def\kb{{\bf k}_\bot}
	\def\pb{{\bf p}_\bot}
	\def\e{\epsilon}
	\def \d {{\rm d}}
	\def\plus{\texttt{+}}
	\def\minus{\texttt{-}}
    
	\long\def\symbolfootnote[#1]#2{\begingroup%
		\def\thefootnote{\fnsymbol{footnote}}\footnote[#1]{#2}\endgroup}
	\def\lsim{ {\ \lower-1.2pt\vbox{\hbox{\rlap{$<$}\lower5pt\vbox{\hbox{$\sim$}
			}}}\ } }
	\def\gsim{ {\ \lower-1.2pt\vbox{\hbox{\rlap{$>$}\lower5pt\vbox{\hbox{$\sim$}
			}}}\ } }
	
	\font\el=cmbx10 scaled \magstep2{\obeylines\hfill \today}
	\vskip 1.5 cm
	
	\centerline{\large\bf Weak decays of $\bm{B_s}$ meson in self-consistent covariant light-front approach}
	
	\small
	\vskip 1.0 cm
	
	\centerline{\bf Thejus Mary S.$^{1}$\symbolfootnote[1]{\href{mailto:thejusmarys@gmail.com}{thejusmarys@gmail.com}} and Rohit Dhir$^{1}$\symbolfootnote[4]{Corresponding author: \href{mailto:dhir.rohit@gmail.com}{dhir.rohit@gmail.com}}}
	
	\centerline{\it $^{1}$Department of Physics and Nanotechnology,}
	\centerline{\it SRM Institute of Science and Technology, Kattankulathur 603203, India.}
	
    \bigskip
    \begin{center}{\large \bf Abstract}\end{center}
   We present a comprehensive study of weak transition form factors and decay observables of the $B_s$ meson in transitions to pseudoscalar ($P$) and vector ($V$) mesons. The $B_s \to P(V)$ form factors are calculated using the self-consistent covariant light-front quark model, with a model-independent $z$-series expansion calibrated to lattice QCD and phenomenological inputs for quark masses and $\beta$ parameters. This enables reliable $q^2$-resolved predictions across the full kinematic range. Based on these form factors, we predict branching ratios and angular observables, including forward-backward asymmetries, polarization fractions, and leptonic convexity parameters for semileptonic decays. Nonleptonic two-body decay rates for $B_s \to PP$ and $B_s \to PV$ modes are also computed within the standard factorization framework using the same dynamical input. Comparisons with results from lattice QCD, light-cone sum rules, and other approaches are presented, highlighting both theoretical consistency and persistent tensions with experiment.

    \newpage 	
    \section {Introduction} \label{S1}
    The study of $b-$decays is a cornerstone of contemporary flavor physics, offering stringent tests of the Standard Model (SM) and exceptional sensitivity to new physics (NP). These decays enable precise determinations of the Cabibbo-Kobayashi-Maskawa (CKM) matrix elements such as $|V_{cb}|$ and $|V_{ub}|$~\cite{PDG:2024cfk}, while deviations from the SM predictions in the unitarity of the CKM or the lepton flavor universality(LFU) would be compelling evidence for NP~\cite{HFLAV:2024ctg}. Experimental efforts at LHCb, BABAR, and Belle II continue to investigate semileptonic and nonleptonic $B_{(s)}$ decays, measuring branching fractions, differential distributions, and angular observables, including forward-backward asymmetry ($A_{\rm FB}$) and polarization fractions ($F_L$)~\cite{HFLAV:2024ctg, Harrison:2023dzh}. The LFU ratios $\mathcal{R}_D$ and $\mathcal{R}_{D^*}$ have shown tensions with SM predictions in global fits, with deviations at the $\sim 3\sigma$ level~\cite{HFLAV:2024ctg, Harrison:2023dzh,Harrison:2021tol}, though recent individual measurements demonstrate improved agreement with SM expectations~\cite{Belle:2019rba, Harrison:2021tol,Bernlochner:2017jka}. The ratios $\mathcal{R}_{D_s^{(*)}}$ remain experimentally challenging to measure precisely but provide enhanced NP sensitivity~\cite{Paolucci:2022mpj,HFLAV:2024ctg}, while $\mathcal{R}_{J/\psi}$ has been constrained by LHCb measurements~\cite{PDG:2024cfk}. A persistent tension ($<3\sigma$) exists between exclusive and inclusive $|V_{cb}|$ determinations, while exclusive extractions from $B_s \to D_s^{(*)} \ell \nu$ decays generally favor lower values~\cite{HFLAV:2024ctg, Harrison:2023dzh, LHCb:2020cyw, Martinelli:2022xir}. Angular observables, such as the longitudinal polarization fraction $F_L^{D^*}$, differ by up to $2.2\sigma$ from theoretical predictions~\cite{Bordone:2019guc, Faustov:2022ybm, Harrison:2023dzh}. Planned upgrades at LHCb and Belle II will enhance sensitivity to rare decay modes and improve access to key LFU observables, strengthening the search for beyond-SM physics~\cite{Bordone:2019guc, Cui:2023jiw}.

    Theoretical predictions for $B_s$ decays face significant challenges due to nonperturbative QCD effects in hadronic form factors. Various approaches\footnote{Initial efforts using flavor-symmetry breaking are  available in studies \cite{Gershtein:1976mv,Khlopov:1978id}.}, such as Light-Cone Sum Rules (LCSR)~\cite{Khodjamirian:2023wol, Bordone:2020gao}, Heavy Quark Effective Theory (HQET)~\cite{Zhang:2022opp, Banerjee:2016fid}, perturbative QCD (pQCD)~\cite{Hu:2019bdf, Fan:2013kqa}, Relativistic Quark Models (RQM)~\cite{Faustov:2022ybm}, Covariant Light-Front Quark Models (CLFQM)~\cite{Enqi:2025vdp, Zhang:2020dla}, Covariant Confined Quark Model (CCQM)~\cite{Pandya:2023ldv, Soni:2021fky} and Lattice QCD (LQCD)~\cite{Witzel:2020msp, Flynn:2023nhi, McLean:2019qcx, Harrison:2021tol, Harrison:2023dzh}, have been developed to compute form factors across different $q^2$ regimes, though none yet provide full kinematic coverage without model-dependent uncertainties. These methods have been further extended to investigate semileptonic $B_s$ decays, enabling predictions of not only branching ratios but also angular observables such as $A_{\rm FB}(q^2)$, $F_L(q^2)$, lepton convexity parameter ($C_{F}^\ell(q^2)$), lepton polarization ($P_{L,T}^\ell(q^2)$), and asymmetry parameter ($\alpha^*(q^2)$), all of which are sensitive probes of NP~\cite{Becirevic:2016hea, Faustov:2022ybm, Cui:2023jiw}. Despite this progress, several persistent discrepancies remain between theory and experiments. 

   Beyond semileptonic channels, the nonleptonic dynamics of $B_s$ decays have been investigated using a variety of theoretical frameworks, including Nonrelativistic Constituent Quark Models (NCQM)~\cite{Albertus:2014bfa}, RQM~\cite{Faustov:2012mt}, QCD Factorization (QCDF)~\cite{ Cai:2021mlt, Azizi:2008ty}, LCSR~\cite{Piscopo:2023opf, Li:2009wq}, pQCD~\cite{Yu:2013pua,Zou:2009zza,Li:2008ts}, CCQM~\cite{Ivanov:2011aa} and symmetry-based/hybrid approaches employing SU(3) flavor symmetry~\cite{Davies:2024vmv, Endo:2021ifc, Colangelo:2005hh}. Despite this theoretical diversity, substantial discrepancies persist between theoretcial predictions and experimental data. These tensions, especially in color-allowed channels like $\overline{B}_s^0 \to D_s^{(*)+} \pi^-$, have motivated recent efforts to incorporate nonfactorizable contributions~\cite{Flynn:2023qmi, Bordone:2021oof} and explore NP interpretations using Effective Field Theory (EFT)~\cite{Greljo:2021npi, Bordone:2021cca}, while LQCD calculations~\cite{Flynn:2023qmi, Harrison:2023dzh} provide improved input for hadronic form factors. These advancements collectively seek to enhance theoretical precision and align with experimental observations, highlighting the crucial role of hadronic $B_{(s)}$ decays in flavor physics.

   We implement the first self-consistent CLFQM (Type-II) analysis of $B_s$ decay observables. This work addresses fundamental theoretical inconsistencies in conventional CLFQM, while providing improved phenomenological precision. The self-consistent framework advances beyond traditional approach by resolving long-standing issues of Lorentz covariance violation and zero-mode artifacts in form factor calculations~\cite{Jaus:1999zv, Cheng:2003sm, Choi:2013mda}. Unlike the conventional Type-I formulation, Type-II systematically replaces mass-dependent quantities with kinetic invariant masses ($M_0$) and employs vertex functions consistent with light-front dynamics. This eliminates spurious $\omega^\mu$-dependent contributions in form factors such as $A_0(q^2)$ and $A_1(q^2)$, ensuring numerically identical results for longitudinal and transverse polarization states, thereby restoring manifest covariance and achieving theoretical self-consistency~\cite{Choi:2013mda, Chang:2019mmh, Chang:2020wvs}. Our approach contributes significantly to precision flavor physics through: (i) First unified $q^2$-resolved predictions for semileptonic and nonleptonic $B_s$ decays, with a focus on heavy-to-heavy transitions, including branching ratios for $P \to P(V)$ transitions and decay observables ($A_{\rm FB}$, $F_L$, $\alpha^*(q^2)$, \textit{etc.}). The form factors are computed using a model-independent $z$-series expansion~\cite{Bourrely:2008za}, calibrated with LQCD and phenomenological inputs for quark masses and $\beta$ parameters~\cite{Chakraborty:2014aca, Chang:2019mmh, Cheng:2003sm, Verma:2011yw}. (ii) Improved precision through a LQCD-consistent implementation of the $z$-expansion and a rigorous treatment of uncertainties, with conservative uncertainty estimates to ensure robust sensitivity to potential NP effects, particularly crucial for testing LFU via $\mathcal{R}_{D_s^{(*)}}$ measurements.

    This paper is organized as follows. Sec.~\ref{S2} describes the methodology for form factor calculations in self-consistent CLFQM and $B_s$ decay formalism. Sec.~\ref{S3} presents numerical results and discussion. We summarize our findings in Sec.~\ref{S4}. Analytical form factor expressions are given in Appendix~\ref{aA}.
 \section{Methodology} \label{S2}
	\subsection {Self-consistent covariant light-front approach} \label{S2_A}
	\begin{figure}[h]
		\centering
\includegraphics[width=.4\textwidth]{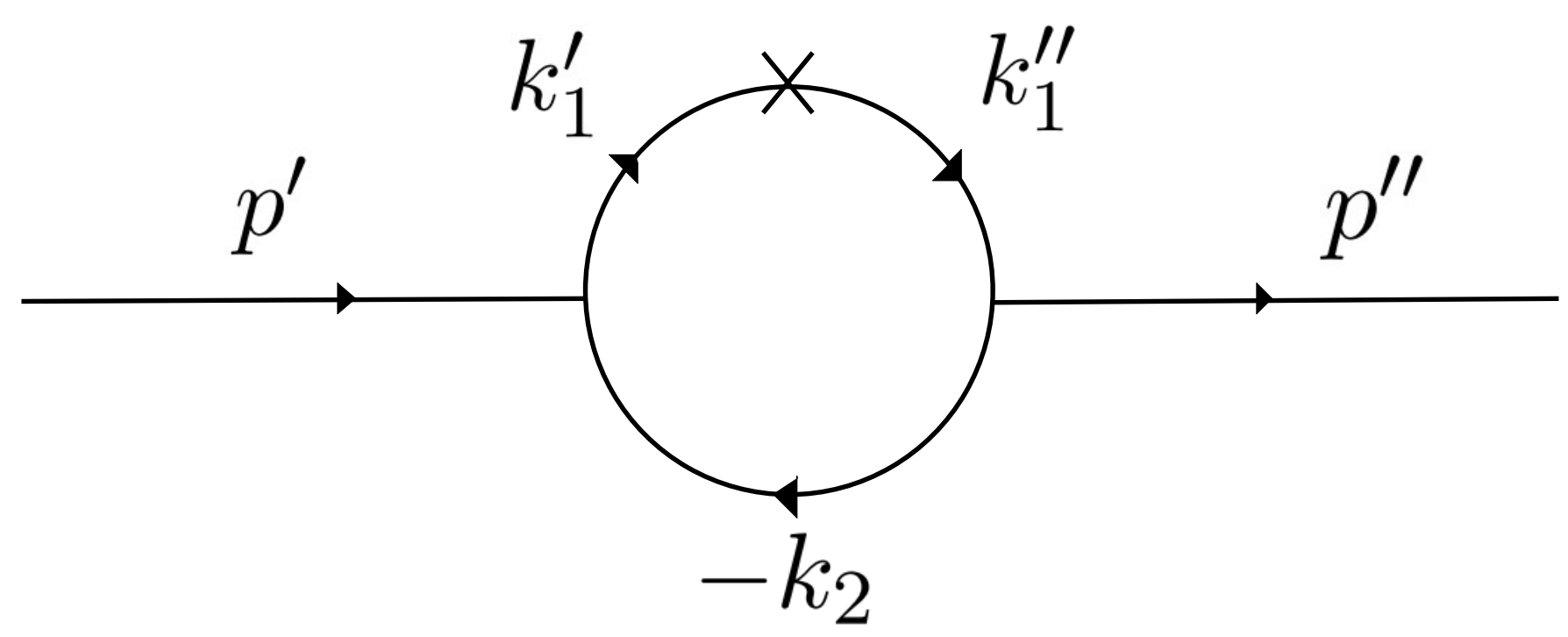}
		\caption{Feynman diagram for meson transition amplitudes, where $\bm{\t}$ denotes the vector or axial-vector current vertex}
		\label{f1}
	\end{figure}
    We employ the self-consistent CLF approach (for details, see Refs. \cite{Jaus:1999zv, Cheng:2003sm, Choi:2013mda, Chang:2019mmh}) to compute the $B_s\to P(V)$ transition form factors. In the CLFQM framework, meson transitions (Fig.~\ref{f1}) are characterized by parent ($p'=k_1'+k_2$) and daughter ($p''=k_1''+k_2$) four-momenta, where $k_1^{\prime(\prime\prime)}$ and $k_2$ are constituent quark momenta with masses $m_1^{\prime(\prime\prime)}$ and $m_2$, respectively. Using light-front coordinates $(x_{1(2)}, \bm{k}_\perp')$, the quark momenta decompose as:
    \begin{equation} \label{e1}
    k_{1(2)}^{\prime +} = x_{1(2)} p^{\prime +}, \quad 
    \bm{k}_{1(2)\perp}^{\prime} = x_{1(2)} \bm{p}_\perp^{\prime} \pm \bm{k}_\perp^{\prime},
    \end{equation}
    with momentum fractions $x_1+x_2=1$. The meson momentum $p'=(p^{\prime -}, p^{\prime +}, \bm{p}_\perp^{\prime})$ satisfies $p^{\prime\pm}=p^{\prime0}\pm p^{\prime3}$ and the mass-shell condition $p^{\prime +}p^{\prime -} - \bm{p}_\perp^{\prime 2} = M^{\prime2}$. Transverse components follow $\bm{k}_\perp^{\prime}=(k^{\prime x},k^{\prime y})$, and $\bm{p}_\perp^{\prime}=(p^{\prime x},p^{\prime y})$. Daughter meson variables are defined analogously (primes $\to$ double primes).

    Conventionally, the meson bound state ($q_1^{\prime}, \bar{q}_2$) in the light-front formalism is given by:
\begin{equation}\label{e2}
|M(p, ^{2S+1}L_J, J_z)\ra = \int \{\d^3\tilde{k}_1\}\{\d^3\tilde{k}_2\} \, 2(2\pi)^3 \delta^3 (\tilde{p}-\tilde{k}_1-\tilde{k}_2) \sum_{h_1,h_2} \Psi^{JJ_z}_{LS} |q_1^{\prime}(k_1^{\prime},h_1) \bar q_2(k_2,h_2)\ra,
\end{equation}
 where $\tilde{p}=(p^+,\bm{p}_\perp)$. $\tilde{k}_{1,2}=(k_{1,2}^{\prime +},\bm{k}^{\prime}_{1,2\perp})$ represents the on-shell light-front momenta of the constituent quarks, with $\{\d^3\tilde{k}\} \equiv \d k^+ \d^2\bm{k}_\perp/[2(2\pi)^3]$. The wave function $\Psi^{JJ_z}_{LS}$ factorizes as:
\begin{equation}\label{e4}
\Psi^{JJ_z}_{LS}(x_1,\bm{k}_\perp^{\prime},h_1,h_2)= R^{SS_z}_{h_1h_2} \psi_{LL_z}(x_1, \bm{k}_\perp^{\prime}),
\end{equation}
with the normalization condition:
\begin{equation}\label{e3}
\sum_{h_1,h_2} \int \frac{\d x_1 \d^2\bm{k}_\perp^{\prime}}{2(2\pi)^3} |\Psi^{JJ_z}_{LS}|^2 = 1.
\end{equation}
The spin-orbital structure $R^{SS_z}_{h_1h_2}$ and complete normalization procedure follow standard light-front conventions \cite{Cheng:2003sm,Jaus:1989au,Cheng:1996if}. For $s$-wave mesons, the radial wave function takes the Gaussian form:
\begin{equation}\label{e5}
\psi(x_1,\bm{k}_\perp^{\prime}) = 4 \frac{\pi^{3/4}}{\beta^{3/2}} \sqrt{\frac{\partial k_z^{\prime}}{\partial x_1}} \exp\left[- \frac{k_z^{\prime 2} + \bm{k}_\perp^{\prime 2}}{2 \beta^2}\right],
\end{equation}
where $\beta \sim \Lambda_{\text{QCD}}$ governs the momentum distribution. The light-front wave function provides the crucial connection between nonperturbative QCD dynamics at $\Lambda_{\text{QCD}}$ scales and observable hadronic properties. It incorporates essential QCD features while ensuring Lorentz covariance and phenomenological consistency in hadronic structure descriptions \cite{Brodsky:1997de}. 
     
     The relative momentum $k_z^{\prime}$ (in the $z$-direction) is given by
	\begin{equation} \label{e6}
		k_z^{\prime} = \Big(x_1 - \frac{1}{2}\Big)M_0^{\prime} + \frac{m_{2}^{2} - m_{1}^{\p2}}{2M_0^{\prime}},
	\end{equation} 
	hence resulting in \cite{Choi:2013mda} 
	\begin{equation} \label{e7}
		\frac{\partial k_z^{\prime}}{\partial x_1} = \frac{M_0^{\prime}}{4x_1 (1-x_1)}\Big\{1-\Big[\frac{m_1^{\p2} - m_2^2}{M_0^{\p2}}\Big]^2\Big\},
	\end{equation}
	where the kinetic invariant mass of the incoming (outgoing) meson is defined as
\begin{equation}
M_{0}^{\prime(\prime\prime)} = \sqrt{\frac{m_{1}^{\prime(\prime\prime)2} + \bm{k}_{\perp}^{\prime(\prime\prime)2}}{x_1} + \frac{m_{2}^{2} + \bm{k}_{\perp}^{\prime(\prime\prime)2}}{x_2}},
\label{e8e9}
\end{equation}
and the transverse momentum of the outgoing quark satisfies ${\bm{k}}_{\perp}^{\prime\prime} = {\bm{k}}_{\perp}^{\prime} - x_2 {\bm{q}}_{\perp}$, for details see Refs.~\cite{Jaus:1999zv, Cheng:2003sm, Choi:2013mda}.

Generally, the transition form factors $B_s \to M^{\prime\prime}$ (where $M^{\prime\prime} = P,~V$) are obtained from explicit expressions for matrix elements of currents between meson states, corresponding to the Feynman diagram in Fig.~\ref{f1}, is given by \cite{Jaus:1989au}, 
	\begin{align} \label{e10}
		{\cal B} ~\equiv~ \la M^{\prime\prime}(p^{\prime\prime}) | V_{\mu}-A_{\mu}|B_s(p^{\prime})\ra,
	\end{align} 
	where $V_{\mu}$ and $A_{\mu}$ are the vector and axial-vector currents, respectively. The transition form factors for $B_s$ meson to $P$ and $V$ final states are defined by the following matrix elements \cite{Cheng:2003sm},
	\begin{align} \label{e11} \la P(p^{\prime\prime})|V_{\mu}|B_s(p^{\prime})\ra & = p_{\mu}f_{+}(q^{2}) + q_{\mu}f_{-}(q^{2}),\\
		\label{e12}
		\la V(p^{\prime\prime}, \varepsilon^{\prime\prime})|V_{\mu}|B_s(p^{\prime})\ra & = \epsilon_{\mu \nu \alpha \beta}\varepsilon^{\prime\prime*\nu}p^{\alpha}q^{\beta}g(q^{2}),\\
		\label{e13}
		\la V(p^{\prime\prime}, \varepsilon^{\prime\prime})|A_{\mu}|B_s(p^{\prime})\ra & = -i\{\varepsilon^{\prime\prime*}_{\mu}f(q^{2}) + \varepsilon^{\prime\prime*}\cdot \bm{p}[p_{\mu}a_{+}(q^{2}) + q_{\mu}a_{-}(q^{2})]\},
	\end{align}
	where $p_\mu = p^{\prime} + p^{\prime\prime}$ and $q_{\mu} = p^{\prime} - p^{\prime\prime}$. The polarization of the outgoing vector meson is denoted by $\varepsilon_{\mu}$ and the convention $\epsilon_{0123} = 1$ is employed. The matrix elements given in Eqs.~\eqref{e11}-\eqref{e13}, are conventionally expressed in terms of the Bauer-Stech-Wirbel (BSW) \cite{Wirbel:1985ji} form factors as follows,
	\begin{align} \label{e14}
		\la P(p^{\prime\prime})|V_{\mu}|B_s(p^{\prime})\ra & = (p_{\mu} - \frac{M_{B_s}^{2}-M_{P}^{2}}{q^{2}}q_{\mu})F^{B_sP}_{1}(q^{2}) + \frac{M_{B_s}^{2}-M_{P}^{2}}{q^{2}}q_{\mu}F^{B_sP}_{0}(q^{2}),\\
		\label{e15}
		\la V(p^{\prime\prime}, \varepsilon^{\prime\prime})|V_{\mu}|B_s(p^{\prime})\ra & = -\frac{1}{M_{B_s}+M_{V}}\epsilon_{\mu \nu \alpha \beta}\varepsilon^{\prime\prime*\nu}p^{\alpha}q^{\beta}V^{B_sV}(q^{2}), \\
		\label{e16}
		\la V(p^{\prime\prime}, \varepsilon^{\prime\prime})|A_{\mu}|B_s(p^{\prime})\ra & = i\{(M_{B_s} + M_{V})\varepsilon_{\mu}^{\prime\prime*}A^{B_sV}_{1}(q^{2}) - \frac{\varepsilon^{\prime\prime*} \cdot \bm{p}}{M_{B_s} + M_{V}}p_{\mu}A^{B_sV}_{2}(q^{2}) \nonumber\\
		& - 2M_{V}\frac{\varepsilon^{\prime\prime*} \cdot \bm{p}}{q^{2}}q_{\mu}[A^{B_sV}_{3}(q^{2})- A^{B_sV}_{0}(q^{2})]\},
	\end{align}
	where $M_{B_s}$ and $M_{P(V)}$ are the masses of the corresponding mesons. The BSW-type form factors can be related to the CLFQM form factors through the following expressions \cite{Cheng:2003sm},
	\begin{equation} \label{e17}
		\begin{gathered}
			F^{B_sP}_{1}(q^{2}) = f_{+}(q^{2}),\hspace{0.5cm}
			F^{B_sP}_{0}(q^{2}) = f_{+}(q^{2}) + \frac{q^{2}}{q\cdot p}f_{-}(q^{2}),\\
			V^{B_sV}(q^{2}) = -(M_{B_s} +M_{V})g(q^{2}),\hspace{0.5cm}
			A^{B_sV}_{1}(q^{2}) = -\frac{f(q^{2})}{M_{B_s} +M_{V}},\\
			A^{B_sV}_{2}(q^{2}) = (M_{B_s} +M_{V})a_{+}(q^{2}),\hspace{0.5cm}
			A^{B_sV}_{3}(q^{2}) - A^{B_sV}_{0}(q^{2}) = \frac{q^{2}}{2M_{V}}a_{-}(q^{2}),
		\end{gathered}
	\end{equation}
	with
	\begin{equation} \label{e18}
		\begin{gathered}
			F^{B_sP}_{1}(0) = F^{B_sP}_{0}(0), \\
			A^{B_sV}_{3}(0) = A^{B_sV}_{0}(0), ~~\text{and}\\
			A^{B_sV}_{3}(q^{2}) = \frac{M_{B_s} +M_{V}}{2 M_{V}}A^{B_sV}_{1}(q^{2}) - \frac{M_{B_s} -M_{V}}{2 M_{V}}A^{B_sV}_{2}(q^{2}).
		\end{gathered}
	\end{equation}

In the CLFQM, constituent quarks are treated off-shell, and zero-mode contributions are systematically included to eliminate spurious terms proportional to the light-like vector $\omega^\mu = (0,2,0_\perp)$, thereby restoring covariance of current matrix elements~\cite{Jaus:1999zv}. This allows for the manifestly covariant calculation of transition amplitudes via one-loop Feynman integrals. For the $B_s(p^\prime) \to M^{\prime\prime}(p^{\prime\prime})$ transition, we adopt the Drell–Yan–West frame ($q^+ = 0$), restricting direct evaluation to the space-like region ($q^2 = -{\bm{q}}_\perp^2 \leq 0$), with extrapolation required for the time-like domain. The light-front kinematics are specified by setting $\bm{p}_\perp^\prime = 0$ and $\bm{p}_\perp^{\prime\prime} = -\bm{q}_\perp$, leading to $\bm{k}_\perp^{\prime\prime} = \bm{k}_\perp^\prime - x_2 \bm{q}_\perp$.

The hadronic matrix element (Eq.~\eqref{e10}) is computed as
\begin{equation}
\mathcal{B} = N_c \int \frac{d^4k_1^\prime}{(2\pi)^4} \frac{H_{M^\prime}H_{M^{\prime\prime}}}{N_1^\prime N_1^{\prime\prime} N_2} \, iS_\mathcal{B},
\end{equation}
where $H_{M^{\prime(\prime\prime)}}$ are vertex functions, and $S_{\mathcal{B}}$ is the Dirac trace \cite{Chang:2019mmh, S:2024adt}. The vertex structures are $i\Gamma_P = -i\gamma_5$ and $i\Gamma_V = i[\gamma^\mu - (k_1 - k_2)^\mu/D_V]$, with $D_V$ involving either the meson or kinetic mass.

The covariant approach by Jaus~\cite{Jaus:1999zv} evaluates hadronic matrix elements via one-loop Feynman integrals and light-front decomposition. By integrating out $k_1^-$ and using regular vertex functions, it eliminates $\omega^\mu$-dependent spurious terms and bridges the gap between the covariant Bethe–Salpeter and standard light-front frameworks. To relate this to the light-front formalism, one performs contour integration over $k_1^{\prime-}$, yielding the light-front amplitude. In the \textit{Type-I correspondence}, this involves the substitutions:
\begin{equation}
N_1^{\prime(\prime\prime)} \to \hat{N}_1^{\prime(\prime\prime)} = x_1(M^{\prime(\prime\prime)2} - M_0^{\prime(\prime\prime)2}), \quad D_V^{\prime(\prime\prime)} \to D_{V,\rm LF}^{\prime(\prime\prime)} = M_0^{\prime(\prime\prime)} + m_1^{\prime(\prime\prime)} + m_2.
\end{equation}
However, the Type-I correspondence introduces residual $\omega$-dependent contributions, particularly through the coefficients $B_j^{(i)}$, which result in inconsistencies between longitudinal ($\lambda = 0$) and transverse ($\lambda = \pm$) helicity amplitudes. These discrepancies are most evident in the form factors, $f(q^2)$ and $a_-(q^2)$, where the matrix elements differ significantly between polarization states, thereby violating the manifest covariance of the theory~\cite{Choi:2013mda, Chang:2019mmh}.

To resolve these issues, the \textit{Type-II correspondence}~\cite{Choi:2013mda, Chang:2019mmh} promotes the replacement $M^{\prime(\prime\prime)} \to M_0^{\prime(\prime\prime)}$ throughout the integrand, including all $M$-dependent terms. The resulting matrix element becomes
\begin{equation}
\hat{\mathcal{B}} = N_c \int \frac{dx_1 d^2\bm{k}_\perp^\prime}{2(2\pi)^3} \frac{h_{M^\prime} h_{M^{\prime\prime}}}{x_2 \hat{N}_1^\prime \hat{N}_1^{\prime\prime}} \hat{S}_{\mathcal{B}},
\end{equation}
with the light-front vertex functions given by
\begin{equation}
\frac{h_{P(V)}}{\hat{N}_1^{\prime(\prime\prime)}} = \frac{1}{\sqrt{2N_c}} \sqrt{\frac{x_2}{x_1}} \frac{\psi(x_1, \bm{k}_\perp^{\prime(\prime\prime)})}{\hat{M}_0^{\prime(\prime\prime)}}, \quad \hat{M}_0^{\prime(\prime\prime)} = \sqrt{M_0^{\prime(\prime\prime)2} - (m_1^{\prime(\prime\prime)} - m_2)^2}.
\end{equation}

This replacement guarantees identical results for all polarization components and eliminates $\omega$-dependence, thereby restoring full Lorentz covariance. Hence, Type-II correspondence offers a self-consistent and numerically reliable framework for computing $B_s \to P(V)$ transition form factors in CLFQM. The expressions for transition form factors are defined in Appendix~\ref{aA}. For technical details and derivations, see Refs.~\cite{Jaus:1999zv, Choi:2013mda, Chang:2019mmh}.

\subsection{$\pmb{q^2}$ dependence of the form factors} \label{S2_B}
In our previous works~\cite{Hazra:2023zdk, S:2024adt}, we analyzed the momentum dependence of form factors under multiple parameterizations. In this analysis, we adopt the $z$-series expansion to describe the full $q^2$ behavior of $B_s \to P(V)$ form factors within the self-consistent CLFQM (Type-II). While light-front computations are naturally carried out in the Drell–Yan–West frame ($q^+ = 0$), which restricts calculations to the space-like region ($q^2 \leq 0$)~\cite{Jaus:1989au, Jaus:1996np, Jaus:1999zv}, physical decays occur in the time-like region ($q^2 \geq 0$)~\cite{Cheng:2003sm}. Therefore, a reliable extrapolation from the space-like to time-like region is essential for obtaining physical observables.

 To achieve this, we employ the $z$-series parameterization. This approach reformulates the form factors in terms of a complex variable $z$, which arises from the analytic continuation of $q^2$ into the complex plane \cite{HFLAV:2024ctg}. Specifically, the form factor is expanded as a power series around the value $q^2 = t_0$, leading to the following expression \cite{S:2024adt, Bourrely:2008za},
	\begin{equation} \label{e29}
		F(q^2) = \frac{1}{1-\frac{q^2}{M_{pole}^2}}\sum_{k=0}^{K} a'_{k}\big[z(q^2)-z(0)\big]^k,
	\end{equation}
	where $M_{pole}$ is the transition pole mass, $a_k$ are real coefficients and $z(q^2) \equiv z(q^2, t_0)$ is the function defined as,
	\begin{equation} \label{e30}
		z(q^2) = \frac{\sqrt{t_+ - q^2}- \sqrt{t_+ - t_0}}{\sqrt{t_+ - q^2}+ \sqrt{t_+ - t_0}}~.
	\end{equation}
    The $t$-variables are defined by,
        \begin{equation} \label{e31}
            t_{\pm} = (M_{B_s} \pm M_{P(V)})^2;~~~t_0 = t_+\Bigg(1-\sqrt{1-\frac{t_-}{t_+}}\Bigg).
        \end{equation}
    It may be noted that the coefficients $a'_k$ do not carry direct physical interpretation \cite{HFLAV:2024ctg}. Furthermore, since higher-order terms in the expansion given in Eq.~\eqref{e29} contribute negligibly, the series is truncated at $K=2$, retaining only three free parameters $a'_0~(\approx F(0))$, $a'_1$, and $a'_2$.

It is important to emphasize that the kinematic range of $q^2$ depends on the final-state meson. For $B_s \to D_s^{(*)}$ decays, the allowed range is $0 \leq q^2 \lesssim 12~\text{GeV}^2$, while for $B_s \to K^{(*)}$ transitions, it extends up to $q^2 \lesssim 24~\text{GeV}^2$. These ranges correspond to $b \to c$ and $b \to u$ transitions, respectively, and lie below the pole positions $M_{\text{pole}}^2$, indicating that the physical region does not encompass the nearest resonances. Therefore, accurate characterization of the full $q^2$ dependence of the form factors is essential for reliable predictions of decay amplitudes~\cite{Melikhov:2001zv}. To achieve this, we calculate $P \to P (V)$ form factors at five distinct $q^2$ points ($-0.01$, $-0.1$, $-5.0$, $-10.0$, $-25.0\text{ for }P~(-20.0\text{  for }V)~\text{GeV}^2$) to fit Eq.~\eqref{e29} and extract the parameters $F(0)$, $a_1^\prime$, and $a_2^\prime$. Furthermore, the analytic structure of the form factors in Eq.~\eqref{e29} reflects contributions from resonant states with definite spin-parity quantum numbers~\cite{Bauer:1988fx, Melikhov:2000yu}; accordingly, we adopt the pole masses listed in Table~\ref{t1} for each transition channel.
	\subsection{Semileptonic decay widths and other physical observables} \label{S2_C}	
	The differential decay width of $B_s$ to $P(V)$ semileptonic decays is expressed in terms of the helicity components as \cite{Faustov:2022ybm, Korner:1989qb},
	\begin{equation} \label{e32}
		\frac{d \Gamma(B_s^{0} \to P(V)\ell^+ \nu_\ell)}{dq^2} = \frac{G_F^2}{(2\pi)^3}|V_{q_1^\prime q_1^{\prime\prime}}|^2 \frac{q^2\sqrt{\lambda}}{24 M_{B_{s}}^3} (1-\frac{m_\ell^2}{q^2})^2 \Big((\mathcal{H}_U + \mathcal{H}_L)(1+\frac{m_\ell^2}{2q^2})+\frac{3m_\ell^2}{2q^2}\mathcal{H}_S\Big),
	\end{equation}
	where $G_F$ is the Fermi constant and $V_{q_1^\prime q_1^{\prime\prime}}$ is the relevant CKM matrix element for $q_1^\prime \to q_1^{\prime\prime}$ transition. The term $\lambda \equiv \lambda (M_{B_s}^2,M_{P(V)}^2,q^2) = (M_{B_s}^2 + M_{P(V)}^2 + q^2)^2 - 4M_{B_s}^2M_{P(V)}^2$ is the K\"all\'en function, and $m_\ell$ is the lepton mass ($\ell = e,~\mu,~\tau$). The helicity components $\mathcal{H}_U$, $\mathcal{H}_L$, and $\mathcal{H}_S$ can be defined as,
	\begin{equation}\label{e33}
		\begin{gathered}
			\mathcal{H}_U = |{H}_+|^2 + |{H}_-|^2, \hspace{0.5cm}
			\mathcal{H}_L = |{H}_0|^2,~~\text{ and }~~ \mathcal{H}_S = |{H}_t|^2,
		\end{gathered}
	\end{equation}
	where ${H}_{\pm}$, ${H}_{0}$, and ${H}_{t}$ are the helicity amplitudes. These helicity amplitudes are related to the corresponding invariant form factors by the following relations:
	\begin{itemize}
		\item[(i)] For $B_s$ to $P$ meson transitions,
		\begin{equation}\label{e34}
			\begin{gathered}
				{H}_{\pm}(q^2) = 0, \hspace{0.5cm}
				{H}_{0}(q^2) = \frac{\sqrt{\lambda}}{\sqrt{q^2}}F_1(q^2),~~\text{ and }~~ {H}_{t}(q^2) = \frac{1}{\sqrt{q^2}}(M_{B_s}^2-M_P^2)F_0(q^2).
			\end{gathered}
		\end{equation}
		
		\item[(ii)] For $B_s$ to $V$ meson transitions,
		\begin{align}
			\label{e35}
			{H}_{\pm}(q^2) = & (M_{B_s} + M_V) A_1(q^2) \mp \frac{\sqrt{\lambda}}{M_{B_s} + M_V}V(q^2), \\
			\label{e36}
			{H}_{0}(q^2) = &\frac{1}{2 M_V \sqrt{q^2}}(M_{B_s} + M_V)(M_{B_s}^2 - M_V^2 -q^2) A_1(q^2) - \frac{\lambda}{M_{B_s} + M_V}A_2(q^2), \\
			\label{e37}
			{H}_{t}(q^2) = & \frac{\sqrt{\lambda}}{\sqrt{q^2}}A_0(q^2).
		\end{align}
	\end{itemize}
	
	Following Eq.~\eqref{e32}, the longitudinal and transverse differential decay widths are given by 
	\begin{align} 
		\label{e38}
		\frac{d \Gamma_{L}(B_s^{0} \to P(V)\ell^+ \nu_\ell)}{dq^2} = & \frac{G_F^2}{(2\pi)^3}|V_{q_1^\prime q_1^{\prime\prime}}|^2 \frac{q^2\sqrt{\lambda}}{24 M_{B_{c}}^3} (1-\frac{m_\ell^2}{q^2})^2 [\mathcal{H}_L(1+\frac{m_\ell^2}{2q^2})+\frac{3m_\ell^2}{2q^2}\mathcal{H}_S], \text{ and} \\
		\label{e39}
		\frac{d \Gamma_{T}(B_s^{0} \to V\ell^+ \nu_\ell)}{dq^2} = & \frac{G_F^2}{(2\pi)^3}|V_{q_1^\prime q_1^{\prime\prime}}|^2 \frac{q^2\sqrt{\lambda}}{24 M_{B_{c}}^3} (1-\frac{m_\ell^2}{q^2})^2 [\mathcal{H}_U(1+\frac{m_\ell^2}{2q^2})],
	\end{align} 
	respectively. The branching ratio is calculated from the decay rate obtained using Eq.~\eqref{e32} by multiplying with $\frac{\tau_{B_s}}{\hbar}$. 

Comprehensive analysis of semileptonic decays requires examining lepton mass effects beyond branching ratios; therefore, it is essential to include other experimentally accessible observables. These observables provide valuable insight into helicity dynamics and possible deviations from the SM, and are expressed in terms of helicity structure functions as follows~\cite{Faustov:2022ybm, Ivanov:2005fd,Becirevic:2016hea}:
\begin{align}
\text{Forward-Backward Asymmetry:} \quad 
A_{\rm FB}(q^2) &= \frac{3}{4} \frac{\mathcal{H}_P - 2\frac{m_\ell^2}{q^2} \mathcal{H}_{SL}}{\mathcal{H}_{\rm total}}, \label{e40} \\[4pt]
\text{Leptonic Convexity Parameter:} \quad 
C_F^\ell(q^2) &= \frac{3}{4} \left(1 - \frac{m_\ell^2}{q^2} \right) \frac{\mathcal{H}_U - 2 \mathcal{H}_L}{\mathcal{H}_{\rm total}}, \label{e41} \\[4pt]
\text{Longitudinal Lepton Polarization:} \quad 
P_L^\ell(q^2) &= \frac{(\mathcal{H}_U + \mathcal{H}_L)\left(1 - \frac{m_\ell^2}{2q^2}\right) - \frac{3m_\ell^2}{2q^2} \mathcal{H}_S}{\mathcal{H}_{\rm total}}, \label{e42} \\[4pt]
\text{Transverse Lepton Polarization:} \quad 
P_T^\ell(q^2) &= -\frac{3\pi m_\ell}{8\sqrt{q^2}} \frac{\mathcal{H}_P + 2\mathcal{H}_{SL}}{\mathcal{H}_{\rm total}}, \label{e43} \\[4pt]
\text{Longitudinal Vector Polarization:} \quad 
F_L(q^2) &= \frac{\mathcal{H}_L\left(1 + \frac{m_\ell^2}{2q^2}\right) + \frac{3m_\ell^2}{2q^2} \mathcal{H}_S}{\mathcal{H}_{\rm total}}, \label{e44e} \\[4pt]
\text{Asymmetry Parameter:} \quad 
\alpha^*(q^2) &= \frac{\mathcal{H}_U + \widetilde{\mathcal{H}}_U - 2(\mathcal{H}_L + \widetilde{\mathcal{H}}_L + 3\widetilde{\mathcal{H}}_S)}{\mathcal{H}_U + \widetilde{\mathcal{H}}_U + 2(\mathcal{H}_L + \widetilde{\mathcal{H}}_L + 3\widetilde{\mathcal{H}}_S)}. \label{e44}
\end{align}
Here, $\widetilde{\mathcal{H}}_i = \frac{m_\ell^2}{2q^2} \mathcal{H}_i$ $(i=U,L,S)$ represents the helicity-flip component. The interference terms are given by $\mathcal{H}_P = |H_+|^2 - |H_-|^2$ and $\mathcal{H}_{SL} = \mathrm{Re}(H_0 H_t^\dagger)$. These observables, less sensitive to hadronic uncertainties, are powerful probes of NP effects~\cite{Becirevic:2016hea}. Furthermore, we compute the $q^2$-averaged values of these observables by integrating their numerators and denominators separately, weighted by $q^2\sqrt{\lambda}\left(1 - \frac{m_\ell^2}{q^2}\right)^2$, where the factor reflects the lepton velocity. For $\overline{B}_s^0 \to P(V)\ell^- \overline{\nu}_\ell$ decays, $A_{\rm FB}$, $P_L^\ell$, and $P_T^\ell$ reverse sign due to the parity-odd nature of the leptonic current, while other observables remain unchanged~\cite{Faustov:2022ybm}.

	\subsection{Nonleptonic decay widths} \label{S2_D}
	The QCD modified weak Hamiltonian generating the ${B}_s$ decay involving $b\to c(u)$ transitions\footnote{The selection rules for $\ov{B}_s^0$ decay modes governed by the Hamiltonian in Eq.~\eqref{e45} are classified as follows: (i) CKM-enhanced modes: $\Delta b = 1$, $\Delta C = 1$, $\Delta S = 0$; and $\Delta b = 1$, $\Delta C = 0$, $\Delta S = -1$; (ii) CKM-suppressed modes: $\Delta b = 1$, $\Delta C = 1$, $\Delta S = -1$; and $\Delta b = 1$, $\Delta C = 0$, $\Delta S = 0$; (iii) CKM-doubly-suppressed modes: $\Delta b = 1$, $\Delta C = -1$, $\Delta S = 0$; and $\Delta b = 1$, $\Delta C = -1$, $\Delta S = -1$.}
 is expressed as follows \cite{Buras:1995iy}:
	\begin{eqnarray}\label{e45}
		H_{w}^{(\Delta b=-1)} = \frac{G_F}{\sqrt{2}}\displaystyle\sum\limits_{Q(q)=u,c}\displaystyle\sum\limits_{q^{'}=d,s}V^*_{Qb}V_{qq^{'}} \Big(a_1(\mu) O^{qq^{'}}_1(\mu) + a_2(\mu) O^{qq^{'}}_2(\mu)\Big) + h.c.,
	\end{eqnarray}
	where $a_{1}$ and $a_{2}$ are the standard perturbative QCD coefficients, evaluated at renormalization scale $\mu \approx m_b^2 $. Local tree-level operators $O_{1,2}$ involving $b \to q$ transition can be expressed as products of color-singlet currents are given below:
	\begin{eqnarray} \label{e46}
		O^{qd}_1 = (\bar b_\alpha q_\alpha)_{V-A} \cdot (\bar q_\beta d_\beta)_{V-A},~~O^{qd}_2 = (\bar b_\alpha q_\beta)_{V-A} \cdot (\bar q_\beta d_\alpha)_{V-A}, \nonumber \\
		O^{qs}_1 = (\bar b_\alpha q_\alpha)_{V-A} \cdot (\bar q_\beta s_\beta)_{V-A},~~O^{qs}_2 = (\bar b_\alpha q_\beta)_{V-A} \cdot (\bar q_\beta s_\alpha)_{V-A}, 
	\end{eqnarray}
	where $(\bar{q}q')_{V-A} \equiv \bar{q} \gamma_{\mu}(1-\gamma_5)q'$, $\alpha$ and $\beta$ are $SU(3)$ color indices.
    
     The scale-independent coefficients $a_1$ and $a_2$, relevant at $\mu \approx m_b^2$, are related to the Wilson coefficients $c_1$ and $c_2$ via:
\begin{align}
\label{e47}
a_1 = c_1 + \frac{1}{N_c}c_2, \quad a_2 = c_2 + \frac{1}{N_c}c_1,
\end{align}
where $N_c$ is the number of QCD colors. These parameters classify nonleptonic decays into: Class I ($a_1$-dominated), Class II ($a_2$-dominated), and Class III (interfering amplitudes)~\cite{Bauer:1986bm}. In the large-$N_c$ limit ($N_c \to \infty$), one sets $a_1 \approx c_1$ and $a_2 \approx c_2$~\cite{Wirbel:1988ft}. 

We adopt $c_1 = 1.12$, $c_2 = -0.26$ at the $b$-quark scale, yielding:
\begin{align}
\label{e48}
a_1 = 1.03, \quad a_2 = 0.11 \quad \text{for } N_c = 3.
\end{align}
The small value of $a_2$ reflects strong color suppression in $B_s$ decays~\cite{Browder:1995gi}. Since these decays are predominantly governed by tree-level amplitudes, subleading penguin and nonfactorizable contributions are neglected in leading-order analysis. However, $N_c$ can be treated as a phenomenological parameter that effectively encodes nonfactorizable dynamics~\cite{Browder:1995gi, Cheng:2001sc, Cheng:2010ry}. Accordingly, we consider both $N_c = \infty$ and $N_c = 3$ in our analysis. Following Ref.~\cite{Cheng:2014rfa}, we also allow $a_i$ to vary within a phenomenologically reasonable range to account for subleading $1/N_c$ effects in bottom-changing decays.

In the factorization framework, the decay amplitude can be expressed as,
\begin{align} \label{eAmp}
\mathcal{A}(B_s \to P_1 P_2 (P V)) & \equiv \frac{G_F}{\sqrt{2}} V_{\rm CKM} \, a_i \, \left[ \langle P_1 | J^\mu | 0 \rangle\, \langle P_2 (V) | J_\mu | B_s \rangle + \langle P_2 (V) | J^\mu | 0 \rangle\, \langle P_1 | J_\mu | B_s \rangle \right],\\ \non
& \equiv \frac{G_F}{\sqrt{2}} V_{\rm CKM} \, a_i \, \mathcal{A}^{(B_s M_1, M_2)},
\end{align}
    where $J^{\mu}$ stands for $V-A$ current. The matrix element of the $J^{\mu}$ between vacuum and final meson ($P$ or $V$) is parameterized by the decay constants $f_{P(V)}$ as,
	\begin{align}
		\label{e51}
		 ~<0|A_{\mu}|P(p^{\prime})> =~ if_{P}p^{\prime}_{\mu},~~~ <0|V_{\mu}|V(p^{\prime},\varepsilon^{\prime})> ~          =~ M^{\prime}_{V}f_{V}\varepsilon^{\prime}_{\mu}.
	\end{align}
Thus, the factorizable hadronic amplitude $\mathcal{A}^{(B_s M_1, M_2)}$ is defined as follows: i) For $B_s \to PP$ decays: $
\mathcal{A}^{(B_s P_1, P_2)} = f_{P_2} (M_{B_s}^2 - M_{P_1}^2) F_0^{B_s \to P_1}(M_{P_2}^2),$ with an analogous contribution for $P_1 \leftrightarrow P_2$. ii) For $B_s \to PV$ decays: $\mathcal{A}^{(B_s P, V)} = 2 f_V\, m_V\, F_1^{B_s \to P}(m_V^2)\, (\varepsilon_V \cdot p_{B_s}),~ \text{and}~\mathcal{A}^{(B_s V, P)} = 2 f_P\, m_V\, A_0^{B_s \to V}(m_P^2)\, (\varepsilon_V \cdot p_{B_s}).$ A more detailed and systematic discussion on factorization can be found in Ref. \cite{Bauer:1986bm, Beneke:2000ry}.

The decay rate for the $B_s \to P P$ and $B_s \to P V$ decay is given by \cite{Beneke:2000ry}
    \begin{equation} 
		\label{e54}
		\Gamma(B_s \to P_1 P_2)= \frac{{\textbf{k}}}{8 \pi M_{B_s}^{2}} |\mathcal{A}(B_s \to P_1 P_2)|^2,
	\end{equation}
	\begin{equation} 
		\label{e55}
		\Gamma(B_s \to P V)= \frac{{\textbf{k}}^3}{8 \pi M_{V}^{2}} |\mathcal{A}(B_s \to P V)|^2,
	\end{equation}
	where \textbf{k} is the three-momentum of the final-state particle in the rest frame of $B_s$ meson and is expressed as,
	\begin{equation} 
		\label{e56}
		\textbf{k} = \frac{1}{2 M_{B_s}}\sqrt{[M_{B_s}^{2}-(M_1 + M_2)^2][M_{B_s}^{2}-(M_1 - M_2)^2]}.
	\end{equation}
	The numerical results for semileptonic and nonleptonic weak decays of $B_s$ meson are discussed in the following section. 
    
	\section {Numerical Results and discussions} \label{S3}
	In this study, we evaluate the bottom-changing transition form factors for $B_s$ to $P$ and $V$ within the framework of the self-consistent CLFQM, with particular attention to heavy-to-heavy transitions. These form factors are obtained using the following constituent quark masses (in GeV) and $\beta$ values (in GeV)~\cite{Verma:2011yw, Chang:2019mmh, Cheng:2003sm}:
   \[
\centerline{
\parbox{0.9\linewidth}{
\centering
$m_u = m_d = 0.26 \pm 0.04$, \quad $m_s = 0.40 \pm 0.05$, \quad $m_c = 1.27 \pm 0.20$, \quad $m_b = 4.50 \pm 0.30$, \\
$\beta_{K} = 0.3479 \pm 0.0029$, \quad $\beta_{D_s} = 0.5432 \pm 0.0095$, \quad $\beta_{B_s} = 0.6019 \pm 0.0074$, \\
$\beta_{K^*} = 0.2926 \pm 0.0047$, \quad $\beta_{D_s^*} = 0.3932 \pm 0.0393$.
}}
\]
The heavy quark masses employed in this analysis are consistent with LQCD determinations~\cite{Chakraborty:2014aca}. The phenomenological $\beta$ parameters, which characterize the meson wave functions in CLFQM and are typically determined through fits to experimental decay constants, align with established values from recent self-consistent CLFQM determinations~\cite{Chang:2020wvs, Chang:2019mmh, Chang:2018zjq}. Nonetheless, we adopt broader parameter ranges (particularly for quark masses) to enable a more robust systematic uncertainty analysis and sensitivity study of the theoretical predictions. We systematically analyze the dependence of transition form factors and their slope parameters on constituent quark masses and $\beta$ values, especially concerning their $q^2$ dependence. Using the $z-$series expansion (Eq.~\eqref{e29}) with pole masses of Table~\ref{t1}, we compute the form factors $B_s \to P$ and $B_s \to V$, listed in Table~\ref{t2} and illustrated over the physical region ($0 \leq q^2 \leq q^2_{\rm max}$) in Figs.~\ref{f2},~\ref{f3} and \ref{flqcd}. Using these results, we calculate the semileptonic branching ratios for $B_s$ decays (Table~\ref{t3}), incorporating lepton masses, CKM elements, and the $B_s$ lifetime from PDG~\cite{PDG:2024cfk}. Beyond branching ratios, we evaluate additional observables, $A_{\rm FB}(q^2)$, $C_F^\ell(q^2)$, $P_{L(T)}^\ell(q^2)$, $F_L(q^2)$, and $\alpha^*(q^2)$, with their integrated values given in Table~\ref{t4} and differential $q^2$ profiles shown in Figs.~\ref{f4}–\ref{f11}. Finally, we predict nonleptonic branching ratios for $B_s \to PP/PV$ decays involving charmed mesons in the final state, using the calculated form factors and decay constants (in MeV)\footnote{We use LQCD and LCSR results (in parentheses) in scarcity of experimental data.}: 
    \begin{equation}
\begin{aligned} \label{e48D}
&f_{\pi} = 130.56~\text{\cite{PDG:2024cfk}}, \quad 
f_{K} = 155.7~\text{\cite{PDG:2024cfk}}, \quad 
f_{\eta} = (181.14)~\text{\cite{Dowdall:2013rya}}, \quad 
f_{D} = 203.8~\text{\cite{PDG:2024cfk}}, \\
&f_{D_s} = 250.1~\text{\cite{PDG:2024cfk}}, \quad 
f_{\eta_c} = 335~\text{\cite{PDG:2024cfk}}, \quad 
f_{\rho} = (210)~\text{\cite{Bharucha:2015bzk}}, \quad 
f_{K^{*}} = (204)~\text{\cite{Bharucha:2015bzk}}, \\
&f_{\phi} = (228.5)~\text{\cite{Chakraborty:2017hry}}, \quad 
f_{D^{*}} = (223.5)~\text{\cite{Lubicz:2017asp}}, \quad 
f_{D_s^*} = 213~\text{\cite{BESIII:2023zjq}}, \quad 
f_{J/\psi} = 416~\text{\cite{PDG:2024cfk}},
\end{aligned}
\end{equation}
     as listed in Tables~\ref{t5}-\ref{t10}. Uncertainties from meson and lepton masses and other associated parameters are neglected, as they are subdominant compared to those from quark masses and $\beta$ values. We compare our branching ratio predictions with existing experimental measurements and other theoretical results. We discuss our numerical results in the following subsections.

\subsection{Form factors}
In this subsection, we analyze the $B_s \to P(V)$ transition form factors within the self-consistent CLFQM framework, as detailed in Table~\ref{t2}. These form factors are evaluated at both $q^2 = 0$ and $q^2 = q^2_{max}$. The quoted uncertainties in the form factors and slope parameters originate from variations in constituent quark masses and $\beta$ parameters, respectively. Graphical representations in Figs.~\ref{f2} and~\ref{f3} further illustrate their $q^2$ dependence. Our key findings are summarized below:

\begin{itemize}
     \item[(i)] For $b \to u$ transitions, $B_s \to K^{(*)}$ transition form factors show significant $q^2$ dependence across a broad range ($0 \le q^2 \lesssim 24$ GeV$^2$). This extended range is crucial for analyzing the influence of $q^2$ on form factors and highlights the significance of resonance pole contributions below the threshold. Figures~\ref{f2} and~\ref{f3} illustrate the $q^2$ variation of $B_s \to P$ and $B_s \to V$ form factors, respectively. Numerically, $F_1^{B_sK}$ exhibits a sharper change at $q^2_{max}$ compared to $F_0^{B_sK}$, indicating strong curvature across the $q^2$ range (see Table~\ref{t2}). This suggests that distinct pole structures influence their behaviors, as the $a_2^\prime$ parameter shows opposing signs for $F_0$ (negative) and $F_1$ (positive). Similar $q^2$ dependent behavior is observed for $B_s \to K^{*}$ transition form factors. Interestingly, a clear hierarchy, $A_0(0) > V(0) > A_1(0) > A_2(0)$, is observed among $B_s \to K^{*}$ form factors, consistent with LCSR expectations\footnote{Our $B_s \to K^*$ results agree with LCSR: $V(0) = 0.296(30)$, $A_0(0) = 0.314(48)$, $A_1(0) = 0.230(25)$~\cite{Bharucha:2015bzk,Khodjamirian:2023wol}. For $B_s \to D_s^*$, LCSR~\cite{Bordone:2019guc} (soft-collinear eﬀective theory sum rules~\cite{Khodjamirian:2023wol,Cui:2023jiw}) gives $V(0) \approx 0.73(0.671)$, $A_0(0) \approx 0.68(0.582)$, $A_1(0) \approx 0.63(0.661)$, $A_2(0) \approx 0.52(0.765)$.}. The $A_1(q^2)$ form factor exhibits an anomalously flat behavior, characterized by weak momentum transfer dependence, as evidenced by its small $|a_1^\prime|$ and $|a_2^\prime|$ coefficients in the $z$-expansion.
    
    \item[(ii)] In contrast, for $b \to c$ transitions, the $V^{B_s D_s^*}$ form factor displays significant curvature. Its larger slope parameters, particularly $a_2^\prime(V) \approx 11.5$, point to a steeper $z$-expansion and possible influence of nearby resonant poles. Furthermore, we observe deviation from heavy-quark symmetry (HQS) in $B_s \to D_s^{(*)}$ transitions, with $V(0)$ differing from $A_0(0) \simeq A_1(0) \simeq A_2(0)$ by approximately $10\%$, consistent with $\mathcal{O}(1/m_Q)$ symmetry breaking and QCD corrections~\cite{Neubert:1993mb}. In $B_s \to D_s^*$ transitions, the $A_1$ form factor remains suppressed and shows minimal $q^2$ dependence, with the ratio $F(q^2_{max})/F(0) \approx 1.4$. Quantitatively, the slope parameters span $a_1^\prime, a_2^\prime \in (-3,~6)$ for $B_s \to P$ and $a_1^\prime, a_2^\prime \in (-4,~12)$ for $B_s \to V$ transitions, as summarized in Table \ref{t2}.

    \item[(iii)] Variations in quark masses and $\beta$ parameters (as shown in Table \ref{t2}) yield mild form factor sensitivities, with total uncertainties of $\sim 12\%$ ($B_s \to P$) and $\sim 27\%$ ($B_s \to V$). In contrast, the associated slope parameters $a_1^\prime$ and $a_2^\prime$ exhibit larger uncertainties, indicating higher sensitivity to input parameters. Furthermore, the uncertainties are predominantly governed by nonperturbative effects, primarily through the meson wave functions \cite{Hazra:2023zdk}, contributing more than $50\%$ to the total uncertainty at $q^2_{max}$.

    \item[(iv)] We emphasize that our numerical results for the $B_s \to K$ and $B_s \to D_s$ transition form factors at both $q^2 = 0$ and $q^2 = q^2_{max}$ are in excellent agreement with LQCD \cite{Flynn:2023nhi, McLean:2019qcx}. Specifically, the LQCD reports $F_{0[1]}^{B_sK}(0) = 0.25 \pm 0.11$ \cite{Flynn:2023nhi} and $F_{0}^{B_sD_s}(q^2_{max}) = 0.907 \pm 0.016$ ~\cite{McLean:2019qcx}; our corresponding results differ by $\sim 4\%$ and $\sim 1\%$, respectively, indicating strong consistency. The $q^2$ dependence of our form factors for $B_s \to K$, $B_s \to D_s$, and $B_s \to D_s^*$ closely follows the steadily rising behavior observed in LQCD studies \cite{Flynn:2023nhi, McLean:2019qcx, Harrison:2021tol}, as shown in Figs.~\ref{f2} and~\ref{f3}.

    \item[(v)] Figure~\ref{flqcd} demonstrates excellent agreement between our $B_s \to D_s^*$ transition form factor calculations and LQCD predictions~\cite{Harrison:2021tol, Yang:2025ydp} across the entire kinematic $q^2$ range, thereby confirming consistency that extends well beyond mere endpoint agreement. This corroborates that our $z$-series parameterization accurately captures nonperturbative QCD dynamics, ensuring reliable extrapolation and reflecting true physical behavior, even when $V(q^2)$ slightly deviates ($<2\sigma$) from the central value yet remains within LQCD uncertainties. This validates our CLFQM approach with LQCD-calibrated heavy quark inputs as a robust tool for predicting form factors. 
    
    \item[(vi)] The self-consistent CLFQM successfully reproduces HQS predictions across kinematic regimes for $B_s$ decays. For $B_s \to D_s$ (pseudoscalar-to-pseudoscalar) transitions, at zero recoil ($q^2 = q_{\text{max}}^2$), Luke's theorem requires $F_0 = F_1$ up to $\mathcal{O}(\Lambda_{\text{QCD}}^2/m_Q^2)$ corrections~\cite{Falk:1990pz}; our calculated values ($F_0=0.92$, $F_1=1.23$) deviate as expected (due to subleading corrections) \cite{Neubert:1996wg}. For $B_s \to D_s^*$ transitions, the vector form factor $V(q^2)$ exhibits scaling consistent with HQET predictions. Specifically, the ratio of the form factor at zero and maximum recoil, $V(q_{\text{max}}^2)/V(0) \approx 1.86$, quantitatively indicate the theoretical scaling factor, $\ln(m_Q)/\sqrt{m_Q} [1+\mathcal{O}(\alpha_s, \Lambda_{\text{QCD}}/m_Q)]$~\cite{Falk:1990pz,Neubert:1996wg} constituting both leading logarithmic and power corrections. Crucially, this HQS structure, including dynamical breaking, emerges naturally, demonstrating intrinsic nonperturbative QCD dynamics and heavy-quark spin-flavor conformation. This validation enables reliable predictions for unmeasured processes (\textit{e.g.}, $B_s \to K^*$) and BSM-sensitive observables.

\end{itemize}

\subsection{Semileptonic decays}
In this subsection, we study the branching ratios of semileptonic $B_s$ meson decays using the transition form factors provided in Table~\ref{t2}. Our predictions for the branching ratios of $B_s^0 \to P\ell^+ \nu_{\ell}$ and $B_s^0 \to V\ell^+ \nu_{\ell}$ are presented in Table~\ref{t3}. Additionally, Table~\ref{t4} summarizes the relative decay widths and mean values of other observables for $B_s$ transitions, including $\langle A_{\rm FB} \rangle$, $\langle C_F^\ell \rangle$, $\langle P_{L(T)}^\ell \rangle$, $\langle F_{L} \rangle$, and $\alpha^*$. Furthermore, Figs.~\ref{f4}$-$\ref{f11} depict the $q^2$ dependence of the differential decay rates and other physical observables for these decays. Our observations are listed as follows:

\begin{itemize}
    \item[(i)] The decay dynamics of $B_s^0 \to P(V)\ell^+ \nu_\ell$ remain structurally identical across lepton flavors; however, the partial widths are sensitive to the lepton mass due to phase-space suppression, leading to significantly reduced branching ratios for heavier leptons such as the $\tau$. The branching ratios for $B_s^0 \to P\ell^+ \nu_\ell$ and $B_s^0 \to V\ell^+ \nu_\ell$ span the ranges $\mathcal{O}(10^{-2})$ to $\mathcal{O}(10^{-5})$ and $\mathcal{O}(10^{-2})$ to $\mathcal{O}(10^{-4})$, respectively (see Table~\ref{t3}). Among these, the $B_s^0 \to D_s^{(*)-}\ell^+ \nu_\ell$ channels yield the highest rates, primarily due to the CKM-favored $b \to c$ transition, which benefits from both a sizable CKM factor $|V_{cb}|$ and favorable phase space. 

    \item[(ii)] We find that semileptonic branching ratios are more sensitive to constituent quark mass variations than to those in the $\beta$ parameter. In $B_s^0 \to P\ell^+ \nu_\ell$ decays, quark mass variations cause up to $\sim 25\%$ deviation, while $\beta$ variations remain within $\sim 3\%$. For $B_s^0 \to V\ell^+ \nu_\ell$, combined uncertainties reach $\sim 32\%$. A similar trend is observed in the $q^2$ dependent decay rates (Fig.~\ref{f4}), where uncertainty bands reflect quadratically combined errors from both sources\footnote{Total uncertainty is computed as $\delta_{\rm total} = \sqrt{\delta_1^2 + \delta_2^2}$.}.

    \item[(iii)] Our predicted branching ratios for semileptonic decays $B_s^0 \to P(V) \, \ell^+ \nu_\ell$ are in excellent agreement with existing experimental data, as summarized in Table~\ref{t3}. In particular, our result (with total uncertainties) $\mathcal{B}(B_s^0 \to K^- \mu^+ \nu_\mu) = (1.07^{+0.27}_{-0.23}) \times 10^{-4}$ show excellent agreement with the LHCb measurement $(1.06 \pm 0.09) \times 10^{-4}$~~\cite{LHCb:2020ist}. Currently, experimental measurements for $B_s^0 \to P(V) \, \ell^+ \nu_\ell$ are limited to muon channels only~\cite{PDG:2024cfk}. Similarly, our predictions $\mathcal{B}(B_s^0 \to D_s^- \mu^+ \nu_\mu) = (2.00^{+0.05}_{-0.07}) \times 10^{-2}$ and $\mathcal{B}(B_s^0 \to D_s^{*-} \mu^+ \nu_\mu) = (4.97^{+1.17}_{-1.13}) \times 10^{-2}$ match well with LHCb results $(2.31 \pm 0.21) \times 10^{-2}$ and $(5.20 \pm 0.50) \times 10^{-2}$~\cite{LHCb:2021qbv} within uncertainties, respectively. Moreover, we observe that the branching ratios for $B_s^0 \to D_s^{*-} \ell^+ \nu_\ell$ decays are approximately twice as large as those for $B_s^0 \to D_s^- \ell^+ \nu_\ell$, indicating a dominant contribution from transverse helicity amplitudes. This helicity structure is further quantified in the partial width decomposition provided in Table~\ref{t4}. A comparative analysis with other theoretical models, namely, RQM~\cite{Faustov:2022ybm}, Type-I CLFQM~~\cite{Soni:2021fky}, and CCQM~\cite{Zhang:2020dla}, reveals that our results are consistent in both trend and magnitude. Furthermore, our predictions fall within the broader range reported by the BS framework~\cite{Chen:2011ut}. Note that the variation of $q^2$ in the differential rates for $B_s^0 \to K^{-} \ell^+ \nu_{\ell}$ and $B_s^0 \to D_s^{(*)-} \ell^+ \nu_{\ell}$ (see Fig. \ref{f4}), exhibits a trend consistent with that observed in LQCD \cite{Flynn:2023nhi, McLean:2019qcx, Harrison:2023dzh, Harrison:2021tol}.

    \item[(iv)] The impact of self-consistency within CLFQM on theoretical predictions for form factors, weak decays, and observables is well-established \cite{Choi:2013mda, Chang:2018zjq, Chang:2019mmh, Chang:2020wvs, Hazra:2023zdk, S:2024adt}. Although a comparison with Type-I formulations is not the primary focus of this work, our findings reveal significant self-consistency effects. Particularly, the branching ratio for $B_s \to K^* \ell \nu_\ell$ can vary by $\sim 30\%$ or more when considering different Type-I form factors and $q^2$ parameterizations \cite{Verma:2011yw, Cheng:2003sm, Jaus:1999zv}, highlighting the crucial role of these choices. A significant discrepancy between $A_0(0)$ and $A_0(q^2_{\text{max}})$ highlights how $q^2$ parameterization\footnote{Note that the numerical results of the Type-I correspondence for Eq. \eqref{e29} for $B_s \to K^*$ form factors are $A_{0}^{B_sK^*}(0)=0.27$, $a_1^\prime(A_{0}^{B_sK^*})=-6.92$, $a_2^\prime(A_{0}^{B_sK^*})=28.07$, $A_{0}^{B_sK^*}(q^2_{max})=8.42$.}, and Type-I inconsistencies alter predictions, emphasizing the need for a self-consistent CLFQM framework. In contrast, for heavy-to-heavy transitions like $B_s \to D_s^* \ell \nu_\ell$, the impact of self-consistency is found to be more moderate, generally remaining within $20\%$. These findings collectively reinforce the imperative of employing self-consistent CLFQM for constructing reliable and precise predictions in semileptonic decays.
    
    \item[(v)] Our LFU results show strong agreement with recent LQCD predictions, within reported uncertainties~\cite{Flynn:2023nhi, McLean:2019qcx}. These ratios serve as precision probes of lepton universality. The LFU ratios for $B_s \to P$ semileptonic decays are given below:
\begin{table}[h!]
    \centering
    \begin{tabular}{ c c c }
    & This Work & LQCD\\ 
    \(\displaystyle \mathcal{R}_K=\frac{\mathcal{B}{(B_s^0 \to K^- \tau^+ \nu_{\tau}})}{\mathcal{B}{(B_s^0 \to K^- \ell^+ \nu_{\ell}})}=\) & $~0.760^{+0.234+0.001}_{-0.325-0.001}~$ & $~0.77 \pm 0.16~$\cite{Flynn:2023nhi}\\
    &&\\
    \(\displaystyle \mathcal{R}_{D_s}=\frac{\mathcal{B}{(B_s^0 \to D_s^- \tau^+ \nu_{\tau}})}{\mathcal{B}{(B_s^0 \to D_s^- \ell^+ \nu_{\ell}})}=\) & $~0.2995^{+0.0092+0.0005}_{-0.0106-0.0005}~$ & $~0.2993 \pm 0.0046~$\cite{McLean:2019qcx}\\
    \end{tabular}
    \end{table}
    
In both cases, the agreement with LQCD is excellent, specifically for $\mathcal{R}_{D_s}$. We also evaluate the ratio of branching fractions for $B_s^0 \to K^- \mu^+ \nu_\mu$ to $B_s^0 \to D_s^- \mu^+ \nu_\mu$:\begin{table}[h!]
    \centering
    \begin{tabular}{ c c c }
    & This Work & LQCD \cite{Flynn:2023nhi}\\ 
    \(\displaystyle \frac{\mathcal{B}{(B_s^0 \to K^- \mu^+ \nu_{\mu}})}{\mathcal{B}{(B_s^0 \to D_s^- \mu^+ \nu_{\mu}})}=\) & $~(5.35^{+1.15+0.00}_{-1.85-0.00}) \t 10^{-3}~$ & $~(4.89 \pm 0.21^{+0.24}_{-0.25}) \t 10^{-3}~$\\
    \end{tabular}
    \end{table}   

Our prediction differs by $\sim 9\%$ from the LQCD result, well within expected uncertainties.

\item[(vi)] Furthermore, $B_s \to V$ semileptonic decays yield:
 \begin{table}[h!]
    \centering
    \begin{tabular}{ c c c }
    & This Work & LQCD\\ 
    \(\displaystyle \mathcal{R}_{K^*}=\frac{\mathcal{B}{(B_s^0 \to K^{*-} \tau^+ \nu_{\tau}})}{\mathcal{B}{(B_s^0 \to K^{*-} \ell^+ \nu_{\ell}})}=\) & $~0.587^{+0.087+0.063}_{-0.017-0.059}~$ & $~-~$\\
    &&\\
    \(\displaystyle \mathcal{R}_{D_s^*}=\frac{\mathcal{B}{(B_s^0 \to D_s^{*-} \tau^+ \nu_{\tau}})}{\mathcal{B}{(B_s^0 \to D_s^{*-} \ell^+ \nu_{\ell}})}=\) & $~0.258^{+0.064+0.045}_{-0.071-0.036}~$ & $~0.266 \pm 0.009~$\cite{Harrison:2023dzh}\\
    \end{tabular}
    \end{table}
    
Our result for $\mathcal{R}_{D_s^*}$ shows a good agreement with LQCD~\cite{Harrison:2023dzh} and LCSR~\cite{Bordone:2019guc} predictions within uncertainties. It may be noted that the inclusion of experimental data (Lattice $+$ Expt.) shifts $\mathcal{R}_{D_s^*}$ to $0.2459 \pm 0.0034$ \cite{Harrison:2023dzh}, representing a ($\approx 2\sigma$) suppression from the lattice-only value of $0.266 \pm 0.009$ and a significant reduction in uncertainty. In addition, preliminary LHCb simulations suggest $\mathcal{R}_{D_s^*} \approx 0.249$~\cite{Paolucci:2022mpj}. Additionally, for $\mathcal{B}(B_s^0 \to D_s^{*-} \mu^+ \nu_{\mu})$, our result shows good agreement within $7\%$ of LQCD prediction $(5.34 \pm 0.56) \times 10^{-2}$ \cite{Harrison:2021tol}. As of now, no experimental results are available for semileptonic $B_s^0 \to P(V) e^+ \nu_e$ or $B_s^0 \to P(V) \tau^+ \nu_{\tau}$ decays. Therefore, our predictions in Table~\ref{t3} provide essential benchmarks for future precision tests of LFU. Moreover, the ratio of the branching fractions,
    \begin{table}[h!]
    \centering
    \begin{tabular}{ c c c c }
    & This Work & LQCD \cite{Harrison:2021tol} & Experiment \cite{LHCb:2020cyw}\\ 
    \(\displaystyle \frac{\mathcal{B}{(B_s^0 \to D_s^- \mu^+ \nu_{\mu}})}{\mathcal{B}{(B_s^0 \to D_s^{*-} \mu^+ \nu_{\mu}})} =\) & $~0.402^{+0.075+0.053}_{-0.086-0.042}~$ & $~0.443 \pm 0.040 \pm 0.004~$ & $~0.464 \pm 0.013 \pm 0.043~$\\
    \end{tabular}
    \end{table}
    
exhibit good agreement with both LQCD and experimental measurement. 

    \item[(vii)] The recent LHCb analysis establishes the quadratic dependence of $B_s^0 \to D_s^{(*)-} \mu^+ \nu_{\mu}$ branching fractions on $|V_{cb}|$ ($\mathcal{B} \propto |V_{cb}|^2$) \cite{LHCb:2020cyw}. Using exclusive determinations of $|V_{cb}|$ through, Caprini-Lellouch-Neubert (CLN), and Boyd-Grinstein-Lebed (BGL) parameterizations. LHCb reports: $|V_{cb}|^{\text{CLN}} = (41.4 \pm 0.6 \pm 0.9 \pm 1.2) \times 10^{-3}$ and $|V_{cb}|^{\text{BGL}} = (42.3 \pm 0.8 \pm 0.9 \pm 1.2) \times 10^{-3}$ yielding the branching ratios, $\mathcal{B}(B_s^0 \to D_s^- \mu^+ \nu_{\mu}) = (2.49 \pm 0.12 \pm 0.14 \pm 0.16) \times 10^{-2} $, and $\mathcal{B}(B_s^0 \to D_s^{*-} \mu^+ \nu_{\mu}) = (5.38 \pm 0.25 \pm 0.46 \pm 0.30) \times 10^{-2}$, respectively \cite{LHCb:2020cyw}. Using $|V_{cb}|^{\text{BGL}} = 42.3 \times 10^{-3}$, our calculations show remarkable consistency with LHCb's results: $\mathcal{B}(B_s^0 \to D_s^- \mu^+ \nu_{\mu}) = 2.15 \times 10^{-2} \quad (\sim 14\% \text{ below central value}) $ and $\mathcal{B}(B_s^0 \to D_s^{*-} \mu^+ \nu_{\mu}) = 5.34 \times 10^{-2} \quad (\sim 1\% \text{ below central value})$. The strong correlation between these results demonstrates that precise determination of $|V_{cb}|$ is critical for accurate theoretical predictions of $B_s^0 \to D_s^{(*)-}\ell^+\nu_{\ell}$ branching fractions and for identifying potential NP.

 \item[(viii)] Further, we evaluate both longitudinal $ \Gamma_L$ and transverse $\Gamma_T $ decay widths. The $B_s^0 \to P\ell^+ \nu_\ell$ transitions exhibit purely longitudinal polarization, while $B_s^0 \to V\ell^+ \nu_\ell$ decays demonstrate a characteristic transverse dominance $\Gamma_T > \Gamma_L$ that emerges from the vector meson's spin structure. We calculate $\langle A_{FB} \rangle$ (Eq.~\eqref{e40}) and plot their $q^2$ behavior (Fig.~\ref{f5}) for crucial dynamical information beyond decay widths. It should be noted that our results (column 4 of Table~\ref{t4}) correspond to relations with positive lepton charge and the direction opposite to the final meson momentum in the rest frame of the initial $B_s$ meson (see Sec.~\ref{S2_C}) \cite{Faustov:2022ybm, Ivanov:2005fd, Becirevic:2016hea}. All calculated $A_{FB}$ values are negative (due to distinct sources) and highly sensitive to NP effects through interference-driven contributions. In $B_s^0 \to P\ell^+\nu_\ell$ decays, the asymmetry arises from scalar–vector interference (in the $\mathcal{H}_{SL}$ term), whereas in $B_s^0 \to V\ell^+\nu_\ell$ transitions, it is predominantly governed by the parity-violating $\mathcal{H}_P$, with a substantial contribution from the ${H}_{-}$ amplitude. Note that the $A_{FB}$ values for $B_s^0 \to P \ell^+ \nu_\ell$ and $\overline{B}_s^0 \to P \ell^- \overline{\nu}_\ell$ transitions are identical due to the vanishing $\mathcal{H}_P$. This consistent negative-value pattern provides an important signature test for theoretical models. Comparison with LQCD results ($A_{FB}(\overline{B}_s^0 \to K^+ \tau^- \overline{\nu}_{\tau})=0.2821 \pm 0.0089$ and $A_{FB}(\overline{B}_s^0 \to K^+ \mu^- \overline{\nu}_{\mu})= 0.0057 \pm 0.0018$ \cite{Flynn:2023nhi}) reveals good agreement for $B_s \to K$ transitions, where our $A_{FB}(B_s^0 \to K^-\tau^+\nu_\tau)$ shows excellent consistency with LQCD, and the extremely small $A_{FB}$ for the muonic channel aligns with uncertainties. For $\overline{B}_s^0 \to D_s^{*+}\tau^-\overline{\nu}\tau$ decay, LQCD reports $-0.092 \pm 0.024$ \cite{Harrison:2021tol} while we obtain $0.067$\footnote{Note that sign difference arises from convention: LQCD defines the forward direction along the final meson momentum, while we use the opposite convention, giving $A_{FB}^\text{LQCD} = -A_{FB}^\text{Our}$.}. The agreement in magnitude ($|A_{FB}^\text{LQCD}| \approx |A_{FB}^\text{Our}|$) confirms consistency within uncertainties, validating both approaches. Our predictions for $\langle A_{FB} \rangle$ also show good agreement with results from other theoretical approaches~\cite{Faustov:2022ybm, Soni:2021fky, Zhang:2020dla}.

    \item[(ix)] In addition, we compute $\langle C_F^\ell \rangle$, $\langle P_L^\ell \rangle$, $\langle P_T^\ell \rangle$, $\langle F_L \rangle$, and $\langle \alpha^* \rangle$ as shown in columns $5-9$ of Table~\ref{t4}, respectively. The corresponding $q^2$ dependent distributions are shown in Figs.~\ref{f6}, \ref{f7}, \ref{f8}, \ref{f9}, and \ref{f11}. It is important to highlight that all the semileptonic decay modes exhibit a negative value for $C_F^\ell$. The negative sign in $B_s^0 \to V\ell^+ \nu_{\ell}$ decays arises because the $\mathcal{H}_L$ component, despite being smaller in magnitude than the unpolarized component $\mathcal{H}_U$, is enhanced by a factor of two in its contribution to the observable (see Eq.~\eqref{e41}). Moreover, $P_T^\ell$ is found to be extremely small in electron decay modes, typically of the order of $\mathcal{O}(10^{-3}-10^{-4})$. Our determinations of $\langle F_L \rangle$ and $\langle P_L^\ell \rangle$ (identical to $\langle A_{\lambda_\ell} \rangle^\text{LQCD}$ \cite{Harrison:2023dzh}, given in Table~\ref{t4}), show excellent agreement with LQCD predictions. For the $B_s^0 \to D_s^{*-} \tau^+ \nu_\tau$ channel, we obtain $F_L = 0.44$, consistent with LQCD values of $(0.440 \pm 0.016)$ \cite{Harrison:2021tol} and $(0.420 \pm 0.012)$ \cite{Harrison:2023dzh}, and $P_L^\tau = 0.52$, in strong agreement with $(0.520 \pm 0.012)$ \cite{Harrison:2021tol} and $(0.5331 \pm 0.0091)$ \cite{Harrison:2023dzh}. The sub-$5\% $ deviations validate the reliability of our helicity amplitude framework. Our $\langle P_L^\tau \rangle$ and $\langle F_L \rangle$ for $\overline{B}_s^0 \to D_s^{*+} \tau^- \overline{\nu}_\tau$ agree with LCSR results\footnote{Note the sign convention $\langle A_{\lambda_\ell} \rangle^\text{LQCD}\equiv-\langle P_L^\ell \rangle$ \cite{Harrison:2023dzh} for comparison with other works.}, within $\sim 7\%$ from central values~\cite{Bordone:2019guc}. Furthermore, the universal negativity of $\alpha^*$ (Fig.~\ref{f11}) quantitatively reflects the longitudinal dominance predicted by HQS, with the $\tau$ channel showing enhanced suppression ($|\alpha^*_\tau/\alpha^*_\mu| \sim 0.7-1.0$) due to $\mathcal{H}_S$ interference. 
\end{itemize}

\subsection{Nonleptonic decays}
In this subsection, we predict the branching ratios of nonleptonic $B_s \to PP/PV$ decays (involving charmed mesons) using the factorization scheme, with decay constants from Eq.~\eqref{e48D} and form factors from Table~\ref{t2}. These results, presented in Tables~\ref{t5},~\ref{t6},~\ref{t7},~\ref{t8},~\ref{t9}, and \ref{t10}, are categorized by CKM hierarchy (favored, suppressed, doubly-suppressed) and, within each, by color topology (Class-I, followed by Class-II) and are compared with other theoretical models therein. As observed in semileptonic $B_s$ decays, uncertainties from input quark masses have a larger impact on branching ratios than those from $\beta$ parameters. Accordingly, we combine these uncertainties in quadrature for the nonleptonic analysis. The key results of our analysis are listed as follows.
\begin{itemize}
    \item[(i)] 
    The branching ratios for two-body nonleptonic $B_s$ meson decays span from $\mathcal{O}(10^{-2})$ to $\mathcal{O}(10^{-8})$. Among $B_s \to PP$ and $B_s \to PV$ modes, the most dominant are the CKM- and color-favored (Class I) decays $\overline{B}_s^0 \to D_s^- D_s^+$ and $\overline{B}_s^0 \to D_s^{+} \rho^-$, with branching ratios of $(1.62^{+0.03}_{-0.02}) \times 10^{-2}$ and $(1.16^{+0.03}_{-0.03}) \times 10^{-2}$, respectively, largely driven by the $B_s \to D_s$ transition form factors and coefficient $a_1$ (see Tables~\ref{t5},~\ref{t6},~\ref{t7},~\ref{t8},~\ref{t9}, and \ref{t10}). Within CKM-favored $B_s \to PP$ decays, $\overline{B}_s^0 \to D^0 K^{0}$ is a notable exception as a color-suppressed (Class-II) mode. At $N_c = \infty$, our predictions for $B_s \to PP$ decays are generally larger than experimental data, except for $\overline{B}_s^0 \to D^0 K^0$. Using $N_c = 3$, where $a_1$ and $a_2$ are smaller, improves agreement for the CKM- and color-favored mode $\overline{B}_s^0 \to D_s^+ \pi^-$. However, a significant discrepancy persists for $\overline{B}_s^0 \to D_s^- D_s^+$ and $\overline{B}_s^0 \to D^0 K^{0}$ decays.

   \item[(ii)] Comparison with experimental branching ratios necessitates the use of an effective color-suppressed amplitude parameter, $a_2^{\mathcal{E}}$, with a significantly larger magnitude, alongside a mildly suppressed color-favored amplitude parameter, $a_1^{\mathcal{E}}$. Detailed analyses of $B \to D\pi$ decays suggest a ratio $|a_2^{\mathcal{E}}/a_1^{\mathcal{E}}| \sim (0.45 - 0.65)e^{\pm i60^\circ}$~\cite{Cheng:2001sc}, while Cheng~\cite{Cheng:2006dm} reports values of $a_1^{\mathcal{E}} = 0.88$ and $a_2^{\mathcal{E}} = -0.47$, corresponding to $|a_2^{\mathcal{E}}/a_1^{\mathcal{E}}| = 0.53$. These results are corroborated by the global analysis of Endo~\textit{et al.}~\cite{Endo:2021ifc}, which finds significant deviations from QCDF predictions, including an $\mathcal{O}(10\%)$ suppression of $a_1$ that can be explained by subleading nonfactorizable corrections. In their analysis, $|a_2^{\text{eff}}|$ was treated as a free parameter, with the fit excluding rescattering effects yielding $|a_2^{\text{eff}}| \sim 0.22 - 0.56$ (see Fig.~3 of Ref.~\cite{Endo:2021ifc}). Additionally, Fleischer \textit{et al.}~\cite{Fleischer:2021cct} observed systematic $|a_1|$ suppression in $B_{(s)}$ decays, with $\overline{B}^0 \to D^+ K^-$ showing the most significant deviation: $|a_1^{DK}| = 0.83 \pm 0.05$ versus the QCDF prediction $1.07 \pm 0.02$ ($4.8\sigma$ tension). Related $b \to c$ transitions exhibit similar suppression: $|a_1^{D\pi}| = 0.83 \pm 0.07$ ($3.3\sigma$) and $|a_1^{D_s\pi}| = 0.87 \pm 0.06$ ($3.2\sigma$), while $b \to u$ modes yield $|a_1^{\pi D_s}| = 0.78 \pm 0.05$. Consequently, we adopt $a_1^{\mathcal{E}} = 0.88$ and $a_2^{\mathcal{E}} = -0.47$ for our analysis~\cite{Cheng:2006dm}. This choice yields $\mathcal{B}(\overline{B}_s^0 \to D_s^+ \pi^-) = (2.87^{+0.07}_{-0.07}) \times 10^{-3}$, in strong agreement with experiment~\cite{PDG:2024cfk}, and improves consistency with experiment for the Class-II decay $\overline{B}_s^0 \to D^0 K^0$. Furthermore, for $B_s \to PP$ decays, we observed that the uncertainties have a significantly greater impact on decay processes involving $F_0^{B_sK}$ form factor ($\sim 27\%$) compared to those involving $F_0^{B_sD_s}$ form factor, where the effect is less than $3\%$.
   
    \item[(iii)] For CKM-suppressed Class-I decays, such as $\overline{B}_s^0 \to D^- D_s^+$ and $\overline{B}_s^0 \to D_s^+ K^-$, the predicted branching ratios are of order $\mathcal{O}(10^{-4})$ in both the large-$N_c$ limit and at $N_c = 3$ (see Table~\ref{t6}). For measured decay $\overline{B}_s^0 \to D_s^+ K^-$, $N_c = 3$ prediction initially deviate from experimental data by $\sim 32\%$~\cite{PDG:2024cfk}; however, this tension is significantly reduced with the inclusion of $a_1^{\mathcal{E}}$, indicating sizable nonfactorizable effects. Our refined prediction, $\mathcal{B}(\overline{B}_s^0 \to D_s^+ K^-) = (2.16^{+0.05}_{-0.05}) \times 10^{-4}$, shows good agreement with experiment. As expected, CKM- and Color-suppressed mode $\overline{B}_s^0 \to \eta_c K^0$ have smaller branching ratio, typically of order $\mathcal{O}(10^{-5})$ or lower (Table~\ref{t6}). Among CKM-doubly-suppressed decays, $\overline{B}_s^0 \to D_s^- K^+$ dominates with a branching ratio of $\mathcal{O}(10^{-5})$, while others fall to $\mathcal{O}(10^{-8})$ (Table~\ref{t7}). These predictions also show good agreement with experiment when using $a_{1}^{\mathcal{E}}$. We emphasize that several theoretical models~\cite{Albertus:2014bfa, Faustov:2012mt, Azizi:2008ty, Li:2009wq, Cai:2021mlt, Endo:2021ifc, Fleischer:2021cct} primarily focus on channels with charmed mesons in the final state, and most, including ours, tend to overestimate branching ratios. While our predictions show good agreement with pQCD results~\cite{Zou:2009zza}, they generally fall within the broader range of theoretical expectations. 

    \item[(iv)] Our predictions for $B_s \to PV$ decays are presented in Tables~\ref{t8}, \ref{t9}, and~\ref{t10}. Decays involving $B_s \to D_s^{(*)}$ transitions, driven by both $F_1(q^2)$ and $A_0(q^2)$ form factors, are color-favored, with branching ratios ranging from $\mathcal{O}(10^{-2})$ to $\mathcal{O}(10^{-4})$. Analogous to $B_s \to PP$ decays, most $B_s \to PV$ transitions are color-allowed, except for color-suppressed modes like $\overline{B}_s \to D^{(*)}K^{(*)}$ and $\overline{B}_s \to K^{(*)} J/\psi(\eta_c)$. Theoretical predictions in the large-$N_c$ limit systematically overestimate the branching ratios of dominant color-allowed and CKM-favored decays, $\overline{B}_s^0 \to D_s^{+}\rho^-$, $\overline{B}_s^0 \to \pi^- D_s^{*+}$, and $\overline{B}_s^0 \to D_s^{-} D_s^{*+}$, which typically measure $\mathcal{O}(10^{-3})$. While using reduced coefficients $a_1$ and $a_2$ at $N_c = 3$ provides partial suppression, substantial discrepancies persist between theory and experiment~\cite{PDG:2024cfk}. The effective coefficients $a_1^{\mathcal{E}}$ and $a_2^{\mathcal{E}}$ significantly improve agreement with data, particularly for $\overline{B}_s^0 \to D_s^{+}\rho^-$ ($(7.16^{+0.18}_{-0.17}) \times 10^{-3}$) and $\overline{B}_s^0 \to \pi^- D_s^{*+}$ ($(2.13^{+0.95}_{-0.80}) \times 10^{-3}$).

    \item[(v)] For CKM-suppressed $B_s \to PV$ decays, dominant modes such as $\overline{B}_s^0 \to D_s^+ K^{*-}$, $\overline{B}_s^0 \to D_s^+ D^{*-}$, $\overline{B}_s^0 \to D^- D_s^{*+}$, and $\overline{B}_s^0 \to K^- D_s^{*+}$ exhibit branching ratios of $\mathcal{O}(10^{-4})$, while the remaining channels lie between $\mathcal{O}(10^{-5})$ and $\mathcal{O}(10^{-7})$. At $N_c = 3$, the reduced values of $a_i$ lead to lower branching ratios, improving agreement with experimental data for Class-I decays. As noted earlier, incorporating the effective coefficients $a_1^{\mathcal{E}}$ and $a_2^{\mathcal{E}}$ yields significantly better consistency with experiment. Notably, for the CKM-suppressed Class-I decay modes, our predictions are in good agreement with other theoretical models~\cite{Albertus:2014bfa, Faustov:2012mt}. Our predictions for CKM-doubly-suppressed $B_s \to PV$ decays generally agree well with pQCD results~\cite{Zou:2009zza}. However, a notable exception is the mode $\overline{B}_s^0 \to K^+ D_s^{*-}$, where theoretical predictions show a significant deviation from experimental data. For both CKM-favored and CKM-suppressed $B_s \to PV$ modes, the largest uncertainty, up to $\sim 44\%$, arises in channels dominated by the $A_0^{B_s D_s^*}(q^2)$ form factor (Tables~\ref{t8} and~\ref{t9}), due to its inherent input uncertainties. In contrast, CKM-doubly-suppressed decays involving $B_s \to K^{(*)}$ transitions show smaller uncertainties, limited to $\sim 30\%$, as shown in Table~\ref{t10}.

    \item[(vi)] Experimental results from LHCb~\cite{LHCb:2015jtt} provide crucial benchmarks through precise measurements of branching fraction ratios involving $D_s^*$ final states. Our theoretical framework shows good agreement, as demonstrated by the ratio:
\begin{table}[h!]
\centering
\begin{tabular}{c c c}
& This Work & Experiment~\cite{LHCb:2015jtt} \\ 
\(\displaystyle \frac{\mathcal{B}(\overline{B}_s^0 \to K^- D_s^{*+})}{\mathcal{B}(\overline{B}_s^0 \to \pi^- D_s^{*+})} =\) & $~0.075^{+0.040}_{-0.047}~$ & $~ 0.068 \pm 0.005^{+0.003}_{-0.002}~$
\end{tabular}
\end{table}

   The absolute branching fractions reported by experiments further validate our theoretical predictions. LHCb measures $\mathcal{B}(B_s^0 \to D_s^{*\mp} K^{\pm}) = (16.3 \pm 1.2^{+0.7}_{-0.5} \pm 4.8) \times 10^{-5}$~\cite{LHCb:2015jtt}, while Belle reports $\mathcal{B}(B_s^0 \to D_s^{*-} \pi^+) = (2.4^{+0.5}_{-0.4} \pm 0.3 \pm 0.4) \times 10^{-3}$ and $\mathcal{B}(B_s^0 \to D_s^{-} \rho^+) = (8.5^{+1.3}_{-1.2} \pm 1.1 \pm 1.3) \times 10^{-3}$~\cite{Belle:2010ldr}. Our $N_c = 3$ predictions for these modes initially deviate by approximately 35\%, 23\%, and 16\%, respectively. Remarkably, incorporating the effective coefficients $a_1^{\mathcal{E}}$ and $a_2^{\mathcal{E}}$ brings all predictions within experimental uncertainties, demonstrating the crucial role of nonfactorizable QCD effects in accurately describing these decay processes.
\end{itemize}

Significant discrepancies persist in $B_{(s)} \to D^{(*)}_{(s)} P(V)$ decays between SM predictions and experimental measurements across various theoretical frameworks. For instance, in $\overline{B}_s^0 \to D_s^+ \pi^-$, the QCDF based SM prediction $(4.42 \pm 0.21) \times 10^{-3}$ exceeds the measured value $(3.00 \pm 0.23) \times 10^{-3}$ by $47\%$, while $\overline{B}_s^0 \to D_s^{*+} \pi^-$ shows a $115\%$ deficit~\cite{Bordone:2020gao}. Such tensions reach $4-5\sigma$ significance in color-allowed modes including $\overline{B}^0 \to D^+ K^-$ and $\overline{B}_s^0 \to D_s^+ \pi^-$, with broader analyses revealing consistent $(30-70)\%$ deviations across $B_s \to D_s^{(*)}P$ channels \cite{Bordone:2020gao, Cai:2021mlt}. Nonfactorizable contributions, calculated through LCSR, yield positive $\mathcal{O}(\text{few}\%)$ amplitude corrections but carry $\sim 100\%$ uncertainties \cite{Piscopo:2023opf}. The Wilson coefficient $a_1$ ($\sim 1$) predicted within QCDF, is consistently found $\sim10\%$ lower in experimental extractions ($\sim 0.8-0.9$), suggesting a universal suppression possibly hinting at NP in tree-level operators \cite{Iguro:2020ndk}. Similarly, the color-suppressed coefficient $a_2$ suffers from large theoretical uncertainties and is often parameterized through an effective $a_2^{\mathcal{E}}$ to capture nonfactorizable and power-suppressed effects \cite{Endo:2021ifc, Piscopo:2023opf}. This discrepancy is further compounded by near-zero SM predictions for CP asymmetries \cite{Panuluh:2024cxc, Gershon:2021pnc}, which may require strong phases generated by nonfactorizable contributions. While NP effects \cite{Iguro:2020ndk} or QED corrections \cite{Beneke:2021jhp} offer partial explanations, model-independent EFT analyses suggest viable NP solutions with additional tensions \cite{Bordone:2021cca, Fleischer:2021cct}. These challenges are particularly pronounced in decays involving neutral $D^0$ mesons or color-suppressed topologies, where subleading effects such as $W$-exchange, penguin diagrams, and final-state interactions contribute nontrivially. While effective coefficients like $a_1^{\mathcal{E}}$ and $a_2^{\mathcal{E}}$ offer a useful phenomenological handle, a full topological amplitude analysis remains the most systematic strategy, though its implementation is hindered by limited data, especially for suppressed or rare $B_s$ decay channels.

\section{Summary} \label{S4}
In this work, we presented a detailed and self-consistent study of semileptonic and nonleptonic decays of the $B_s$ meson into $P(V)$ mesons within the CLFQM, employing a robust $z$-expansion for form factor parameterization. The computed $B_s \to P$ and $B_s \to V$ form factors exhibit excellent agreement with LQCD predictions at both low and high $q^2$, with minimal sensitivity to quark model input parameters. The resulting semileptonic branching ratios, particularly for $B_s^0 \to D_s^{(*)-} \ell^+ \nu_\ell$ and $B_s^0 \to K^- \ell^+ \nu_\ell$, are in strong agreement with experimental data, LQCD and other theoretical works, supporting the validity of our dynamical framework. Ratios probing LFU, $R_K$, $R_{D_s}$, and $R_{D_s^*}$, show remarkable consistency with LQCD and LCSR predictions, highlighting the model’s utility for testing LFU with high precision. Additionally, angular observables such as $\la A_{FB} \ra$, $\la C_F^\ell \ra$, $\la P_{L(T)}^\ell \ra$, $\la F_L \ra$, and $\la \alpha^* \ra$ exhibit consistent trends across lepton flavors and are in close agreement with theoretical expectations; in particular, $\langle A_{FB} \rangle$, $\langle P_L^\tau \rangle$ and $\langle F_L \rangle$ agree with both LQCD and LCSR within $7\%$. These helicity-dependent quantities enhance sensitivity to potential beyond-SM effects, especially in $\tau$ channels. In the nonleptonic sector, we compute branching ratios across all CKM classes and identify dominant Class-I decays such as $\overline{B}_s^0 \to D_s^{(*)+} \pi^-$ and $\overline{B}_s^0 \to D_s^{+} \rho^-$, which are well reproduced using effective Wilson coefficients $a^{\mathcal{E}}_1 = 0.88$, $a^{\mathcal{E}}_2 = -0.47$. This effective approach accounts for leading nonfactorizable QCD effects and achieves good agreement with experimental data. Nonetheless, persistent discrepancies highlight the ongoing challenges in modeling long-distance dynamics and topological amplitudes, which remain a frontier for both theoretical frameworks and experimental validation.

Collectively, our results demonstrate that the self-consistent CLFQM, calibrated by form factor and helicity inputs, offers high predictive power for both SM tests and NP searches. This framework complements LQCD, particularly where nonperturbative effects and form factor uncertainties are critical. The strong agreement of our predictions with LQCD and experimental data across diverse $B_s$ decays establishes robust SM baselines. Our conservative uncertainty estimates ensure robust sensitivity to potential NP effects, while the direct computation of key observables offers crucial input for interpreting forthcoming high-precision measurements from ongoing and future experiments.

\vspace{1em}
\noindent\textbf{\underline{Note added}}$-$ After the completion of this work, we became aware of Ref.~\cite{Bordone:2025jur}, in which the authors present predictions for $\mathcal{R}_{D_s} = 0.2989 \pm 0.0041$, $\mathcal{R}_{D_s^*} = 0.2508 \pm 0.0029$, $F_L^{\overline{B}_s^0 \to D_s^{*+} e^- \overline{\nu}_{e}} =0.5090 \pm 0.0092$, $F_L^{\overline{B}_s^0 \to D_s^{*+} \mu^- \overline{\nu}_{\mu} }=0.5092 \pm 0.0092$, and $F_L^{\overline{B}_s^0 \to D_s^{*+} \tau^- \overline{\nu}_{\tau}} =0.4453 \pm 0.0062$, using combined LQCD + LCSR within the HQE framework. These results are in excellent agreement with our predictions.

\section*{Acknowledgment}
            The author (RD) gratefully acknowledge the financial support by the Department of Science and Technology (SERB:TAR/2022/000606), New Delhi.

\appendix
\section*{Appendix}
\section{Form factor relations} \label{aA}
	The form factors governing the $B_s \to P(V)$ transitions are explicitly expressed as \cite{S:2024adt}, 
	\begin{equation} \label{e27}
		F(q^2) = N_c \int \frac{\d x_1 \d^2{\bm{k}_{\perp}^{\prime}}}{(2 \pi)^3}\frac{\chi_{B_s}^{\prime}\chi_{P(V)}^{\prime\prime}} {2 x_{2}} \widetilde{F}(x_1, {\bm{k}}_{\perp}^{\prime}, q^2),
	\end{equation}
	where \begin{align} \label{e28}
		\chi_{B_s}^{\prime} = \frac{1}{\sqrt{2 N_c}} \sqrt{\frac{x_2}{x_1}}\frac{\psi(x_1, {\bm{k}}_{\perp}^{\prime})}{\hat{M}_0^{\prime}},~\text{and} ~~~\chi_{P(V)}^{\prime\prime} = \frac{1}{\sqrt{2 N_c}} \sqrt{\frac{x_2}{x_1}}\frac{\psi(x_1, {\bm{k}}_{\perp}^{\prime\prime})}{\hat{M}_0^{\prime\prime}}.
	\end{align}
    It is important to emphasize that the integration in Eq.~\eqref{e27} is performed over the ranges $x_1 \subset [0,1]$ and $\bm{k}_{\perp}^{\prime} \subset [0, \infty]$. The form factor function $\widetilde{F}(x_1, {\bm{k}}_{\perp}^{\prime}, q^2) \equiv \{ \widetilde{f}_{\pm}(x_1, {\bm{k}}_{\perp}^{\prime}, q^2)$, $\widetilde{g}(x_1, {\bm{k}}_{\perp}^{\prime}, q^2)$, $\widetilde{f}(x_1, {\bm{k}}_{\perp}^{\prime}, q^2)$, $\widetilde{a}_{\pm}(x_1, {\bm{k}}_{\perp}^{\prime}, q^2)\}$ corresponding to $B_s \to P(V)$ transitions are defined as follows \cite{S:2024adt},
	\begin{itemize}
		\item[(i)] $B_s$ to $P$ form factors,
		\begin{align}
			\label{eA1}
			\widetilde{f}_{+}(x_1, {\bm{k}}_{\perp}^{\prime}, q^2) = x_1 M_0^{\p 2} + x_1 M_0^{\prime\prime 2} +x_2q^2 - x_1 (m_1^{\prime} -m_2)^2 - x_1 (m_1^{\prime\prime} -m_2)^2 - x_2 (m_1^{\prime} -m_1^{\prime\prime})^2, 
		\end{align}
		\begin{align}
			\label{eA2}
			\widetilde{f}_{-}(x_1, {\bm{k}}_{\perp}^{\prime}, q^2) = & -2 x_1 x_2 M^{\p 2} - 2 k_{\perp}^{\p 2} - 2 m_1^{\prime} m_2 +2 (m_1^{\prime\prime} - m_2)(x_2 m_1^{\prime} +x_1 m_2) \nonumber\\& - 2 \frac{{\bm{k}}_{\perp}^{\prime} \cdot {\bm{q}}_{\perp}}{q^2}\Big[(x_1 - x_2)M^{\p 2} +M^{\prime\prime 2} + x_2 (q^2 +q\cdot p) +2 x_1 M_0^{\p 2} \nonumber\\&- 2(m_1^{\prime} + m_1^{\prime\prime})(m_1^{\prime} -m_2)\Big] + 4 \frac{p\cdot q}{q^2}\Big[k_{\perp}^{\p 2} + \frac{2 ({\bm{k}}_{\perp}^{\prime} \cdot {\bm{q}}_{\perp})^2}{q^2}\Big] + 4 \frac{({\bm{k}}_{\perp}^{\prime} \cdot {\bm{q}}_{\perp})^2}{q^2}.
		\end{align}
		
		\item[(ii)] $B_s$ to $V$ form factors,
		\begin{align} 
			\label{eA3}
			\widetilde{g}(x_1, {\bm{k}}_{\perp}^{\prime}, q^2) = -2\Big\{x_2m_1^{\prime} + x_1m_2 + (m_1^{\prime} - m_1^{\prime\prime})\frac{{\bm{k}}_{\perp}^{\prime}\cdot {\bm{q}}_\perp}{q^2} + \frac{2}{D_{V, con}^{\prime\prime}}\Big[{k}_{\perp}^{\p 2} +\frac{({\bm{k}}_{\perp}^{\prime} \cdot {\bm{q}}_{\perp})^2}{q^2} \Big] \Big\},	\end{align}
		\begin{align}
			\label{eA4}
			\widetilde{f}(x_1, {\bm{k}}_{\perp}^{\prime}, q^2) = & -2\Big\{
			-(m_1^{\prime}+m_1^{\prime\prime})^2 (m_1^{\prime}-m_2) +(x_1m_2-x_2m_1^{\prime}) M^{\p 2} +(x_1m_2+x_2m_1^{\prime}) M^{\prime\prime 2} \nonumber\\& -x_1(m_2-m_1^{\prime} )(M_0^{\p 2}+M_0^{\prime\prime 2}) + 2x_1m_1^{\prime\prime}M_0^{\p 2} -4 (m_1^{\prime}-m_2)\Big({k}_{\perp}^{\p 2} +\frac{({\bm{k}}_{\perp}^{\prime} \cdot {\bm{q}}_{\perp})^2}{q^2}\Big) \nonumber\\& - m_2 q^2 -(m_1^{\prime}+m_1^{\prime\prime})(q^2+q\cdot p)\frac{{\bm{k}}_{\perp}^{\prime}\cdot {\bm{q}}_{\perp}}{q^2} +4 (m_1^{\prime}-m_2) B_1^{(2)} +\frac{2}{D_{V, con}^{\prime\prime}} \Big[ \Big({k}_{\perp}^{\p 2} \nonumber\\& +\frac{({\bm{k}}_{\perp}^{\prime} \cdot {\bm{q}}_{\perp})^2}{q^2}\Big) \Big((x_1-x_2)M^{\p 2} + M^{\prime\prime 2} -2(m_1^{\prime}-m_1^{\prime\prime})(m_1^{\prime}-m_2) +2x_1M_0^{\p 2} \nonumber\\& -q^2 -2 (q^2 +q\cdot p)\frac{{\bm{k}}_{\perp}^{\prime}\cdot {\bm{q}}_{\perp}}{q^2}\Big) - \Big(M^{\p 2} +M^{\prime\prime 2} -q^2 + 2(m_1^{\prime}-m_2)(m_1^{\prime\prime} \nonumber\\& +m_2)\Big) B_1^{(2)} +2 B_3^{(3)}\Big]\Big\}, 	\end{align}
		\begin{align}
			\label{eA5}
			\widetilde{a}_{+}(x_1, {\bm{k}}_{\perp}^{\prime}, q^2) = & 2\Big\{(m_1^{\prime\prime} -2x_1m_1^{\prime} +m_1^{\prime} +2x_1m_2)\frac{{\bm{k}}_{\perp}^{\prime}\cdot {\bm{q}}_{\perp}}{{q}_{\perp}^2} + (x_1-x_2)(x_2m_1^{\prime} +x_1m_2) \nonumber\\& +\frac{2}{D_{V, con}^{\prime\prime}} \frac{{\bm{k}}_{\perp}^{\prime\prime}\cdot {\bm{q}}_{\perp}}{x_2 {q}_{\perp}^2} \Big[{\bm{k}}_{\perp}^{\prime}\cdot{\bm{k}}_{\perp}^{\prime\prime}+ (x_1m_2 - x_2m_1^{\prime\prime})(x_1m_2+ x_2m_1^{\prime})\Big] \Big\},
		\end{align}
		\begin{align}
			\label{eA6}
			\widetilde{a}_{-}(x_1, {\bm{k}}_{\perp}^{\prime}, q^2) = & -2\Big\{(3-2x_1)(x_2m_1^{\prime}+x_1m_2) -\Big[(6x_1-7)m_1^{\prime}+(4 -6x_1)m_2 +m_1^{\prime\prime}\Big]\frac{{\bm{k}}_{\perp}^{\prime} \cdot {\bm{q}}_{\perp}}{q^2}\nonumber\\& + 4(m_1^{\prime} - m_2) \Big[ 2 \Big( \frac{{\bm{k}}_{\perp}^{\prime} \cdot {\bm{q}}_{\perp}}{q^2}\Big)^2 + \frac{{k}_{\perp}^{\p 2}}{q^2}\Big] {- 4 \frac{(m_1^{\prime} - m_2)}{q^2} B_1^{(2)}} + \frac{1}{D_{V, con}^{\prime\prime}} \Big[-2\big(M^{\p 2} \nonumber\\& +M^{\prime\prime 2} -q^2 + 2(m_1^{\prime}-m_2)(m_1^{\prime\prime}+m_2)\big) \big(A_3^{(2)} + A_4^{(2)} - A_2^{(1)} \big) + \Big(2M^{\p 2} -q^2 \nonumber\\& -x_1(M^{\p2} - M_0^{\p2}) +x_1(M^{\prime\prime2} - M_0^{\prime\prime2})- 2(m_1^{\prime} -m_2)^2 +(m_1^{\prime} +m_1^{\prime\prime})^2 \Big)\big(A_1^{(1)}\nonumber\\& + A_2^{(1)} -1\big) +2Z_2\big(2A_4^{(2)} -3A_2^{(1)} +1\big) + 2 \frac{q\cdot p}{q^2} \big(4 A_2^{(1)}A_1^{(2)} - 3A_1^{(2)}\big) \nonumber\\& +\frac{2}{q^2}\Big(\big(M^{\p 2} + M^{\prime\prime 2} -q^2 + 2(m_1^{\prime}-m_2)(m_1^{\prime\prime}+m_2)\big) B_1^{(2)} -2 B_3^{(3)}\Big)\Big] \Big\}.
		\end{align}
	\end{itemize}
	The coefficients $A^{(i)}_{j}$ and $B^{(i)}_{j}$ are given as \cite{S:2024adt},
	\begin{equation} \label{eA7}
		\begin{gathered}
			A_1^{(1)} = \frac{x_1}{2}, \hspace{0.5cm}
			A_1^{(2)} = -{k}_{\perp}^{\p 2} - \frac{({\bm{k}}_{\perp}^{\prime}\cdot{\bm{q}}_{\perp})^2}{q^2}, \hspace{0.5cm}
			A_2^{(1)} = A_1^{(1)} - \frac{{\bm{k}}_{\perp}^{\prime}\cdot{\bm{q}}_{\perp}}{q^2}, \\
			A_3^{(2)} = A_1^{(1)}A_2^{(1)}, \hspace{0.5cm}
			A_4^{(2)} = (A_2^{(1)})^2 - \frac{1}{q^2}A_1^{(2)}, \\
			B_1^{(2)} = A_1^{(1)}Z_2 - A_2^{(1)},\hspace{0.5cm}
			B_3^{(3)} = B_1^{(2)}Z_2 + (p^2 -\frac{(q\cdot p)^2}{q^2})A_1^{(1)}A_1^{(2)}, ~\text{ and}
		\end{gathered}
	\end{equation}
	\begin{equation} \label{eA8}
		Z_2 = x_1(M^{\p2} - M_0^{\p2}) + m_1^{\p 2} -m_2^2 +(1-2x_1)M^{\p 2} + (q^2 + q\cdot p)\frac{{\bm{k}}_{\perp}^{\prime}\cdot{\bm{q}}_{\perp}}{q^2}. 
	\end{equation}
	It is important to emphasize that the above expressions for the form factors are derived within the Type-I scheme. The corresponding Type-II expressions can be obtained by replacing $M^{\p(\p\p)}$ with $M_0^{\p(\p\p)}$, as detailed in Ref.~\cite{Chang:2019mmh, S:2024adt}. Furthermore, these expressions are specific to the longitudinal polarization case ($\lambda = 0$); for transverse polarizations ($\lambda = \pm$), the results can be derived by excluding all terms involving the $B^{(i)}_{j}$ functions.
    \bibliography{BsDs_bibtex_new.bib}
    
\newpage

	\begin{table}[ht]
		\caption{Transition pole masses for $B_s \to P$ and $V$ form factors (in GeV). }
		\label{t1}
		\begin{tabular}{c | c | c | c | c}
			\hline 	\hline
			\multirow{2}{*}{Quark transition}& $F_1(q^2), V(q^2)$ & $F_0(q^2)$	         & $A_0(q^2)$		  & $A_1(q^2), A_2(q^2)$ \\ \cline{2-5}
			& $J^P = 1^-$   & $0^+$	& $0^-$	   & $1^+$ \\ \hline
			$b \to u$	    & $5.325$ & $5.670$ &  $5.279$ & $5.726$ \\
			$b \to c$	    & $6.331$ & $6.712$ & $6.274$  & $6.736$ \\ \hline \hline
		\end{tabular}
	\end{table}

	\begin{table}[ht]
		\caption{$B_s \to P$ and $B_s \to V$ form factors in self-consistent CLFQM (Type-II with $z$-series parameterization using Eq. \eqref{e29})}
		\label{t2}
		\begin{tabular}{c|  c  c  c  c  } \hline 	\hline
		{Form Factor} & $F(0)= a_0^\prime$ & $a_1^\prime$ & $a_2^\prime$ &{$F(q^2_{max})$} \\  \hline
	\multicolumn{5}{c}{$B_s \to P$ transitions} \\ \hline
        $F_0^{B_sK}$              & $0.26^{+0.03+0.00}_{-0.03-0.00}$ & $-0.33^{+0.12+0.02}_{-0.17-0.02}$ & $-0.51^{+0.75+0.02}_{-0.47-0.02}$ & $1.18^{+0.54+0.01}_{-0.40-0.01}$ \\
        $F_1^{B_sK}$              & $0.26^{+0.03+0.00}_{-0.03-0.00}$ & $-1.24^{+0.20+0.01}_{-0.24-0.01}$ & $1.75^{+0.90+0.06}_{-0.66-0.06}$  & $4.73^{+1.07+0.02}_{-0.89-0.03}$ \\ \hline
        $F_0^{B_sD_s}$            & $0.66^{+0.01+0.00}_{-0.01-0.00}$ & $-0.45^{+0.41+0.08}_{-0.49-0.08}$ & $-0.72^{+2.31+0.11}_{-1.73-0.10}$ & $0.92^{+0.04+0.00}_{-0.03-0.00}$ \\
        $F_1^{B_sD_s}$            & $0.66^{+0.01+0.00}_{-0.01-0.00}$ & $-3.18^{+0.50+0.05}_{-0.57-0.05}$ & $6.05^{+3.66+0.24}_{-2.96-0.23}$  & $1.23^{+0.05+0.00}_{-0.05-0.00}$ \\ \hline

        \multicolumn{5}{c}{$B_s \to V$ transitions} \\ \hline
        $V^{B_sK^*}$              & $0.27^{+0.06+0.01}_{-0.05-0.01}$ & $-1.73^{+0.39+0.00}_{-0.51-0.00}$ & $3.78^{+1.89+0.10}_{-1.27-0.10}$  & $2.35^{+0.71+0.01}_{-0.55-0.02}$ \\
        $A_0^{B_sK^*}$            & $0.30^{+0.02+0.01}_{-0.05-0.01}$ & $-1.60^{+0.12+0.02}_{-0.05-0.02}$ & $2.70^{+0.70+0.06}_{-0.64-0.06}$  & $2.39^{+0.04+0.04}_{-0.16-0.05}$ \\
        $A_1^{B_sK^*}$            & $0.22^{+0.01+0.01}_{-0.02-0.01}$ & $-0.45^{+0.04+0.03}_{-0.02-0.02}$ & $-0.62^{+0.10+0.08}_{-0.01-0.08}$ & $0.72^{+0.03+0.00}_{-0.04-0.00}$ \\
        $A_2^{B_sK^*}$            & $0.19^{+0.04+0.00}_{-0.03-0.00}$ & $-1.13^{+0.22+0.01}_{-0.26-0.01}$ & $2.11^{+0.78+0.11}_{-0.58-0.10}$  & $1.17^{+0.28+0.00}_{-0.24-0.00}$ \\ \hline
        $V^{B_sD_s^*}$            & $0.73^{+0.06+0.04}_{-0.05-0.04}$ & $-4.60^{+0.92+0.06}_{-1.19-0.01}$ & $11.46^{+7.25+0.13}_{-5.06-0.44}$ & $1.36^{+0.19+0.04}_{-0.16-0.06}$ \\
        $A_0^{B_sD_s^*}$          & $0.59^{+0.11+0.05}_{-0.11-0.05}$ & $-3.40^{+0.11+0.15}_{-0.02-0.09}$ & $8.00^{+2.53+0.10}_{-2.85-0.48}$  & $1.08^{+0.14+0.07}_{-0.15-0.09}$ \\
        $A_1^{B_sD_s^*}$          & $0.58^{+0.05+0.03}_{-0.05-0.04}$ & $-1.03^{+0.05+0.17}_{-0.05-0.13}$ & $-0.31^{+1.57+0.03}_{-1.48-0.22}$ & $0.83^{+0.05+0.03}_{-0.06-0.05}$ \\
        $A_2^{B_sD_s^*}$          & $0.58^{+0.03+0.01}_{-0.03-0.02}$ & $-3.53^{+0.56+0.28}_{-0.65-0.22}$ & $8.41^{+4.03+1.09}_{-3.11-1.29}$  & $1.03^{+0.10+0.00}_{-0.09-0.01}$ \\ \hline \hline
		\end{tabular}
	\end{table}
\begin{figure}
\centering
\begin{subfigure}[b]{0.48\textwidth}
    \centering
    \includegraphics[width=\textwidth]{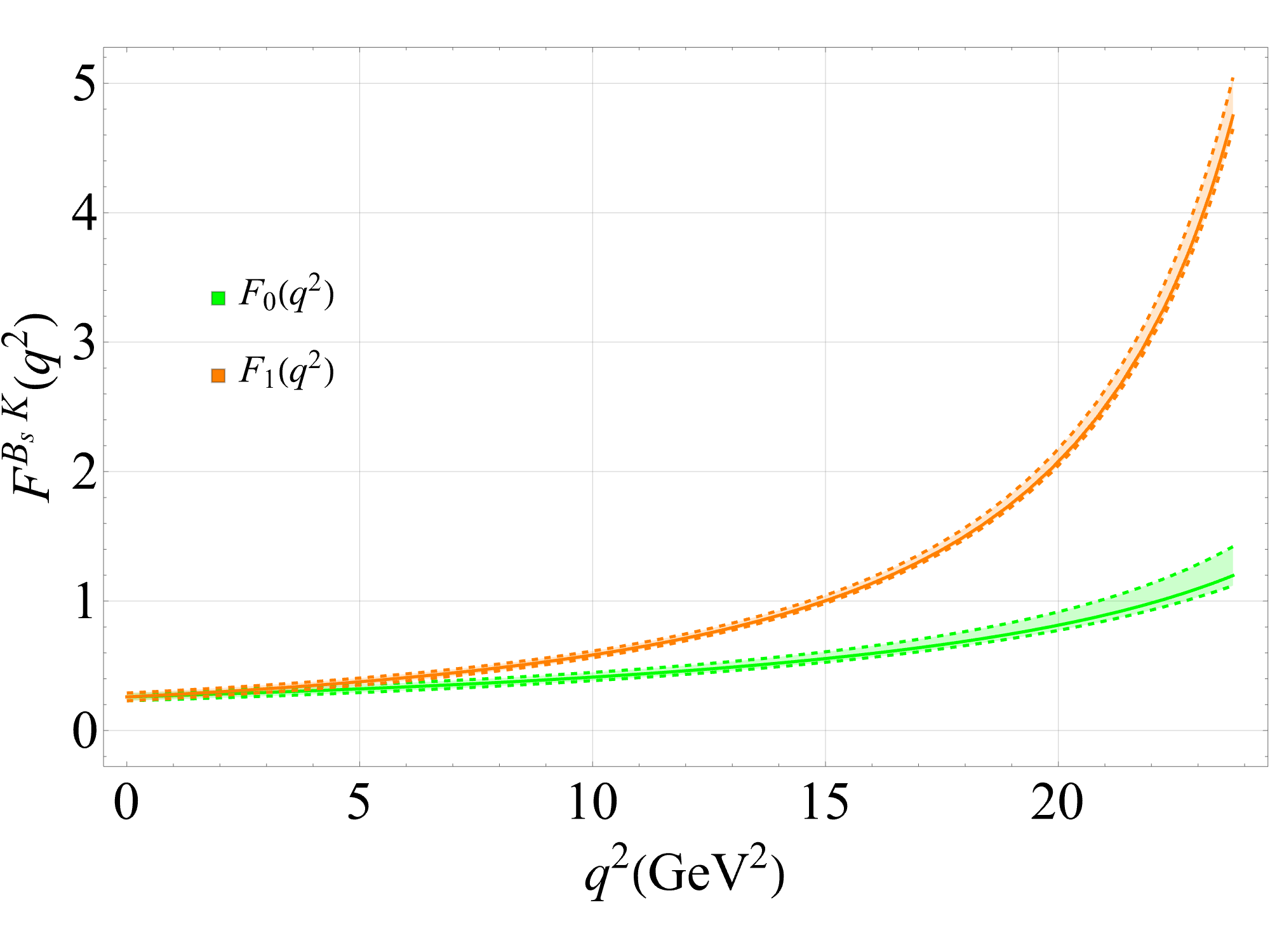}
    \caption{$B_s \to K$ transition} \label{f2a}
\end{subfigure}
\hfill
\begin{subfigure}[b]{0.48\textwidth}
    \centering
    \includegraphics[width=\textwidth]{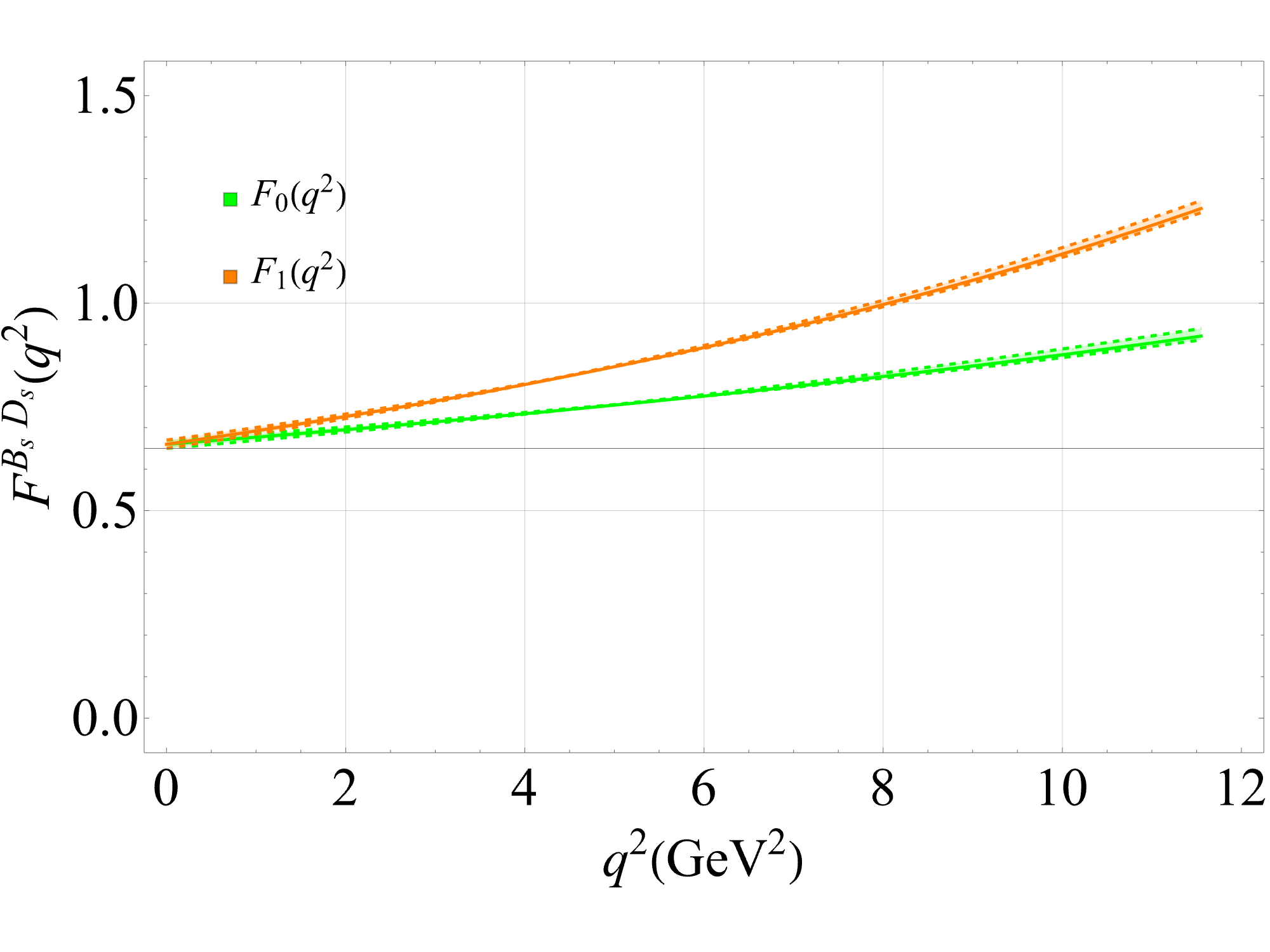}
    \caption{$B_s \to D_s$ transition} \label{f2b}
\end{subfigure}
\caption{$q^2$ dependence of $B_s \to P$ form factors.}\label{f2}
\end{figure}

\begin{figure}
\centering
\begin{subfigure}[b]{0.48\textwidth}
    \centering
    \includegraphics[width=\textwidth]{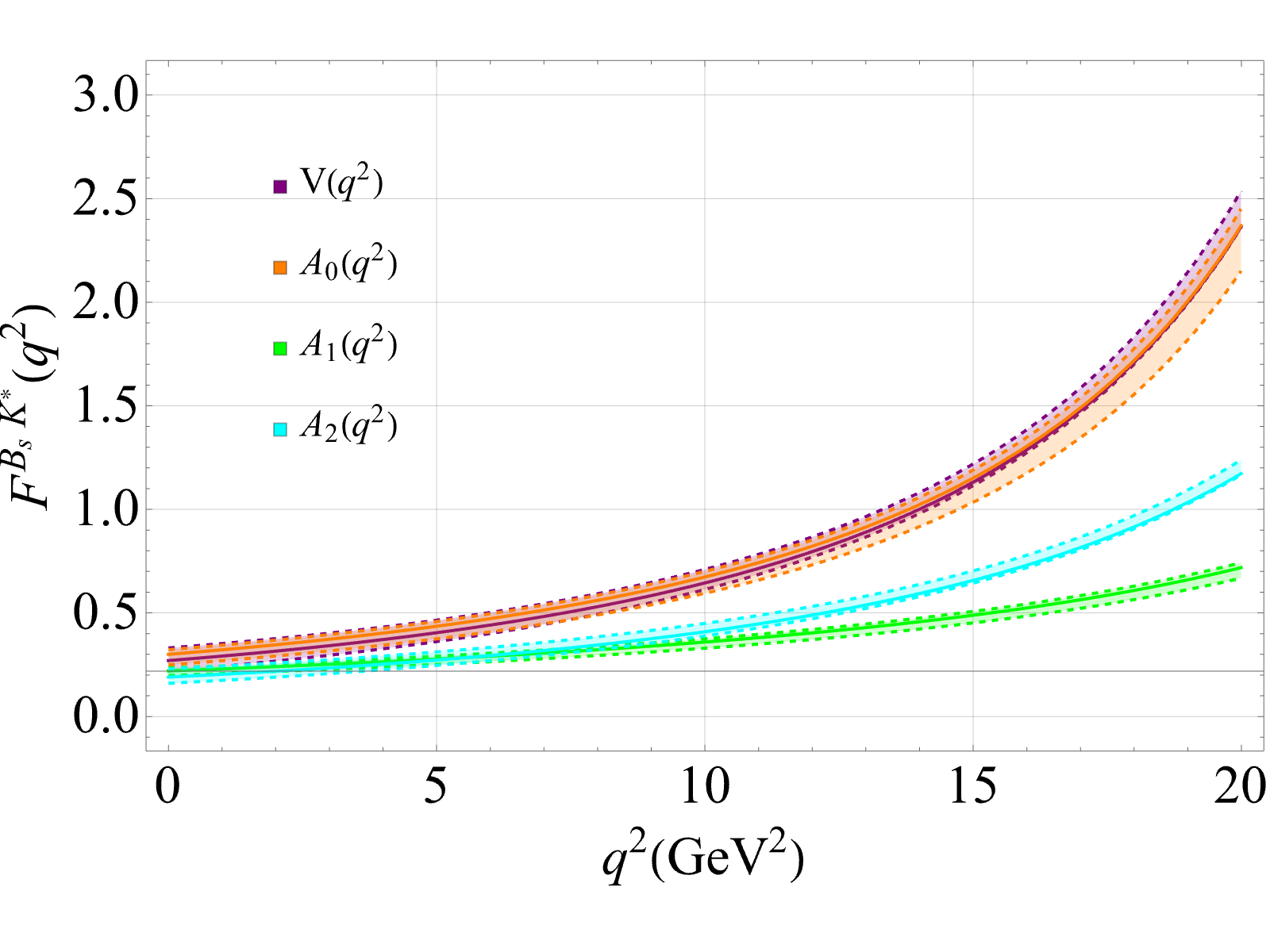}
    \caption{$B_s \to K^*$ transition}
\end{subfigure}
\hfill
\begin{subfigure}[b]{0.48\textwidth}
    \centering
    \includegraphics[width=\textwidth]{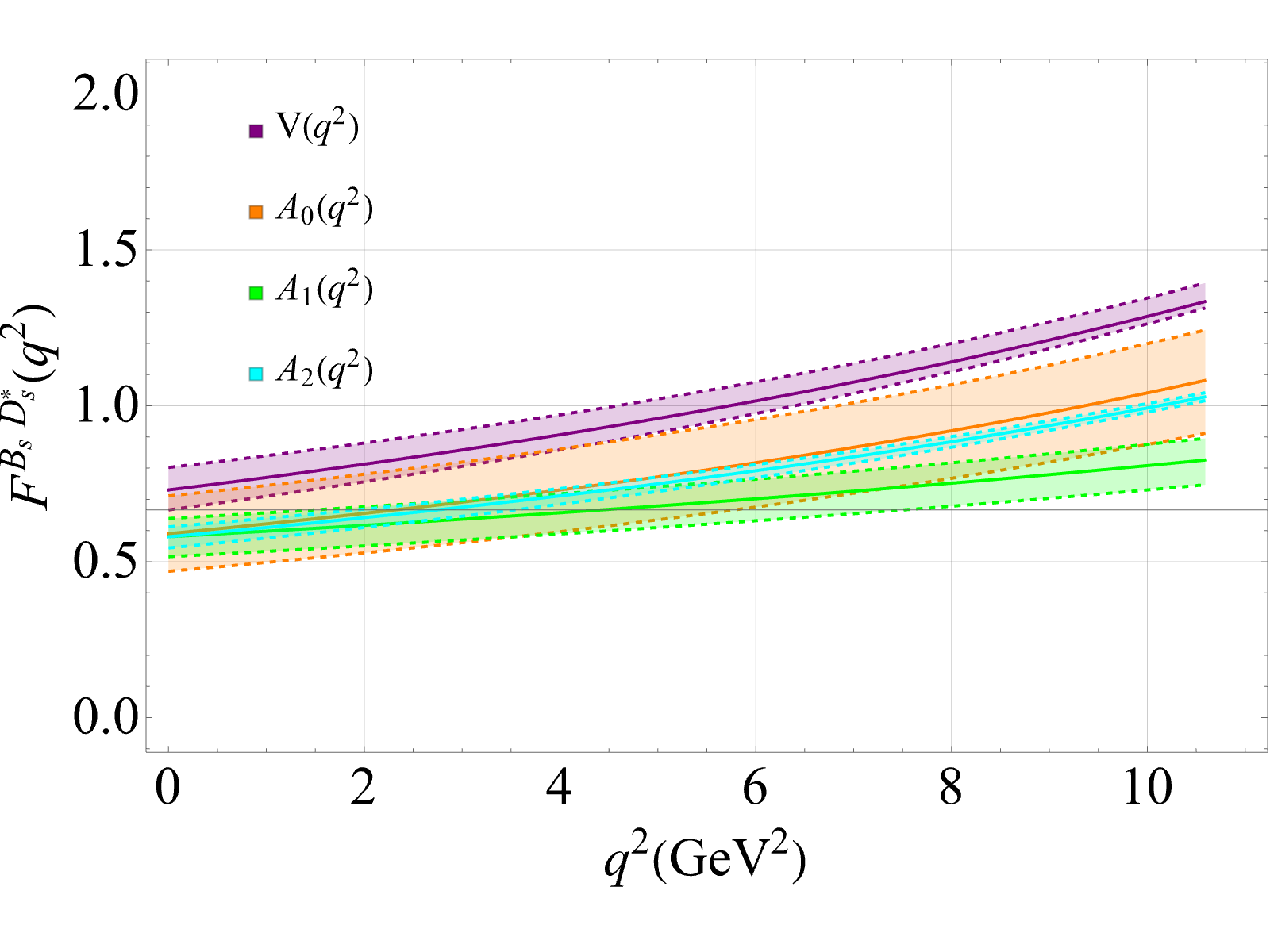}
    \caption{$B_s \to D_s^*$ transition}
\end{subfigure}
\caption{$q^2$ dependence of $B_s \to V$ form factors.} \label{f3}
\end{figure}

\begin{figure}
\centering
\begin{subfigure}[b]{0.48\textwidth}
    \centering
    \includegraphics[width=\textwidth]{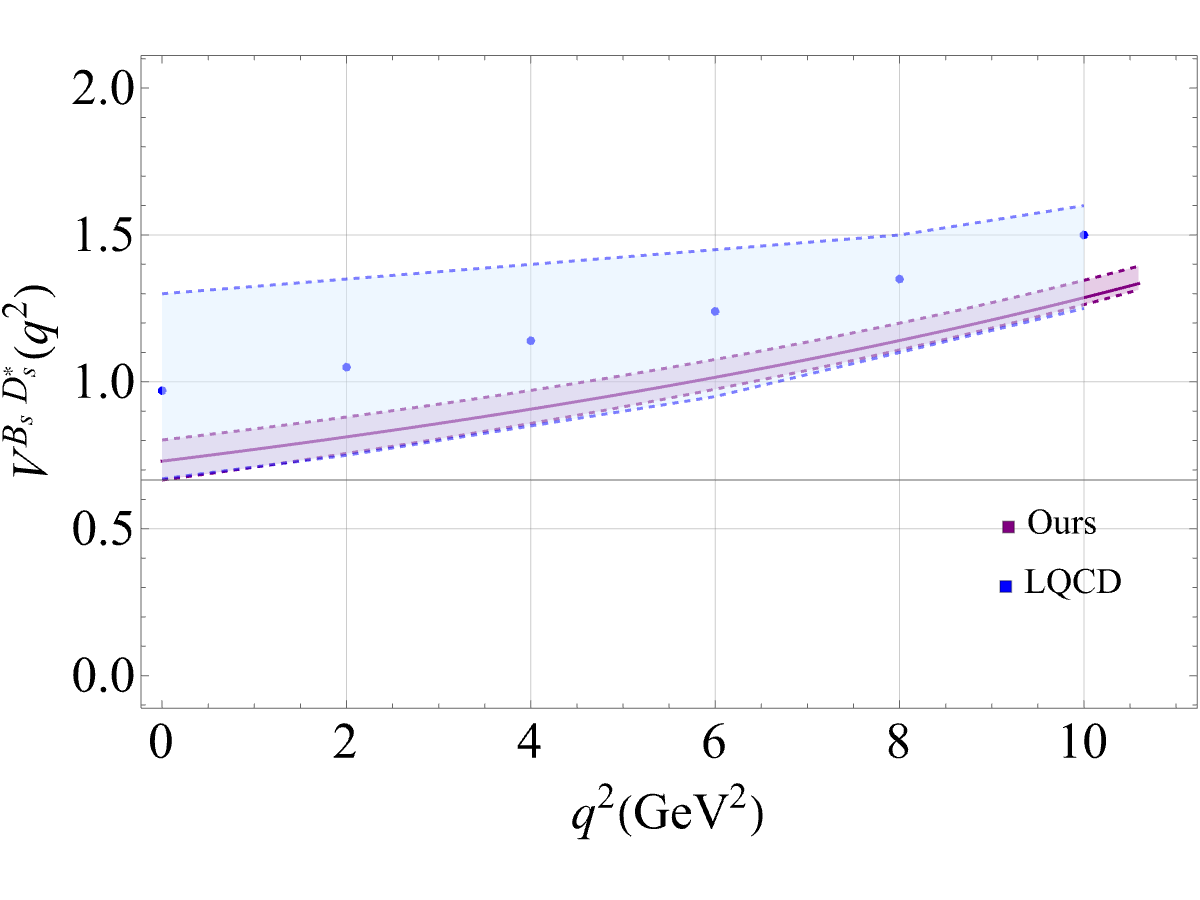}
    \caption{$V^{B_sD_s^*}(q^2)$}
\end{subfigure}
\hfill
\begin{subfigure}[b]{0.48\textwidth}
    \centering
    \includegraphics[width=\textwidth]{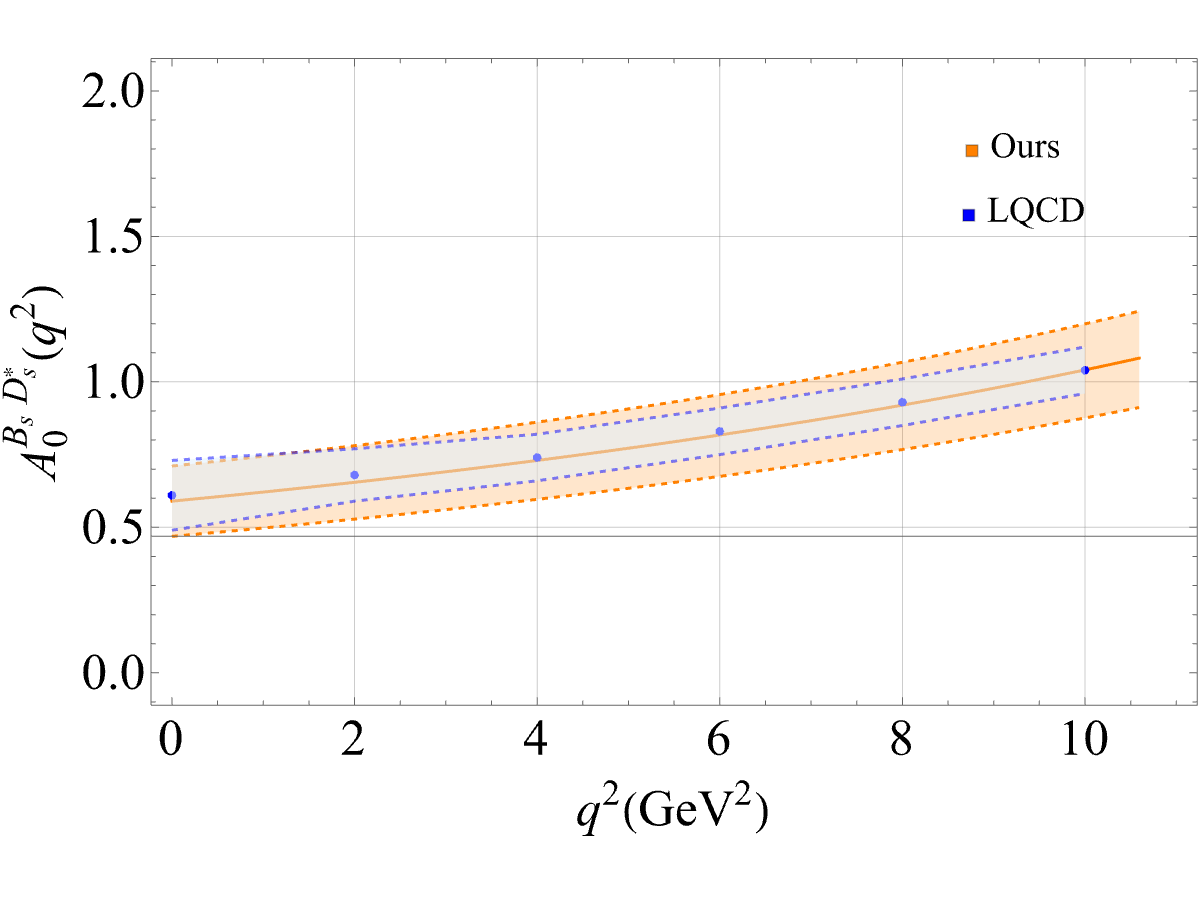}
    \caption{$A_0^{B_sD_s^*}(q^2)$}
\end{subfigure}
\hfill
\begin{subfigure}[b]{0.48\textwidth}
    \centering
    \includegraphics[width=\textwidth]{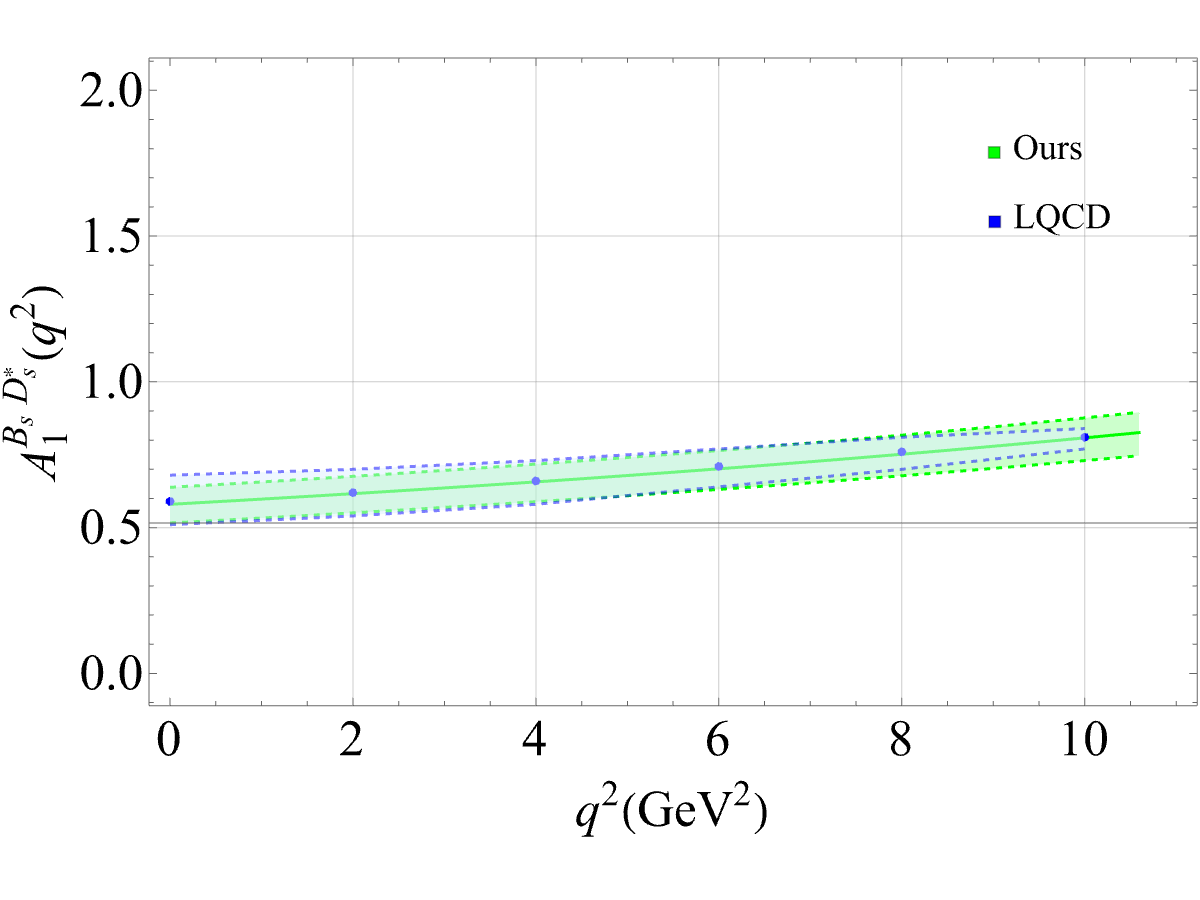}
    \caption{$A_1^{B_sD_s^*}(q^2)$}
\end{subfigure}
\hfill
\begin{subfigure}[b]{0.48\textwidth}
    \centering
    \includegraphics[width=\textwidth]{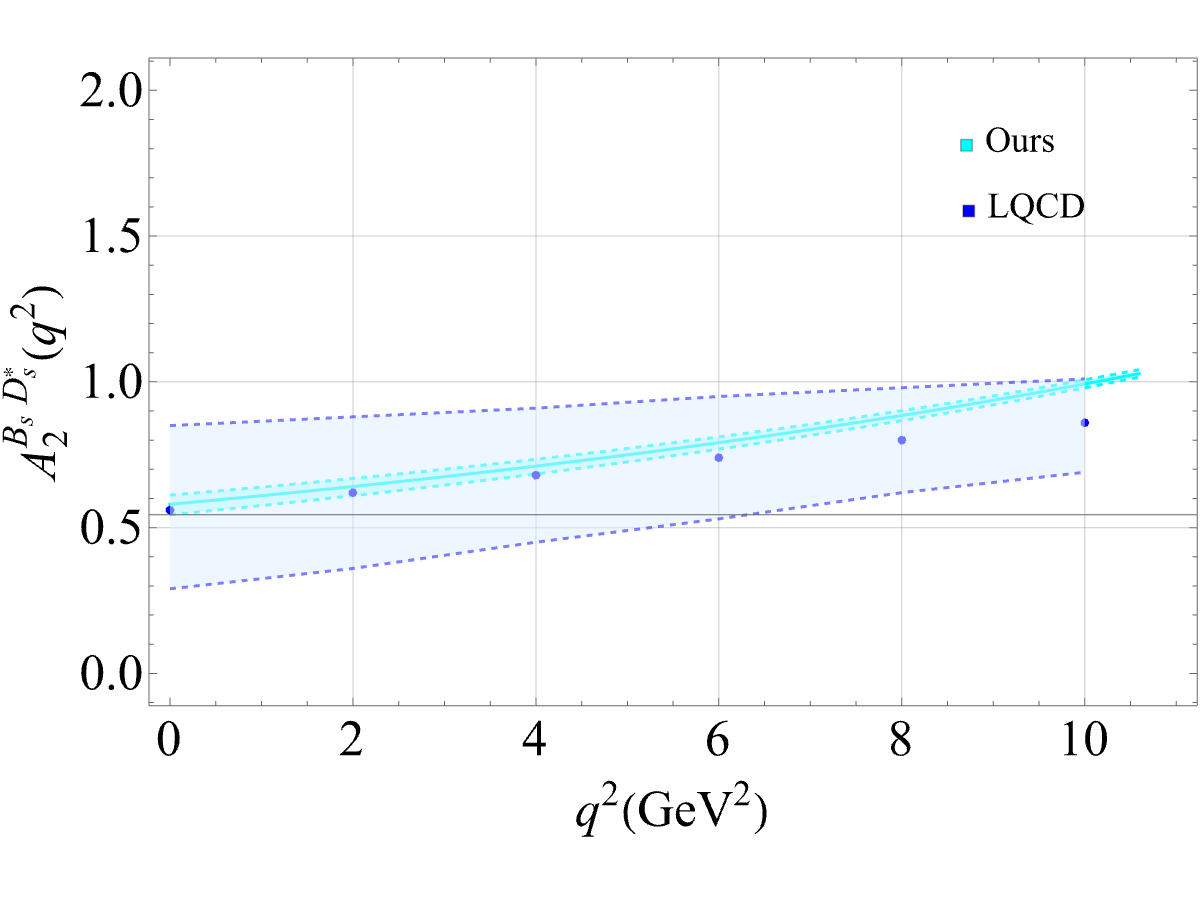}
    \caption{$A_2^{B_sD_s^*}(q^2)$}
\end{subfigure}
\caption{Comparison of $B_s \to D_s^*$ form factors with LQCD results \cite{Harrison:2021tol, Yang:2025ydp}.} \label{flqcd}
\end{figure}

\begin{table}[ht]
\caption{Branching ratios of $B_{s} \to P (V) \ell\nu_\ell$ decays in Type-II CLFQM. }
\label{t3}
\begin{tabular}{ c c c c c c}
\hline 	\hline
 \multirow{2}{*}{Decay} & \multirow{2}{*}{This Work} & {RQM}& {CLFQM} & {CCQM} & {PDG} \\ 
 & & \cite{Faustov:2022ybm} & (Type-I) \cite{Zhang:2020dla} & \cite{Soni:2021fky} & \cite{PDG:2024cfk} \\ \hline
$B_{s}^0 \to K^{-}e^{+} \nu_e$              & $(1.07^{+0.27+0.00}_{-0.23-0.00}) \t 10^{-4}$ & $1.56 \t 10^{-4}$                   & $9.23 \t 10^{-5}$                              & -                                &       -                          \\
$B_{s}^0 \to K^{-}\mu^{+} \nu_{\mu}$        & $(1.07^{+0.27+0.00}_{-0.23-0.00}) \t 10^{-4}$ & $1.55 \t 10^{-4}$                   & $9.23 \t 10^{-5}$                              & -                                & $(1.06 \pm 0.09) \t 10^{-4}$    \\
$B_{s}^0 \to K^{-}\tau^{+} \nu_{\tau}$      & $(8.13^{+2.04+0.01}_{-1.79-0.01}) \t 10^{-5}$ & $9.10 \t 10^{-5}$                   & $6.18 \t 10^{-5}$                              & -                                &         -                        \\ \hline
$B_{s}^0 \to D_{s}^{-}e^{+} \nu_e$          & $(2.01^{+0.05+0.00}_{-0.04-0.06}) \t 10^{-2}$ & $2.12 \t 10^{-2}$                   & $2.41 \t 10^{-2}$                              & $(2.89 \pm 0.50) \t 10^{-2}$     &         -                        \\
$B_{s}^0 \to D_{s}^{-}\mu^{+} \nu_{\mu}$    & $(2.00^{+0.05+0.00}_{-0.04-0.06}) \t 10^{-2}$ & $2.12 \t 10^{-2}$                   & $2.41 \t 10^{-2}$                              & $(2.88 \pm 0.49) \t 10^{-2}$     & $(2.31 \pm 0.21) \t 10^{-2}$    \\
$B_{s}^0 \to D_{s}^{-}\tau^{+} \nu_{\tau}$  & $(5.99^{+0.15+0.01}_{-0.14-0.01}) \t 10^{-3}$ & $6.10 \t 10^{-3}$                   & $7.20 \t 10^{-3}$                              & $(0.78 \pm 0.15) \t 10^{-2}$     &     -                            \\ \hline \hline
$B_{s}^0 \to K^{*-}e^{+} \nu_e$             & $(4.73^{+0.03+0.39}_{-0.50-0.40}) \t 10^{-4}$ & $3.29 \t 10^{-4}$                   & $3.01 \t 10^{-4}$                              & -                                &     -                            \\
$B_{s}^0 \to K^{*-}\mu^{+} \nu_{\mu}$       & $(4.72^{+0.03+0.39}_{-0.50-0.40}) \t 10^{-4}$ & $3.29 \t 10^{-4}$                   & $3.01 \t 10^{-4}$                              & -                                & -                               \\
$B_{s}^0 \to K^{*-}\tau^{+} \nu_{\tau}$     & $(2.77^{+0.08+0.16}_{-0.29-0.18}) \t 10^{-4}$ & $1.82 \t 10^{-4}$                   & $1.56 \t 10^{-4}$                              & -                                &        -                         \\ \hline
$B_{s}^0 \to D_{s}^{*-}e^{+} \nu_e$         & $(4.99^{+1.06+0.52}_{-0.93-0.66}) \t 10^{-2}$ & $5.06 \t 10^{-2}$                   & $5.91 \t 10^{-2}$                              & $(6.42 \pm 0.67) \t 10^{-2}$     &          -                       \\
$B_{s}^0 \to D_{s}^{*-}\mu^{+} \nu_{\mu}$   & $(4.97^{+1.05+0.52}_{-0.92-0.65}) \t 10^{-2}$ & $5.05 \t 10^{-2}$                   & $5.91 \t 10^{-2}$                              & $(6.39 \pm 0.67) \t 10^{-2}$     & $(5.20 \pm 0.50) \t 10^{-2}$    \\
$B_{s}^0 \to D_{s}^{*-}\tau^{+} \nu_{\tau}$ & $(1.28^{+0.23+0.12}_{-0.21-0.15}) \t 10^{-2}$ & $1.23 \t 10^{-2}$                   & $1.46 \t 10^{-2}$                              & $(1.53 \pm 0.15) \t 10^{-2}$     &    -       \\\hline \hline 
\end{tabular}
\end{table}
\begin{table}[ht]
\caption{Relative decay widths and physical observables of $B_{s} \to (P) V\ell\nu_\ell$ decays in CLFQM (Type-II). }
\label{t4}
\begin{tabular}{c  c  c  c  c  c  c  c  c c c}
\hline 	\hline
 \multirow{2}{*}{Decay} & \multirow{2}{*}{$\frac{\Gamma_{L}}{\Gamma}$} & \multirow{2}{*}{$\frac{\Gamma_{T}}{\Gamma}$} & \multirow{2}{*}{$\langle A_{FB}\rangle$}\footnote{The $A_{FB}$ results from LQCD are as follows:  $A_{FB}(\overline{B}_s^0 \to K^+ \mu^- \overline{\nu}_{\mu})= 0.0057 \pm 0.0018$~\cite{Flynn:2023nhi}, $A_{FB}(\overline{B}_s^0 \to K^+ \tau^- \overline{\nu}_{\tau})=0.2821 \pm 0.0089$~\cite{Flynn:2023nhi}, and $A_{FB}(\ov{B}_s^0 \to D_s^{*+} \tau^- \ov{\nu}_{\tau}) = -0.092 \pm 0.024$~\cite{Harrison:2021tol}} & \multirow{2}{*}{$\langle C_{F}^{\ell}\rangle$}& \multirow{2}{*}{$\langle P_{L}^{\ell}\rangle$}\footnote{The $A_{\lambda_l}(\ov{B}_s^0 \to D_s^{*+} \tau^- \ov{\nu}_{\tau})$ results from LQCD~\cite{Harrison:2023dzh, Harrison:2021tol} and LCSR~\cite{Bordone:2019guc} are $0.5331 \pm 0.0091$~\cite{Harrison:2023dzh}, $0.520 \pm 0.012$~\cite{Harrison:2021tol} and $0.486 \pm 0.023$~\cite{Bordone:2019guc}.} & \multirow{2}{*}{$\langle P_{T}^{\ell}\rangle$} & \multirow{2}{*}{$\langle F_{L} \rangle$}\footnote{Similarly, the LQCD~\cite{Harrison:2023dzh, Harrison:2021tol} and LCSR~\cite{Bordone:2019guc} predictions for $F_{L}(\ov{B}_s^0 \to D_s^{*+} \tau^- \ov{\nu}_{\tau})$ are $0.420 \pm 0.012$~\cite{Harrison:2023dzh}, $0.440 \pm 0.016$~\cite{Harrison:2021tol} and $0.471 \pm 0.016$~\cite{Bordone:2019guc}.} & \multirow{2}{*}{$\langle \alpha^* \rangle$}  \\ 
 &  &  &  &  &  &  & &  \\ \hline
$B_{s}^0 \to K^{-}e^{+} \nu_e$              & $1.00$ & $0.00$ & $-4.98 \t 10^{-7}$ & $-1.50$ & $1.00$  & $-7.97 \t 10^{-4}$ & -     & -   \\
$B_{s}^0 \to K^{-}\mu^{+} \nu_{\mu}$        & $1.00$ & $0.00$ & $-7.56 \t 10^{-3}$ & $-1.48$ & $0.98$  & $-0.15$            & -     & -  \\
$B_{s}^0 \to K^{-}\tau^{+} \nu_{\tau}$      & $1.00$ & $0.00$ & $-0.29$            & $-0.44$ & $-0.18$ & $-0.88$            & -       & -  \\ \hline
$B_{s}^0 \to D_{s}^{-}e^{+} \nu_e$          & $1.00$ & $0.00$ & $-1.17 \t 10^{-6}$ & $-1.50$ & $1.00$  & $-1.15 \t 10^{-3}$ & -     & -   \\
$B_{s}^0 \to D_{s}^{-}\mu^{+} \nu_{\mu}$    & $1.00$ & $0.00$ & $-1.53 \t 10^{-2}$ & $-1.45$ & $0.96$  & $-0.21$            & -     & -   \\
$B_{s}^0 \to D_{s}^{-}\tau^{+} \nu_{\tau}$  & $1.00$ & $0.00$ & $-0.35$            & $-0.23$ & $-0.42$ & $-0.81$            & -     & -   \\ \hline \hline
$B_{s}^0 \to K^{*-}e^{+} \nu_e$             & $0.47$ & $0.53$ & $-0.35$            & $-0.31$ & $1.00$  & $-1.45 \t 10^{-4}$ & $0.47$ & $-0.28$   \\
$B_{s}^0 \to K^{*-}\mu^{+} \nu_{\mu}$       & $0.47$ & $0.53$ & $-0.36$            & $-0.30$ & $0.99$  & $-2.73 \t 10^{-2}$ & $0.47$& $-0.28$   \\
$B_{s}^0 \to K^{*-}\tau^{+} \nu_{\tau}$     & $0.46$ & $0.54$ & $-0.43$            & $-0.08$ & $0.60$  & $-8.67 \t 10^{-2}$ & $0.46$& $-0.27$    \\ \hline
$B_{s}^0 \to D_{s}^{*-}e^{+} \nu_e$         & $0.50$ & $0.50$ & $-0.24$            & $-0.37$ & $1.00$  & $-2.55 \t 10^{-4}$ & $0.50$& $-0.33$   \\
$B_{s}^0 \to D_{s}^{*-}\mu^{+} \nu_{\mu}$   & $0.50$ & $0.50$ & $-0.24$            & $-0.36$ & $0.99$  & $-4.55 \t 10^{-2}$ & $0.50$ & $-0.33$  \\
$B_{s}^0 \to D_{s}^{*-}\tau^{+} \nu_{\tau}$ & $0.44$ & $0.56$ & $-0.30$            & $-0.05$ & $0.52$  & $-7.69 \t 10^{-2}$ & $0.44$ & $-0.22$  \\ \hline \hline
\end{tabular}
\end{table}

\begin{table}[ht]
\centering
\caption{Branching ratios of $B_s \to PP$ decays for CKM-favored modes.}
\label{t5}
\begin{tabular}{ccccc}
\hline
\multirow{2}{*}{Decay} & \multicolumn{2}{c}{This Work} & RQM~\cite{Faustov:2012mt} & PDG~\cite{PDG:2024cfk} \\
\cline{2-3}
& $N_c = \infty$ \quad [$N_c=3$] & $a_1^{\mathcal{E}}=0.88,~a_2^{\mathcal{E}}=-0.47$ & \begin{tabular}[c]{@{}c@{}}[QCDF~\cite{Azizi:2008ty}]\end{tabular} & [HFLAV \cite{HFLAV:2024ctg}] \\
\hline
\multicolumn{5}{c}{$\Delta b=1,~\Delta C=1,~\Delta S=0$} \\
\hline
\multirow{2}{*}{$\overline{B}_{s}^{0} \to D_{s}^{+}\pi^{-}$} 
& $(4.65^{+0.12}_{-0.11}) \times 10^{-3}$ & \multirow{2}{*}{~~$(2.87^{+0.07}_{-0.07}) \times 10^{-3}$~~} &~~$3.5 \times 10^{-3}$~~ & $(2.98 \pm 0.14) \times 10^{-3}$ \\
& $[(3.96^{+0.10}_{-0.10}) \times 10^{-3}]$ & & [$(1.42 \pm 0.57) \times 10^{-3}$] & $[(2.83^{+0.18}_{-0.19}) \times 10^{-3}]$ \\
\hline
\multirow{2}{*}{$\overline{B}_{s}^{0} \to D^{0}K^{0}$} 
& $(1.62^{+0.42}_{-0.37}) \times 10^{-4}$ & \multirow{2}{*}{~~ $(5.29^{+1.39}_{-1.20}) \times 10^{-4}$~~} & \multirow{2}{*}{-} & \multirow{2}{*}{$(4.3 \pm 0.9) \times 10^{-4}$} \\
& $[(3.07^{+0.81}_{-0.70}) \times 10^{-5}]$ & & & \\
\hline
\multicolumn{5}{c}{$\Delta b=1,~\Delta C=0,~\Delta S=-1$} \\
\hline
\multirow{2}{*}{$\overline{B}_{s}^{0} \to D_{s}^{-}D_{s}^{+}$} 
& $(1.62^{+0.03}_{-0.02}) \times 10^{-2}$ & \multirow{2}{*}{~~$(1.00^{+0.02}_{-0.01}) \times 10^{-2}$~~} & $1.1 \times 10^{-2}$ & \multirow{2}{*}{$(4.4 \pm 0.5)\times 10^{-3}$} \\
& $[(1.38^{+0.02}_{-0.02}) \times 10^{-2}]$ & & $[(2.17 \pm 0.83)\times 10^{-3}]$ & \\
\hline
\end{tabular}
\end{table}


\begin{table}[ht]
\caption{Branching ratios of $B_s \to PP$ decays for CKM-suppressed modes.}
\label{t6}
\centering
\scalebox{0.85}{
\begin{tabular}{ c c c c c c }
\hline \hline 
\multirow{2}{*}{Decay} & \multicolumn{2}{c}{This Work} & NCQM\cite{Albertus:2014bfa} & RQM\cite{Faustov:2012mt} & {PDG~\cite{PDG:2024cfk}} \\ 
\cline{2-3} 
& $N_c = \infty$ \quad $[N_c = 3]$ & $a_1^{\mathcal{E}}=0.88,~a_2^{\mathcal{E}}=-0.47$ & [QCDF \cite{Azizi:2008ty}] & [LCSR \cite{Li:2009wq}] & [HFLAV \cite{HFLAV:2024ctg}] \\
\hline
\multicolumn{6}{c}{$\Delta b=1,~ \Delta C=1,~ \Delta S=-1$} \\ \hline
\multirow{2}{*}{$\overline{B}_{s}^{0} \to D_{s}^{+}K^{-}$} & $(3.50^{+0.09}_{-0.08}) \times 10^{-4}$ & \multirow{2}{*}{~~$(2.16^{+0.05}_{-0.05}) \times 10^{-4}$~~} & ~~$4.0 \times 10^{-4}$~~ & ~~$2.8 \times 10^{-4}$~~ & $(2.25 \pm 0.12) \times 10^{-4}$\\
& $[(2.98^{+0.07}_{-0.07}) \times 10^{-4}]$ & & $[(1.03 \pm 0.51) \times 10^{-4}]$ & $[(1.3^{+0.5}_{-0.4})\times 10^{-4}]$ & $[(2.12 \pm 0.14) \times 10^{-4}]$ \\ \hline
\multicolumn{6}{c}{$\Delta b=1,~ \Delta C=0,~ \Delta S=0$} \\ \hline

\multirow{2}{*}{$\overline{B}_{s}^{0} \to D^{-}D_{s}^{+}$} & $(5.58^{+0.10}_{-0.09}) \times 10^{-4}$ & \multirow{2}{*}{~~$(3.45^{+0.06}_{-0.05}) \times 10^{-4}$~~} & $4.0 \times 10^{-4}$ & - & \multirow{2}{*}{-} \\
& $[(4.75^{+0.09}_{-0.07}) \times 10^{-4}]$ & & $[(1.20 \pm 0.73) \times 10^{-4}]$ & $[(2.8 \pm 0.5) \times 10^{-4}]$ & \\ \hline

\multirow{2}{*}{$\overline{B}_{s}^{0} \to \eta_{c} K^{0}$} & $(2.71^{+0.74}_{-0.63}) \times 10^{-5}$ & \multirow{2}{*}{~~$(8.87^{+2.42}_{-2.06}) \times 10^{-5}$~~} & \multirow{2}{*}{-} & \multirow{2}{*}{-} & \multirow{2}{*}{-} \\
& $[(5.16^{+1.41}_{-1.20}) \times 10^{-6}]$ & & & & \\ \hline \hline
\end{tabular}}
\end{table}


\begin{table}[ht]
\caption{Branching ratios of $B_s \to PP$ decays for CKM-doubly-suppressed modes.}
\label{t7}
\scalebox{0.93}{
\begin{tabular}{ c  c c  c c}
\hline 	\hline
\multirow{2}{*}{Decay} & \multicolumn{2}{c}{This Work} & pQCD & PDG~\cite{PDG:2024cfk}\\
\cline{2-3}
& $N_c = \infty$ \quad [$N_c=3$] & $a_1^{\mathcal{E}}=0.88,~a_2^{\mathcal{E}}=-0.47$ & \cite{Zou:2009zza} & [HFLAV \cite{HFLAV:2024ctg}] \\
\hline
\multicolumn{5}{c}{$\Delta b=1,~ \Delta C=-1,~ \Delta S=0$} \\ \hline
\multirow{2}{*}{$\overline{B}_{s}^{0} \to D^{-}K^{+}$} 
& $(1.36^{+0.36}_{-0.31}) \times 10^{-6}$ 
& \multirow{2}{*}{~~$(8.38^{+2.20}_{-1.90}) \times 10^{-7}$~~} 
& \multirow{2}{*}{~~$(1.21^{+0.61}_{-0.47}) \times 10^{-6}$~~} & \multirow{2}{*}{-} \\
& $[(1.16^{+0.30}_{-0.26}) \times 10^{-6}]$ && & \\ \hline

\multirow{2}{*}{$\overline{B}_{s}^{0} \to \overline{D}^{0}K^{0}$} 
& $(7.30^{+1.91}_{-1.65}) \times 10^{-8}$ 
& \multirow{2}{*}{~~$(2.39^{+0.63}_{-0.54}) \times 10^{-7}$~~} 
& \multirow{2}{*}{$(2.43^{+2.23}_{-1.48}) \times 10^{-8}$} & \multirow{2}{*}{-} \\
& $[(1.39^{+0.36}_{-0.31}) \times 10^{-8}]$ && & \\ \hline

\multicolumn{5}{c}{$\Delta b=1,~ \Delta C=-1,~ \Delta S=-1$} \\ \hline
\multirow{2}{*}{$\overline{B}_{s}^{0} \to D_{s}^{-}K^{+}$} 
& $(4.03^{+1.06}_{-0.91}) \times 10^{-5}$ 
& \multirow{2}{*}{~~$(2.49^{+0.65}_{-0.56}) \times 10^{-5}$~~} 
& \multirow{2}{*}{$(2.94^{+1.48}_{-1.18}) \times 10^{-5}$} & $(2.25 \pm 0.12) \times 10^{-4}$ \\
& $[(3.43^{+0.90}_{-0.78}) \times 10^{-5}]$ & && $[(2.12 \pm 0.14) \times 10^{-4}]$ \\
\hline \hline
\end{tabular}}
\end{table}


\begin{table}[ht]
\caption{Branching ratios of $B_s \to PV$ decays for CKM-favored modes.}
\label{t8}
\begin{tabular}{ c c c c c}
\hline 	\hline
\multirow{2}{*}{Decay} & \multicolumn{2}{c}{This Work} & NCQM \cite{Albertus:2014bfa} & PDG \cite{PDG:2024cfk} \\
\cline{2-3}
& $N_c = \infty$ \quad $[N_c = 3]$ & $a_1^{\mathcal{E}}=0.88,~a_2^{\mathcal{E}}=-0.47$ & [RQM \cite{Faustov:2012mt}] & [HFLAV \cite{HFLAV:2024ctg}] \\
\hline
\multicolumn{5}{c}{$\Delta b=1,~\Delta C=1,~\Delta S=0$} \\ \hline
\multirow{2}{*}{$\overline{B}_{s}^{0} \to \pi^{-}D_{s}^{*+}$} 
& $(3.46^{+1.52}_{-1.30}) \times 10^{-3}$ 
& \multirow{2}{*}{$(2.13^{+0.95}_{-0.80}) \times 10^{-3}$} & $4.50 \times 10^{-3}$ & $(1.9^{+0.5}_{-0.4}) \times 10^{-3}$ \\
& $[(2.94^{+1.30}_{-1.10}) \times 10^{-3}]$ & & $[2.7 \times 10^{-3}]$ & $[(2.20^{+0.61}_{-0.54}) \times 10^{-3}]$ \\ \hline

\multirow{2}{*}{$\overline{B}_{s}^{0} \to D_{s}^{+}\rho^{-}$} 
& $(1.16^{+0.03}_{-0.03}) \times 10^{-2}$ 
& \multirow{2}{*}{$(7.16^{+0.18}_{-0.17}) \times 10^{-3}$} & $1.26 \times 10^{-2}$ & \multirow{2}{*}{$(6.8 \pm 1.4) \times 10^{-3}$} \\
& $[(9.87^{+0.24}_{-0.23}) \times 10^{-3}]$ & & $[9.4 \times 10^{-3}]$ & \\ \hline

\multirow{2}{*}{$\overline{B}_{s}^{0} \to D^{0}K^{*0}$} 
& $(1.72^{+0.26}_{-0.52}) \times 10^{-4}$ 
& \multirow{2}{*}{$(5.63^{+0.84}_{-1.70}) \times 10^{-4}$} & \multirow{2}{*}{-} & \multirow{2}{*}{$(4.4 \pm 0.6) \times 10^{-4}$} \\
& $[(3.27^{+0.48}_{-0.99}) \times 10^{-5}]$ & &  & \\ \hline

\multirow{2}{*}{$\overline{B}_{s}^{0} \to K^{0} D^{*0}$} 
& $(1.67^{+0.43}_{-0.37}) \times 10^{-4}$ 
& \multirow{2}{*}{$(5.45^{+1.40}_{-1.22}) \times 10^{-4}$} & \multirow{2}{*}{-} & \multirow{2}{*}{$(2.8 \pm 1.1) \times 10^{-4}$} \\
& $[(3.17^{+0.81}_{-0.71}) \times 10^{-5}]$ & &  & \\ \hline

\multicolumn{5}{c}{$\Delta b=1,~\Delta C=0,~\Delta S=-1$} \\ \hline

\multirow{2}{*}{$\overline{B}_{s}^{0} \to D_{s}^{-}D_{s}^{*+}$} 
& $(7.83^{+3.30}_{-2.83}) \times 10^{-3}$ 
& \multirow{2}{*}{$(4.83^{+2.01}_{-1.74}) \times 10^{-3}$} & - & \multirow{2}{*}{-} \\
& $[(6.66^{+2.81}_{-2.41}) \times 10^{-3}]$ & & $[6.1 \times 10^{-3}]$ & \\ \hline

\multirow{2}{*}{$\overline{B}_{s}^{0} \to D_{s}^{+}D_{s}^{*-}$} 
& $(7.26^{+0.10}_{-0.08}) \times 10^{-3}$ 
& \multirow{2}{*}{$(4.48^{+0.06}_{-0.05}) \times 10^{-3}$} & - & \multirow{2}{*}{-} \\
& $[(6.18^{+0.08}_{-0.07}) \times 10^{-3}]$ & & $[1.0 \times 10^{-2}]$ & \\ \hline \hline
\end{tabular}
\end{table}


\begin{table}[ht]
\caption{Branching ratios of $B_s \to PV$ decays for CKM-suppressed modes.}
\label{t9}
\begin{tabular}{ c c c c c}
\hline 	\hline
\multirow{2}{*}{Decay} & \multicolumn{2}{c}{This Work} & NCQM \cite{Albertus:2014bfa} & PDG \cite{PDG:2024cfk} \\
\cline{2-3}
& $N_c = \infty$ \quad $[N_c = 3]$ & $a_1^{\mathcal{E}}=0.88,~a_2^{\mathcal{E}}=-0.47$ & [RQM \cite{Faustov:2012mt}] & [HFLAV \cite{HFLAV:2024ctg}] \\\hline

\multicolumn{5}{c}{$\Delta b=1,~\Delta C=1,~\Delta S=-1$} \\ \hline
\multirow{2}{*}{$\overline{B}_{s}^{0} \to K^{-}D_{s}^{*+}$} 
& $(2.59^{+1.13}_{-0.97}) \times 10^{-4}$ 
& \multirow{2}{*}{$(1.60^{+0.70}_{-0.60}) \times 10^{-4}$} & $8.0 \times 10^{-4}$ & \multirow{2}{*}{$(1.32^{+0.40}_{-0.32}) \times 10^{-4}$} \\
& $[(2.20^{+0.97}_{-0.82}) \times 10^{-4}]$ & & $[2.1 \times 10^{-4}]$ & \\ \hline

\multirow{2}{*}{$\overline{B}_{s}^{0} \to D_{s}^{+}K^{*-}$} 
& $(5.70^{+0.13}_{-0.13}) \times 10^{-4}$ 
& \multirow{2}{*}{$(3.52^{+0.08}_{-0.08}) \times 10^{-4}$} & $4.0 \times 10^{-4}$ & \multirow{2}{*}{-} \\
& $[(4.85^{+0.11}_{-0.11}) \times 10^{-4}]$ & & $[4.7 \times 10^{-4}]$ & \\ \hline

\multicolumn{5}{c}{$\Delta b=1,~\Delta C=0,~\Delta S=0$} \\ \hline

\multirow{2}{*}{$\overline{B}_{s}^{0} \to D^{-}D_{s}^{*+}$} 
& $(2.85^{+1.20}_{-1.03}) \times 10^{-4}$ 
& \multirow{2}{*}{$(1.76^{+0.74}_{-0.63}) \times 10^{-4}$} & \multirow{2}{*}{-} & \multirow{2}{*}{-} \\
& $[(2.43^{+1.03}_{-0.88}) \times 10^{-4}]$ & &  & \\ \hline

\multirow{2}{*}{$\overline{B}_{s}^{0} \to D_{s}^{+}D^{*-}$} 
& $(4.41^{+0.06}_{-0.05}) \times 10^{-4}$ 
& \multirow{2}{*}{$(2.72^{+0.04}_{-0.03}) \times 10^{-4}$} & \multirow{2}{*}{-} & \multirow{2}{*}{-} \\
& $[(3.75^{+0.05}_{-0.05}) \times 10^{-4}]$ & &  & \\ \hline

\multirow{2}{*}{$\overline{B}_{s}^{0} \to \eta_{c} K^{*0}$} 
& $(1.84^{+0.27}_{-0.54}) \times 10^{-5}$ 
& \multirow{2}{*}{$(6.02^{+0.87}_{-1.78}) \times 10^{-5}$} & \multirow{2}{*}{-} & \multirow{2}{*}{-} \\
& $[(3.50^{+0.50}_{-1.03}) \times 10^{-6}]$ & &  & \\ \hline

\multirow{2}{*}{$\overline{B}_{s}^{0} \to K^{0}J/\psi$} 
& $(2.43^{+0.64}_{-0.55}) \times 10^{-5}$ 
& \multirow{2}{*}{$(7.93^{+2.10}_{-1.80}) \times 10^{-5}$} & \multirow{2}{*}{-} & \multirow{2}{*}{-} \\
& $[(4.61^{+1.22}_{-1.05}) \times 10^{-6}]$ & &  & \\ \hline \hline
\end{tabular}
\end{table}


\begin{table}[ht]
\caption{Branching ratios of $B_s \to PV$ decays for CKM-doubly-suppressed modes.}
\label{t10}
\begin{tabular}{ c c c c c}
\hline 	\hline
\multirow{2}{*}{Decay} & \multicolumn{2}{c}{This Work} & pQCD \cite{Zou:2009zza} & \multirow{2}{*}{PDG \cite{PDG:2024cfk}} \\
\cline{2-3}
& $N_c = \infty$ \quad $[N_c = 3]$ & $a_1^{\mathcal{E}}=0.88,~a_2^{\mathcal{E}}=-0.47$ & [QCDF \cite{Azizi:2008ty}] & \\\hline

\multicolumn{5}{c}{$\Delta b=1,~\Delta C=-1,~\Delta S=0$} \\ \hline

\multirow{2}{*}{$\overline{B}_{s}^{0} \to D^{-}K^{*+}$} 
& $(1.44^{+0.21}_{-0.44}) \times 10^{-6}$ 
& \multirow{2}{*}{$(8.91^{+1.31}_{-2.69})\times 10^{-7}$} 
& $(1.42^{+0.65}_{-0.53}) \times 10^{-6}$ & \multirow{2}{*}{-} \\
& $[(1.23^{+0.18}_{-0.37}) \times 10^{-6}]$ & & [-] &  \\ \hline

\multirow{2}{*}{$\overline{B}_{s}^{0} \to K^{+}D^{*-}$} 
& $(1.40^{+0.36}_{-0.31}) \times 10^{-6}$ 
& \multirow{2}{*}{$(8.63^{+2.22}_{-1.93})\times 10^{-7}$} 
& $(1.38^{+0.69}_{-0.56}) \times 10^{-6}$ & \multirow{2}{*}{-} \\
& $[(1.19^{+0.31}_{-0.27}) \times 10^{-6}]$ & & [-] &  \\ \hline

\multirow{2}{*}{$\overline{B}_{s}^{0} \to \overline{D}^{0}K^{*0}$} 
& $(7.78^{+1.15}_{-2.35}) \times 10^{-8}$ 
& \multirow{2}{*}{$(2.54^{+0.38}_{-0.76})\times 10^{-7}$} 
& $(0.60^{+0.44}_{-0.31}) \times 10^{-8}$ & \multirow{2}{*}{-} \\
& $[(1.48^{+0.23}_{-0.45}) \times 10^{-8}]$ & & [-] &  \\ \hline

\multirow{2}{*}{$\overline{B}_{s}^{0} \to K^{0}\overline{D}^{*0}$} 
& $(7.53^{+1.93}_{-1.68}) \times 10^{-8}$ 
& \multirow{2}{*}{$(2.46^{+0.63}_{-0.55})\times 10^{-7}$} 
& $(3.25^{+2.74}_{-1.77}) \times 10^{-8}$ & \multirow{2}{*}{-} \\
& $[(1.43^{+0.37}_{-0.32}) \times 10^{-8}]$ & & [-] &  \\ \hline

\multicolumn{5}{c}{$\Delta b=1,~\Delta C=-1,~\Delta S=-1$} \\ \hline

\multirow{2}{*}{$\overline{B}_{s}^{0} \to D_{s}^{-}K^{*+}$} 
& $(4.18^{+0.61}_{-1.25}) \times 10^{-5}$ 
& \multirow{2}{*}{$(2.58^{+0.38}_{-0.78})\times 10^{-5}$} 
& $(3.31^{+1.59}_{-1.24}) \times 10^{-5}$ & \multirow{2}{*}{-} \\
& $[(3.55^{+0.53}_{-1.08}) \times 10^{-5}]$ & & $[(5.00 \pm 2.20) \times 10^{-5}]$ &  \\ \hline

\multirow{2}{*}{$\overline{B}_{s}^{0} \to K^{+}D_{s}^{*-}$} 
& $(2.44^{+0.63}_{-0.55}) \times 10^{-5}$ 
& \multirow{2}{*}{$(1.51^{+0.39}_{-0.34})\times 10^{-5}$} 
& $(3.62^{+1.96}_{-1.41}) \times 10^{-5}$ & \multirow{2}{*}{$(1.32^{+0.40}_{-0.32}) \times 10^{-4}$} \\
& $[(2.08^{+0.54}_{-0.47}) \times 10^{-5}]$ & & $[(1.59 \pm 0.67) \times 10^{-4}]$ &  \\ \hline \hline
\end{tabular}
\end{table}


\begin{figure}
\centering
    \begin{subfigure}[b]{0.48\textwidth}
    \centering
    \includegraphics[width=\textwidth]{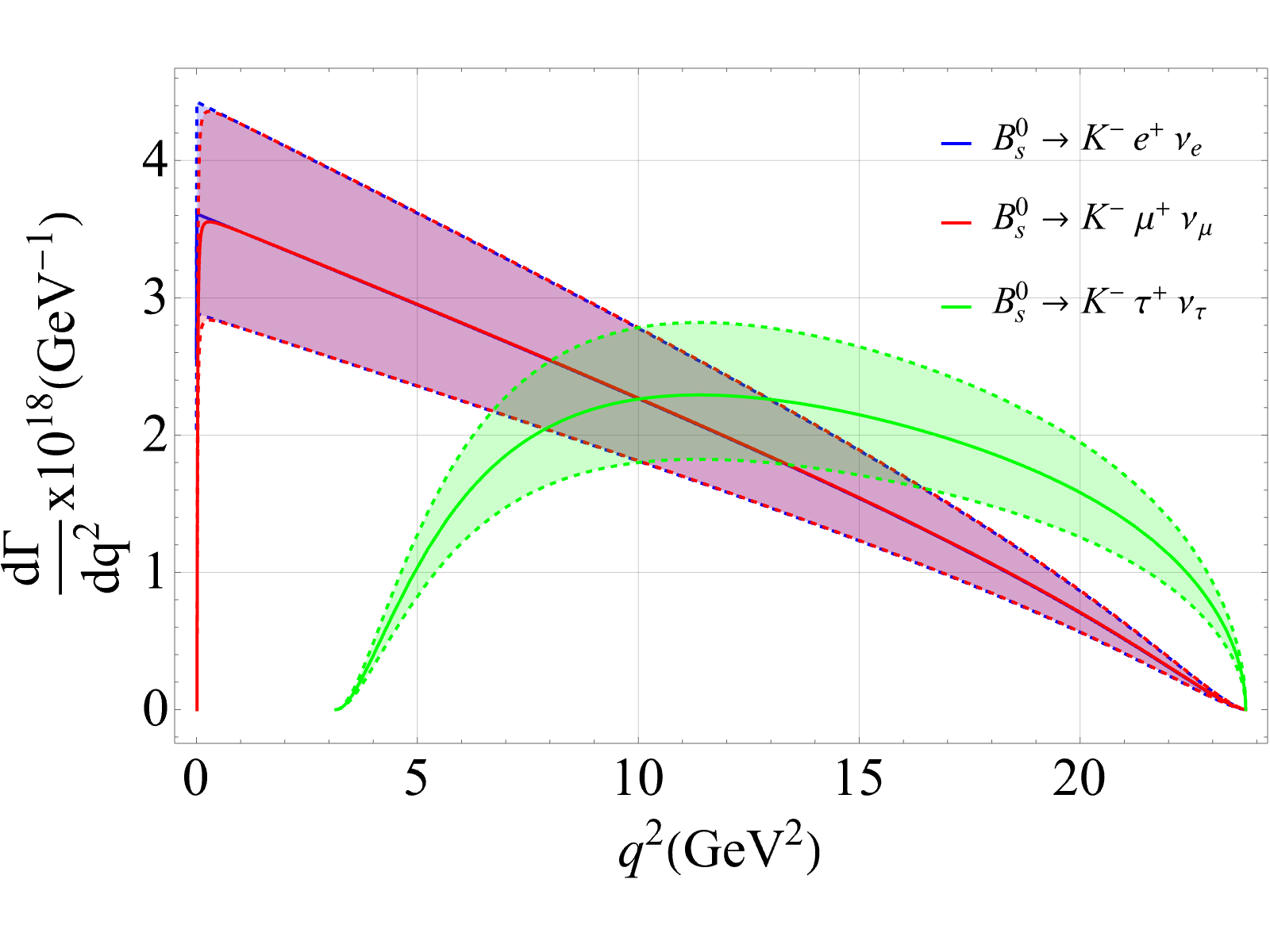}
    \caption{$B_s^0 \to K^- \ell^+ \nu_{\ell}$} \label{f4a}
    \end{subfigure}
	\hfill
    \begin{subfigure}[b]{0.48\textwidth}
        \centering
        \includegraphics[width=\textwidth]{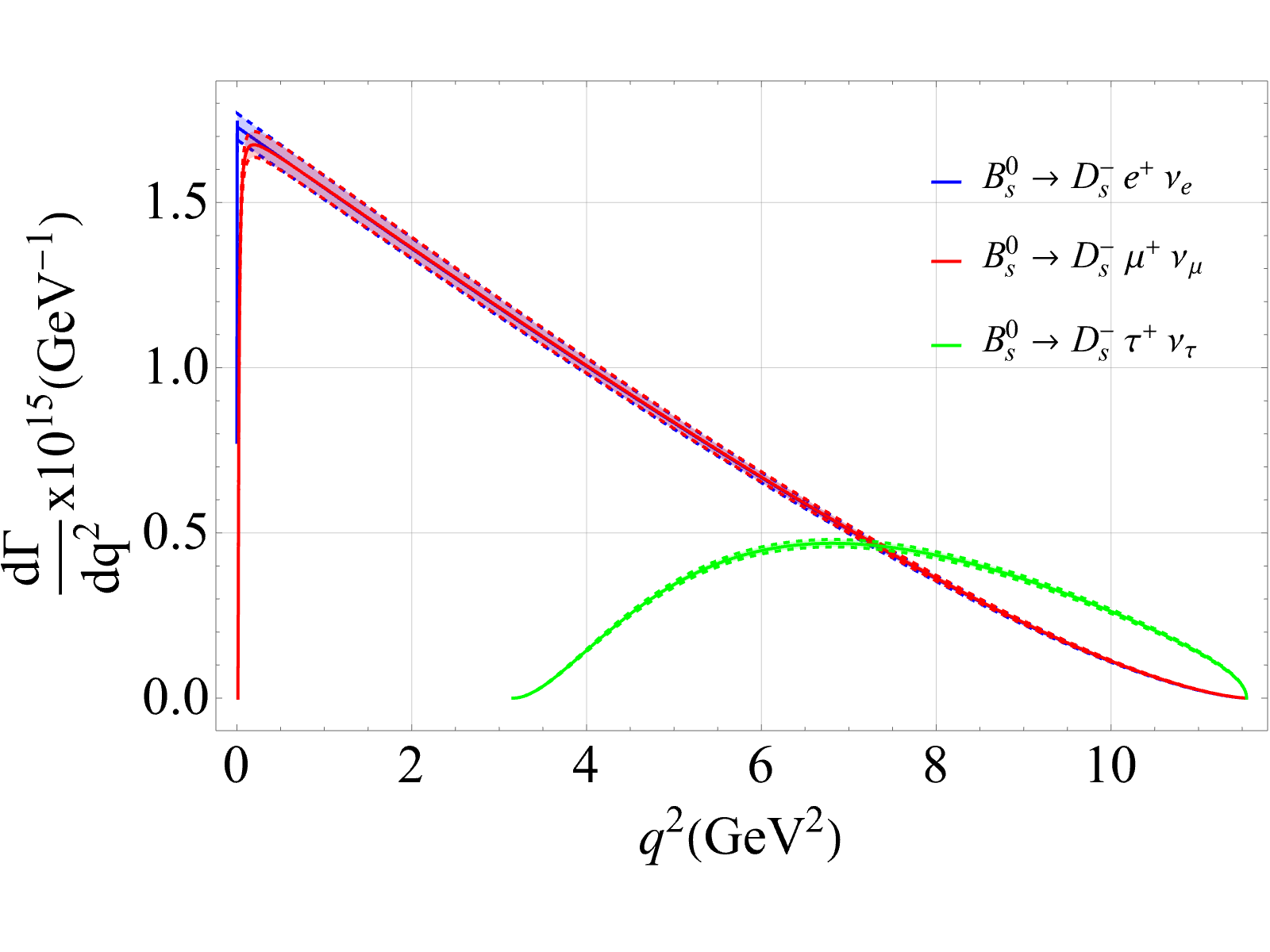}
	\caption{$B_s^0 \to D_s^- \ell^+ \nu_{\ell}$} \label{f4b}
    \end{subfigure}
        \hfill
    \begin{subfigure}[b]{0.48\textwidth}
    \centering
    \includegraphics[width=\textwidth]{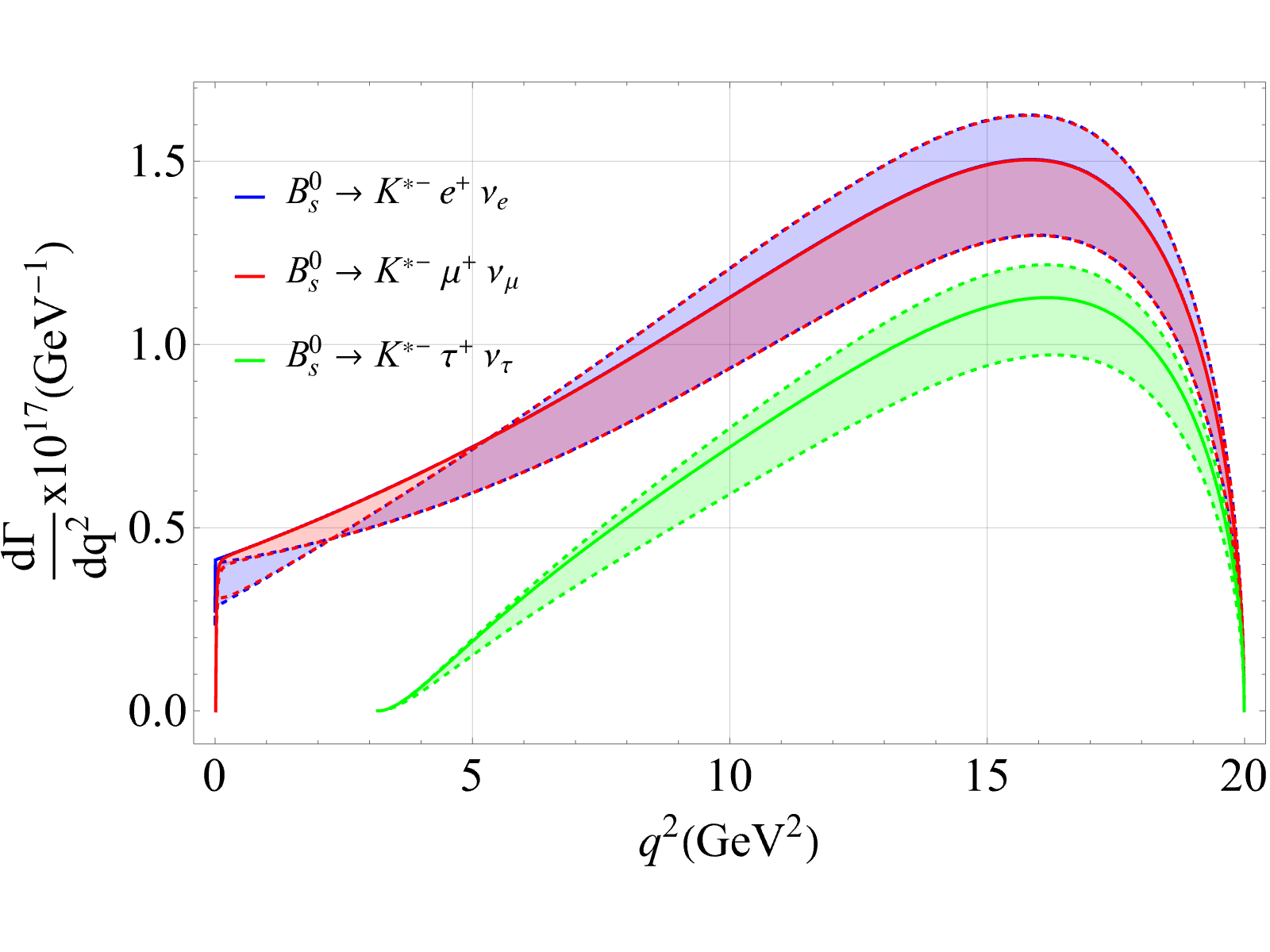}
    \caption{$B_s^0 \to K^{*-} \ell^+ \nu_{\ell}$} \label{f4c}
    \end{subfigure}
	\hfill
    \begin{subfigure}[b]{0.48\textwidth}
        \centering
        \includegraphics[width=\textwidth]{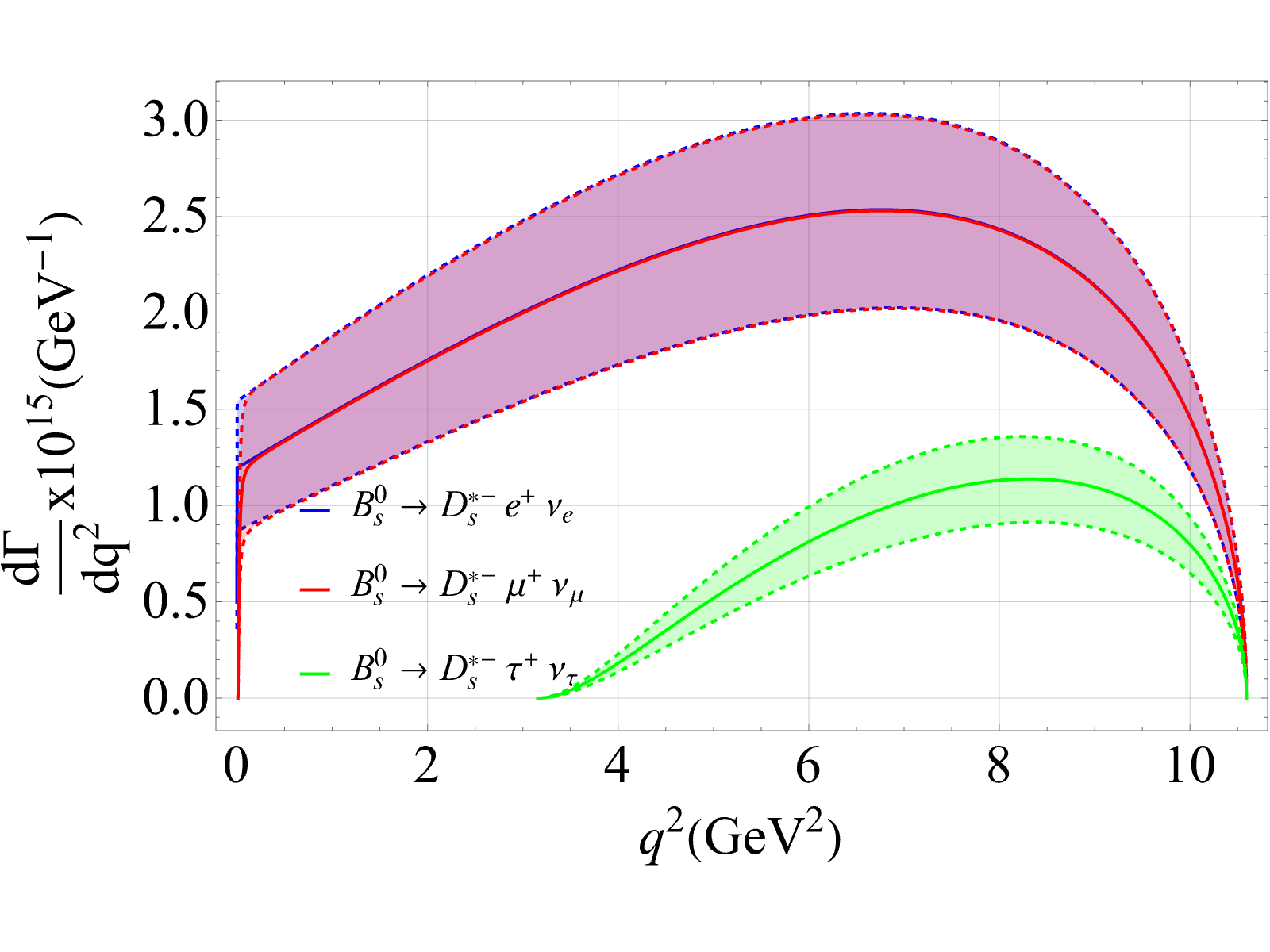}
	\caption{$B_s^0 \to D_s^{*-} \ell^+ \nu_{\ell}$} \label{f4d}
    \end{subfigure}
    \caption{The differential decay rates of $B_s \to P (V) \ell^+ \nu_{\ell}$ decays.} \label{f4}
\end{figure}
\begin{figure}
\centering
    \begin{subfigure}[b]{0.48\textwidth}
    \centering
    \includegraphics[width=\textwidth]{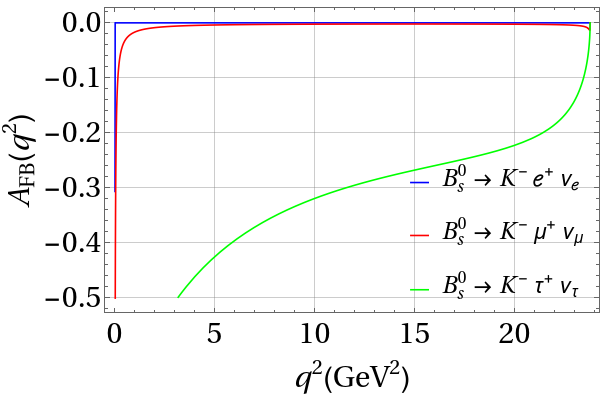}
    \caption{$B_s^0 \to K^- \ell^+ \nu_{\ell}$} \label{f5a}
    \end{subfigure}
	\hfill
    \begin{subfigure}[b]{0.48\textwidth}
        \centering
        \includegraphics[width=\textwidth]{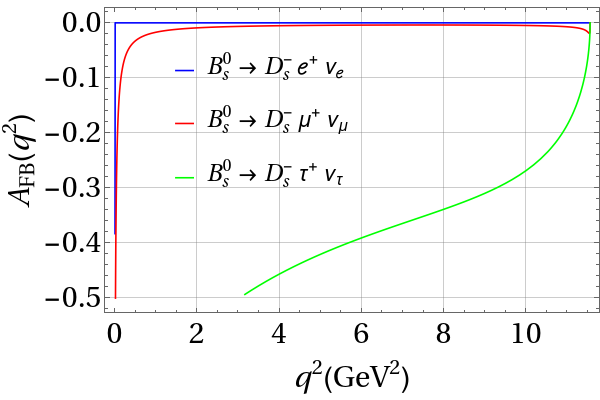}
	\caption{$B_s^0 \to D_s^- \ell^+ \nu_{\ell}$} \label{f5b}
    \end{subfigure}
        \hfill
    \begin{subfigure}[b]{0.48\textwidth}
    \centering
    \includegraphics[width=\textwidth]{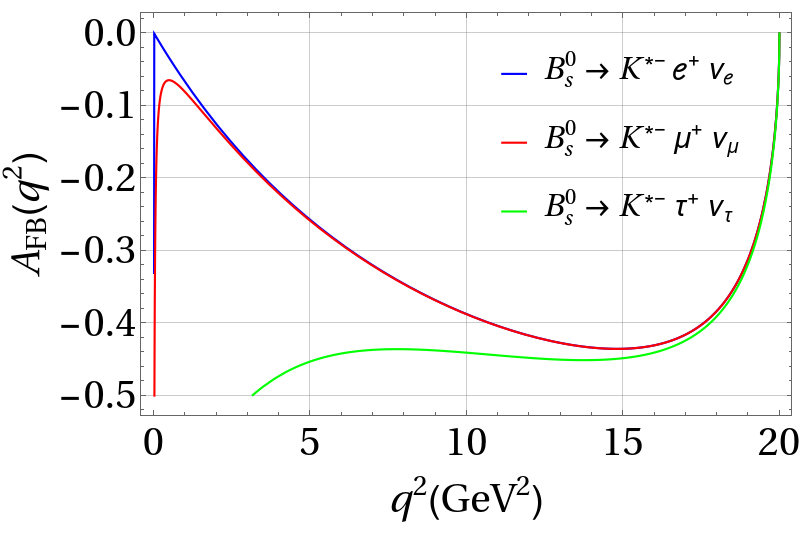}
    \caption{$B_s^0 \to K^{*-} \ell^+ \nu_{\ell}$} \label{f5c}
    \end{subfigure}
	\hfill
    \begin{subfigure}[b]{0.48\textwidth}
        \centering
        \includegraphics[width=\textwidth]{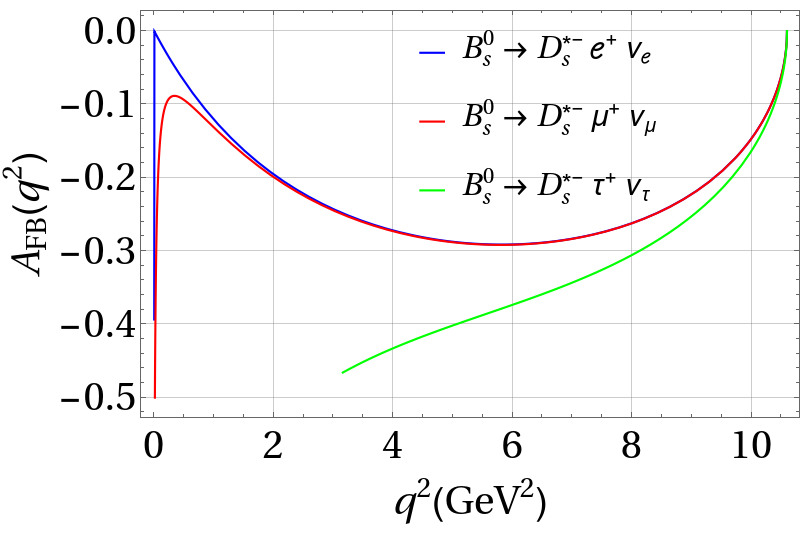}
	\caption{$B_s^0 \to D_s^{*-} \ell^+ \nu_{\ell}$} \label{f5d}
    \end{subfigure}
    \caption{The forward-backward asymmetries of $B_s \to P (V) \ell^+ \nu_{\ell}$ decays.} \label{f5}
\end{figure}
\begin{figure}
\centering
    \begin{subfigure}[b]{0.48\textwidth}
    \centering
    \includegraphics[width=\textwidth]{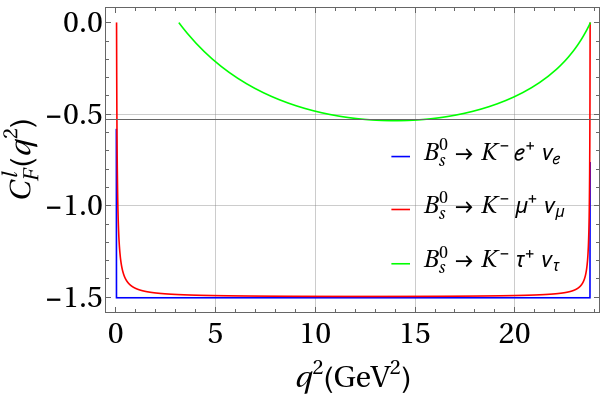}
    \caption{$B_s^0 \to K^- \ell^+ \nu_{\ell}$} \label{f6a}
    \end{subfigure}
	\hfill
    \begin{subfigure}[b]{0.48\textwidth}
        \centering
        \includegraphics[width=\textwidth]{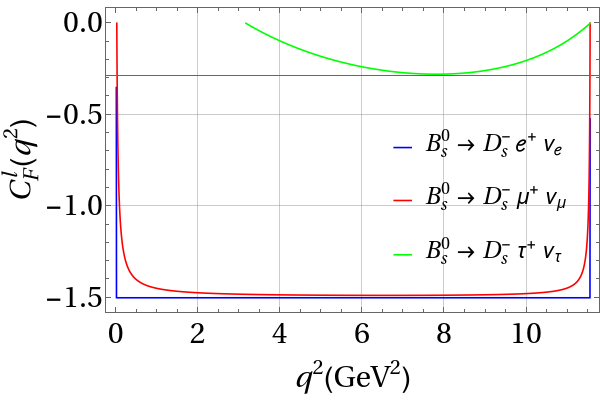}
	\caption{$B_s^0 \to D_s^- \ell^+ \nu_{\ell}$} \label{f6b}
    \end{subfigure}
        \hfill
    \begin{subfigure}[b]{0.48\textwidth}
    \centering
    \includegraphics[width=\textwidth]{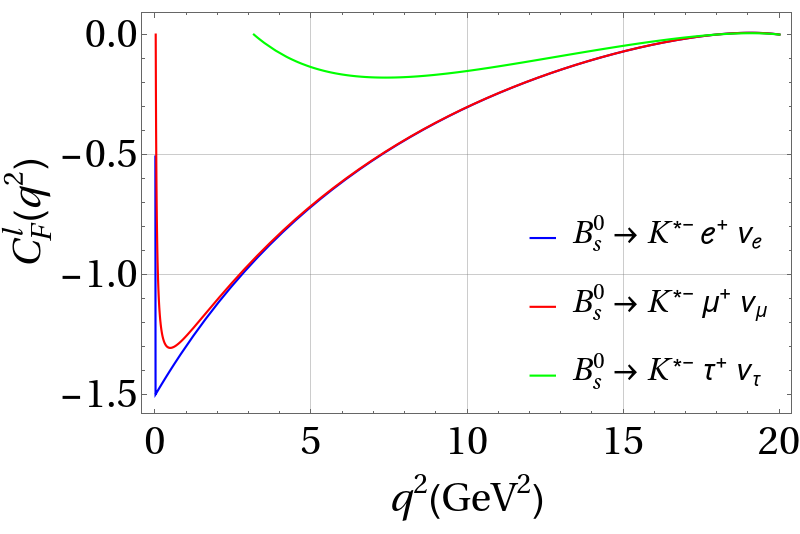}
    \caption{$B_s^0 \to K^{*-} \ell^+ \nu_{\ell}$} \label{f6c}
    \end{subfigure}
	\hfill
    \begin{subfigure}[b]{0.48\textwidth}
        \centering
        \includegraphics[width=\textwidth]{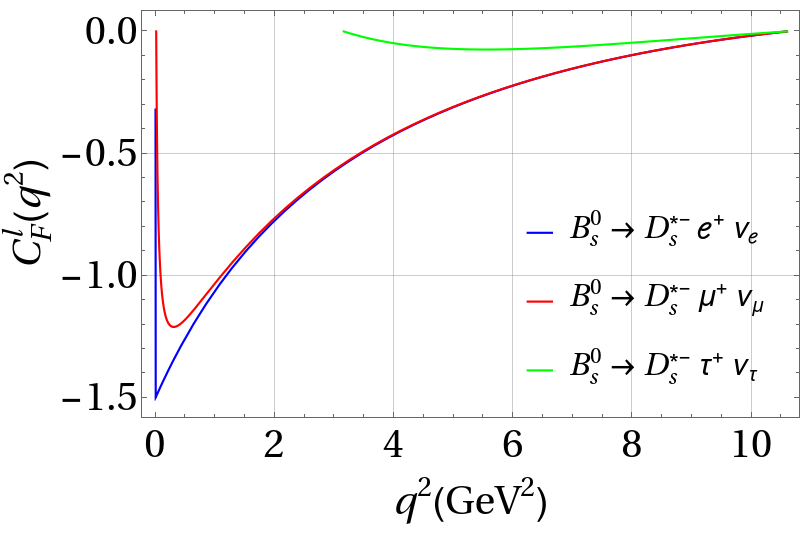}
	\caption{$B_s^0 \to D_s^{*-} \ell^+ \nu_{\ell}$} \label{f6d}
    \end{subfigure}
    \caption{The leptonic convexity parameter of $B_s \to P (V) \ell^+ \nu_{\ell}$ decays.} \label{f6}
\end{figure}
\begin{figure}
\centering
    \begin{subfigure}[b]{0.48\textwidth}
    \centering
    \includegraphics[width=\textwidth]{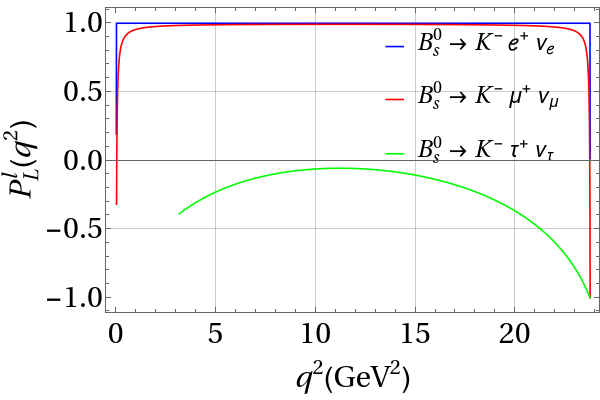}
    \caption{$B_s^0 \to K^- \ell^+ \nu_{\ell}$} \label{f7a}
    \end{subfigure}
	\hfill
    \begin{subfigure}[b]{0.48\textwidth}
        \centering
        \includegraphics[width=\textwidth]{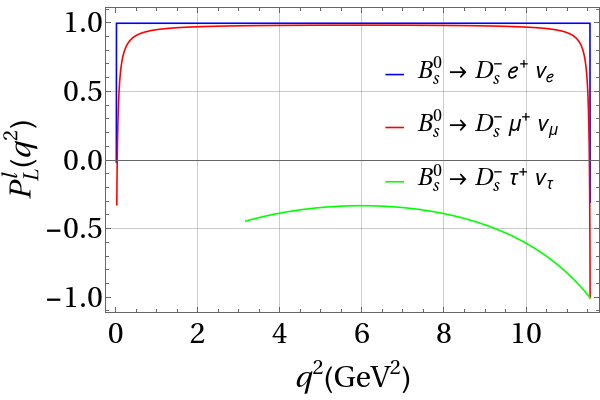}
	\caption{$B_s^0 \to D_s^- \ell^+ \nu_{\ell}$} \label{f7b}
    \end{subfigure}
        \hfill
    \begin{subfigure}[b]{0.48\textwidth}
    \centering
    \includegraphics[width=\textwidth]{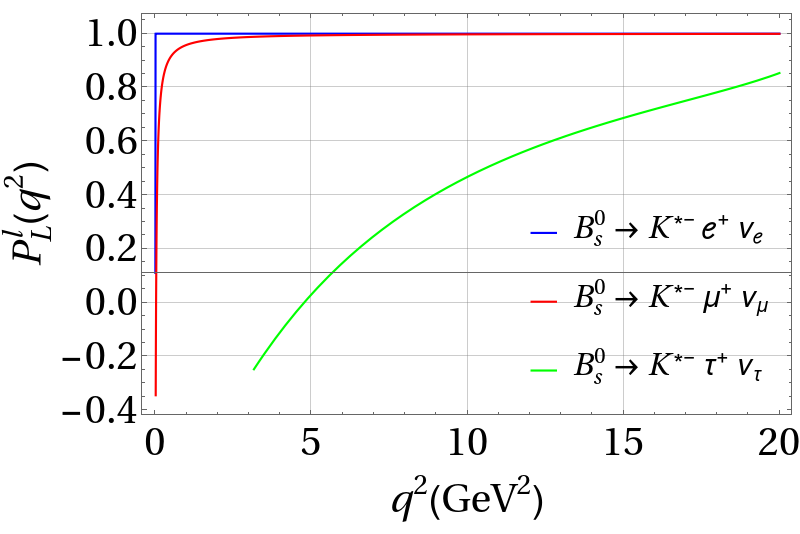}
    \caption{$B_s^0 \to K^{*-} \ell^+ \nu_{\ell}$} \label{f7c}
    \end{subfigure}
	\hfill
    \begin{subfigure}[b]{0.48\textwidth}
        \centering
        \includegraphics[width=\textwidth]{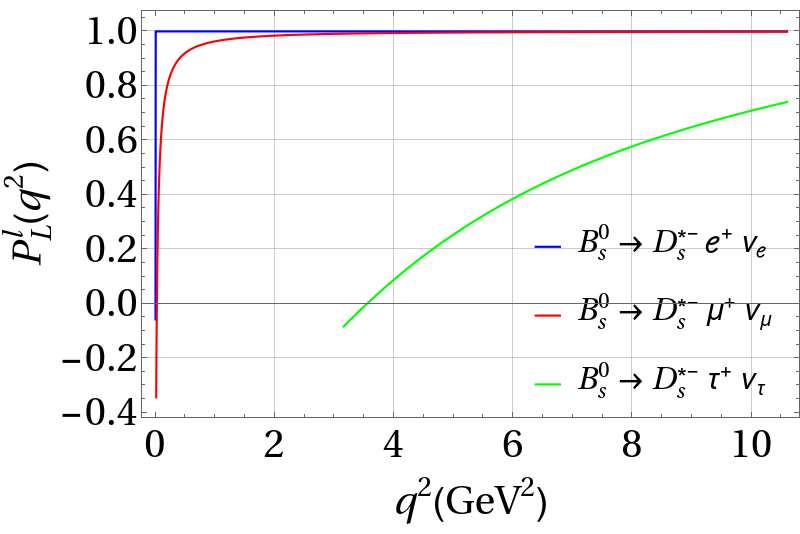}
	\caption{$B_s^0 \to D_s^{*-} \ell^+ \nu_{\ell}$} \label{f7d}
    \end{subfigure}
    \caption{The longitudinal polarization of a charged lepton of $B_s \to P (V) \ell^+ \nu_{\ell}$ decays.} \label{f7}
\end{figure}
\begin{figure}
\centering
    \begin{subfigure}[b]{0.48\textwidth}
    \centering
    \includegraphics[width=\textwidth]{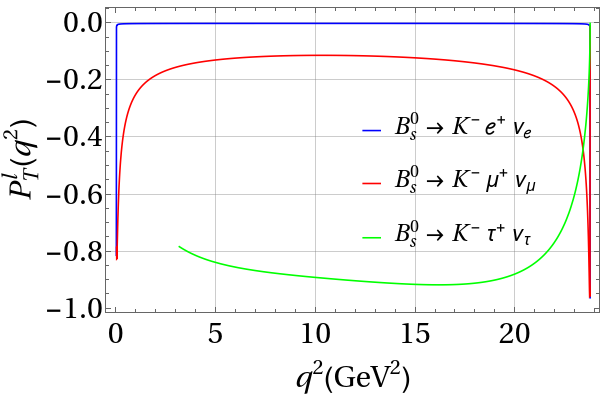}
    \caption{$B_s^0 \to K^- \ell^+ \nu_{\ell}$} \label{f8a}
    \end{subfigure}
	\hfill
    \begin{subfigure}[b]{0.48\textwidth}
        \centering
        \includegraphics[width=\textwidth]{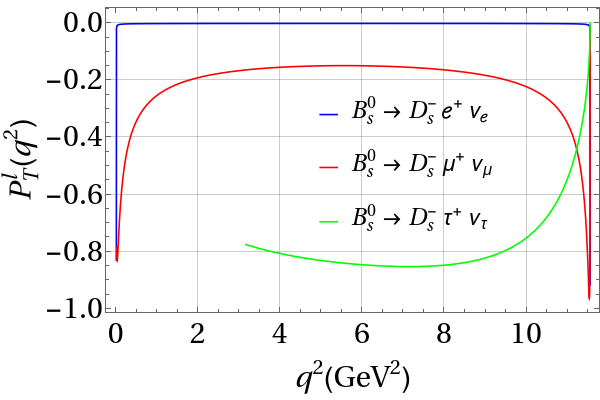}
	\caption{$B_s^0 \to D_s^- \ell^+ \nu_{\ell}$} \label{f8b}
    \end{subfigure}
        \hfill
    \begin{subfigure}[b]{0.48\textwidth}
    \centering
    \includegraphics[width=\textwidth]{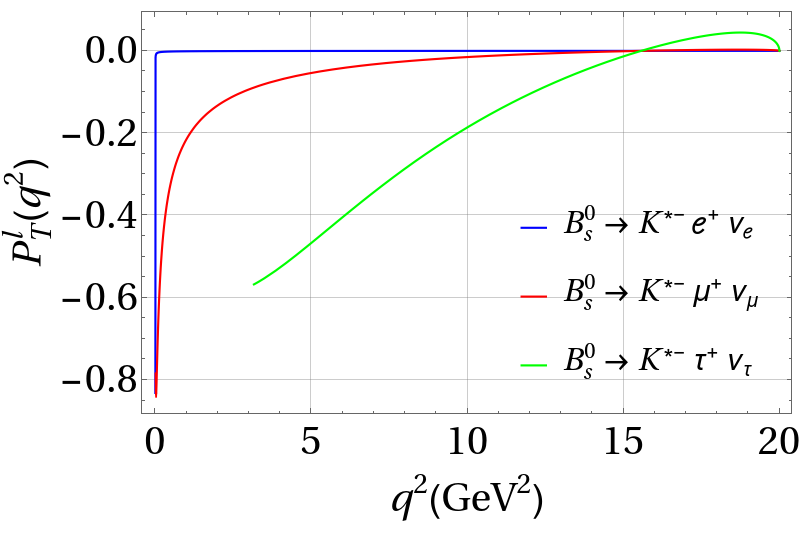}
    \caption{$B_s^0 \to K^{*-} \ell^+ \nu_{\ell}$} \label{f8c}
    \end{subfigure}
	\hfill
    \begin{subfigure}[b]{0.48\textwidth}
        \centering
        \includegraphics[width=\textwidth]{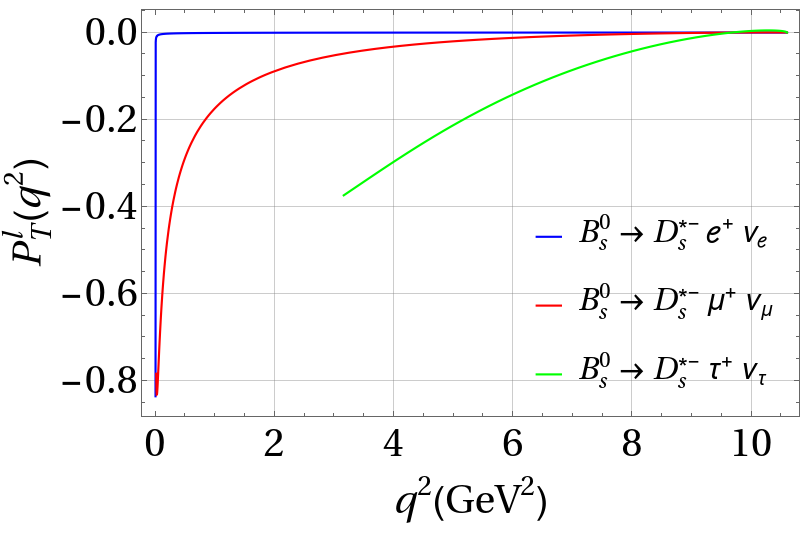}
	\caption{$B_s^0 \to D_s^{*-} \ell^+ \nu_{\ell}$} \label{f8d}
    \end{subfigure}
    \caption{The transverse polarization of a charged lepton of $B_s \to P (V) \ell^+ \nu_{\ell}$ decays.} \label{f8}
\end{figure}
\begin{figure}
\centering
    \begin{subfigure}[b]{0.48\textwidth}
    \centering
    \includegraphics[width=\textwidth]{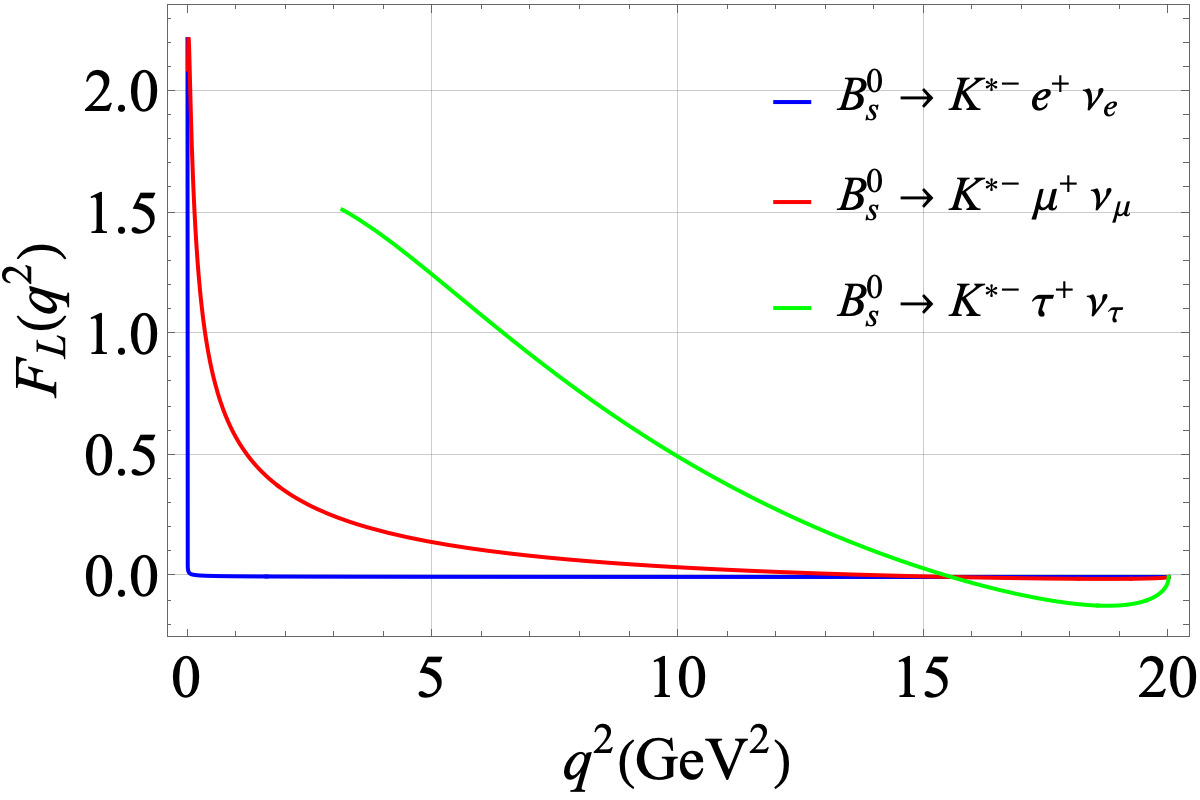}
    \caption{$B_s^0 \to K^{*-} \ell^+ \nu_{\ell}$} \label{f9a}
    \end{subfigure}
	\hfill
    \begin{subfigure}[b]{0.48\textwidth}
        \centering
        \includegraphics[width=\textwidth]{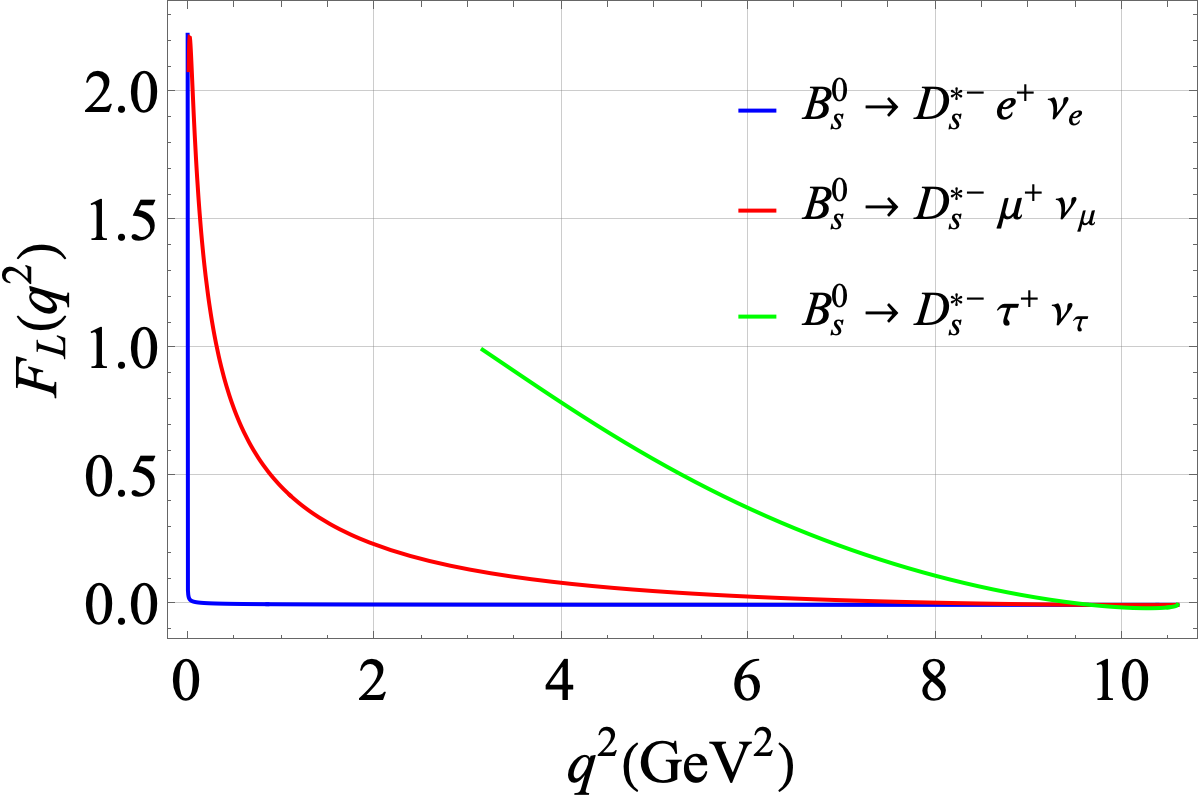}
	\caption{$B_s^0 \to D_s^{*-} \ell^+ \nu_{\ell}$} \label{f9b}
    \end{subfigure}
    \caption{The longitudinal polarization fraction of $B_s \to V \ell^+ \nu_{\ell}$ decays.} \label{f9}
\end{figure}
\begin{figure}
\centering
    \begin{subfigure}[b]{0.48\textwidth}
    \centering
    \includegraphics[width=\textwidth]{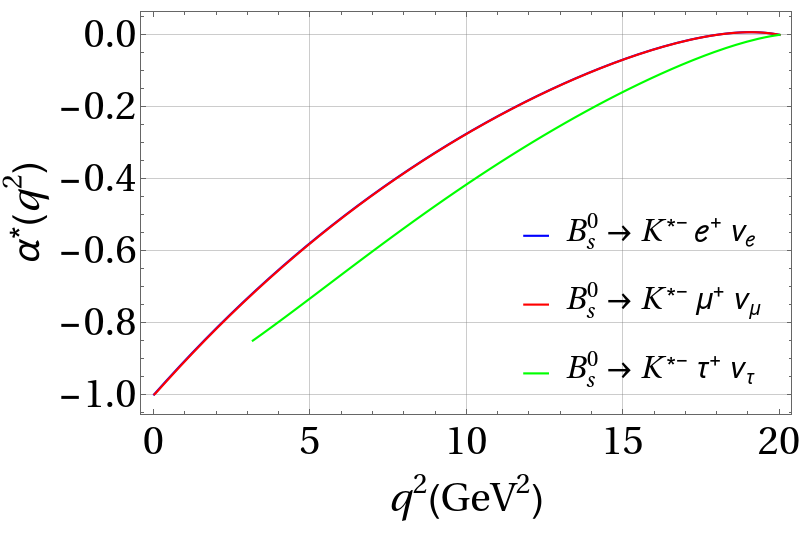}
    \caption{$B_s^0 \to K^{*-} \ell^+ \nu_{\ell}$} \label{f11a}
    \end{subfigure}
	\hfill
    \begin{subfigure}[b]{0.48\textwidth}
        \centering
        \includegraphics[width=\textwidth]{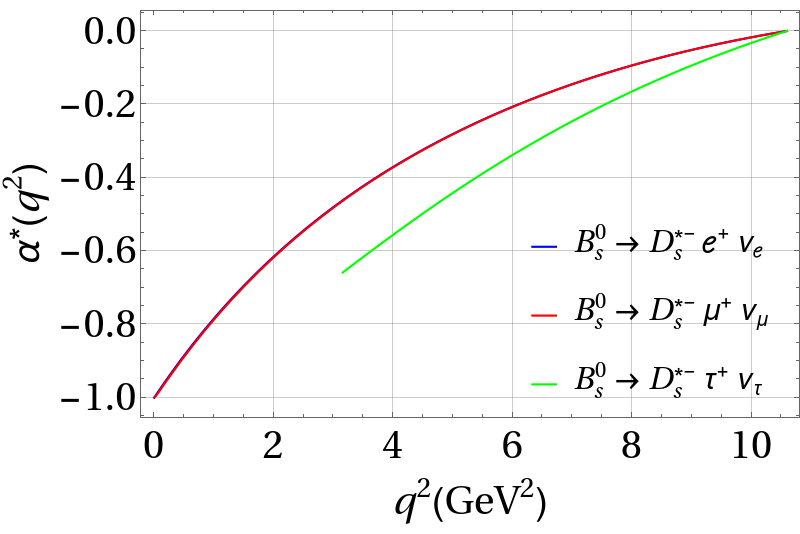}
	\caption{$B_s^0 \to D_s^{*-} \ell^+ \nu_{\ell}$} \label{f11b}
    \end{subfigure}
    \caption{The asymmetry parameter of $B_s \to V \ell^+ \nu_{\ell}$ decays.} \label{f11}
\end{figure}

\end{document}